# The Nilpotent Dirac Equation and its Applications in Particle Physics

## Peter Rowlands

*Department of Physics, University of Liverpool, Oliver Lodge Laboratory, Oxford Street, Liverpool, L69 7ZE, UK.  e-mail prowlandsl@liverpool.ac.uk*

*Abstract*. The nilpotent Dirac formalism has been shown, in previous publications, to generate new physical explanations for aspects of particle physics, with the additional possibility of calculating some of the parameters involved in the Standard Model. The applications so far obtained are summarised, with an outline of some more recent developments.

## 1 The nilpotent Dirac equation

Some aspects of particle physics are more easily understood if we first express the Dirac equation in a more algebraic form than usual, with the gamma matrices replaced by equivalent operators from vector and quaternion algebra.[1-7] Here, we define unit quaternion operators $(1, \boldsymbol{i}, \boldsymbol{j}, \boldsymbol{k})$ according to the usual rules:

$$\boldsymbol{i}^2 = \boldsymbol{j}^2 = \boldsymbol{k}^2 = \boldsymbol{ijk} = -1$$
$$\boldsymbol{ij} = -\boldsymbol{ji} = \boldsymbol{k} \; ; \; \boldsymbol{jk} = -\boldsymbol{kj} = \boldsymbol{i} \; ; \; \boldsymbol{ki} = -\boldsymbol{ik} = \boldsymbol{j} \; ,$$

and also *multivariate* 4-vector operators $(i, \mathbf{i}, \mathbf{j}, \mathbf{k})$, which are isomorphic to complex quaternions or Pauli matrices:

$$\mathbf{i}^2 = \mathbf{j}^2 = \mathbf{k}^2 = 1$$
$$\mathbf{ij} = -\mathbf{ji} = i\mathbf{k} \; ; \; \mathbf{jk} = -\mathbf{kj} = i\mathbf{i} \; ; \; \mathbf{ki} = -\mathbf{ik} = i\mathbf{j} \, .$$

The combination of these two sets of units produces a 32-part algebra (or group of order 64, taking into account both + and − signs), which can be directly related to that of the five $\gamma$ matrices, with mappings of the form:

$$\gamma^0 = -i\boldsymbol{i} \; ; \gamma^1 = \mathbf{i}\boldsymbol{k} \; ; \gamma^2 = \mathbf{j}\boldsymbol{k} \; ; \gamma^3 = \mathbf{k}\boldsymbol{k} \; ; \gamma^5 = i\boldsymbol{j} \, , \tag{1}$$

or, alternatively,

$$\gamma^0 = i\boldsymbol{k} \; ; \gamma^1 = \mathbf{i}\boldsymbol{i} \; ; \; \gamma^2 = \mathbf{j}\boldsymbol{i} \; ; \; \gamma^3 = \mathbf{k}\boldsymbol{i} \; ; \gamma^5 = i\boldsymbol{j} \, . \tag{2}$$

Applying (1) directly to the conventional form of the Dirac equation,

$$\left( \gamma^0 \frac{\partial}{\partial t} + \gamma^1 \frac{\partial}{\partial x} + \gamma^2 \frac{\partial}{\partial y} + \gamma^3 \frac{\partial}{\partial z} + im \right) \psi = 0 \; ,$$

we obtain:

$$\left(-i\boldsymbol{i}\frac{\partial}{\partial t} + k\mathbf{i}\frac{\partial}{\partial x} + k\mathbf{j}\frac{\partial}{\partial y} + k\mathbf{k}\frac{\partial}{\partial z} + im\right)\psi = 0 \quad.$$

Multiplying the equation from the left by $\boldsymbol{j}$ then alters the algebraic representation to (2) and the Dirac equation becomes:

$$\left(i k\frac{\partial}{\partial t} + \boldsymbol{i}\mathbf{i}\frac{\partial}{\partial x} + \boldsymbol{i}\mathbf{j}\frac{\partial}{\partial y} + \boldsymbol{i}\mathbf{k}\frac{\partial}{\partial z} + \boldsymbol{i}\boldsymbol{j}m\right)\psi = 0 \quad.$$

If we now apply a free-particle solution, such as

$$\psi = A\ e^{-i(Et-\mathbf{p}.\mathbf{r})} \quad.$$

to this equation, we find that:

$$(kE + \boldsymbol{i}\mathbf{i}p_x + \boldsymbol{i}\mathbf{j}p_y + \boldsymbol{i}\mathbf{k}p_x + \boldsymbol{i}\boldsymbol{j}\ m)\ A\ e^{-i(Et-\mathbf{p}.\mathbf{r})} = 0 \quad,$$

or, in a more compact form,

$$(kE + \boldsymbol{i}\mathbf{i}\ \mathbf{p} + \boldsymbol{i}\boldsymbol{j}\ m)\ A\ e^{-i(Et-\mathbf{p}.\mathbf{r})} = 0 \quad,$$

where $\mathbf{p}$ is a multivariate vector. The equation is only valid when $A$ is a multiple of $(kE + \boldsymbol{i}\mathbf{i}\ \mathbf{p} + \boldsymbol{i}\boldsymbol{j}\ m)$. In principle, this means that $A$, and hence $\psi$, must be a *nilpotent* or square root of zero. Here, of course, we rely on the fact, that, for a multivariate $\mathbf{p}$, the product $\mathbf{pp}$ becomes identical to the product of the scalar magnitudes $pp = p^2$. It is, additionally, identical to the product of the helicities $(\boldsymbol{\sigma}.\mathbf{p})\ (\boldsymbol{\sigma}.\mathbf{p})$, indicating that the multivariate vector (or equivalent Pauli matrix) representation of $\mathbf{p}$ automatically incorporates the concept of spin.

## 2 Solutions of the Dirac equation

Conventionally, the Dirac equation allows four solutions, corresponding to the four combinations of fermion and antifermion, and spin up and spin down, which may be arranged in a column vector, or Dirac 4-spinor. Here, we identify the solutions as produced by the combinations of $\pm E$, $\pm \mathbf{p}$ (or $\boldsymbol{\sigma}.\mathbf{p}$). In our notation, we could write these terms in the form:

$$\psi_1 = (kE + \boldsymbol{i}\mathbf{i}\ \mathbf{p} + \boldsymbol{i}\boldsymbol{j}\ m)\ e^{-i(Et-\mathbf{p}.\mathbf{r})}$$
$$\psi_2 = (kE - \boldsymbol{i}\mathbf{i}\ \mathbf{p} + \boldsymbol{i}\boldsymbol{j}\ m)\ e^{-i(Et+\mathbf{p}.\mathbf{r})}$$
$$\psi_3 = (-kE + \boldsymbol{i}\mathbf{i}\ \mathbf{p} + \boldsymbol{i}\boldsymbol{j}\ m)\ e^{i(Et-\mathbf{p}.\mathbf{r})}$$
$$\psi_4 = (-kE - \boldsymbol{i}\mathbf{i}\ \mathbf{p} + \boldsymbol{i}\boldsymbol{j}\ m)\ e^{i(Et+\mathbf{p}.\mathbf{r})} \quad,$$

and apply a single differential operator, but it is more useful to remove the variation in the signs of $E$ and $\mathbf{p}$ from the exponential term, by making the differential operator a 4-term row vector, which, in the equation, forms a scalar product with the Dirac 4-spinor. Incorporating all four terms into a single expression, we obtain

$$\left( \pm i\mathbf{k} \frac{\partial}{\partial t} \pm i\boldsymbol{\nabla} + ijm \right)(\pm \mathbf{k}E \pm i\mathbf{i}\ \mathbf{p} + ij\ m)\ e^{-i(Et - \mathbf{p.r})} = 0$$

as the new version of the Dirac equation for a free particle (though it can be shown that an equation of similar form also applies to bound states).[4] Reducing this to the eigenvalue form, and multiplying out, produces the classical relativistic momentum-energy conservation equation:

$$(\pm \mathbf{k}E \pm i\mathbf{i}\ \mathbf{p} + ij\ m)\ (\pm \mathbf{k}E \pm i\mathbf{i}\ \mathbf{p} + ij\ m) = E^2 - p^2 - m^2 = 0\ .$$

It is significant that there are exactly four solutions to the Dirac equation. Both quaternion and complex operators, of course, require an equal representation for + and − signs, suggesting eight possible sign combinations for $\pm \mathbf{k}E \pm i\mathbf{i}\ \mathbf{p} \pm ij\ m$; but only four of these will be independent, since the overall sign for the state vector is an arbitrary scalar factor. So the sign of one of $\mathbf{k}E$, $i\mathbf{i}\ \mathbf{p}$ or $ij\ m$ must behave as if fixed. With only $E$ and $\mathbf{p}$ terms represented in the exponential factor, it becomes evident that the fixed term must be $m$. Four solutions also result from the fact that the quaternionic structure of the state vector can be related to the conventional $4 \times 4$ matrix formulation using quaternionic matrices, and the conventional formulation is itself uniquely determined by the 4-D space-time signature of the equation, a $2n$-D space-time requiring a $2^n \times 2^n$ matrix representation of the Clifford algebra.[5]

In the case of quaternionic matrices, it is also significant that the (hidden) quaternion operators $i, j, k$ applied, along with 1, to the rows and columns, and also to the rows of the Dirac 4-spinor, are *identical* in meaning to the same operators applied to the terms in the nilpotent state vector, as one can be derived from the other.

## 3 Fermions and bosons

There are good reasons for believing that the nilpotent form of the Dirac equation is the most fundamental. It is automatically second quantized, fulfilling all the requirements of a quantum field theory; it removes the infrared divergence in the fermion propagator, and the divergent loop calculation for the self-energy of the non-interacting fermion; and it introduces supersymmetry as a mathematical operation without the need for additional particles. It also allows an easy calculation of parity states, and a simple method of introducing $C$, $P$ or $T$ transformations. In addition, state vectors for fermions, antifermions, bosons and baryons have immediately recognizable forms. A fermion, for example, may be represented by a row (or column) vector, whose components are four creation (or annihilation) operators:

$$(kE + ii\ \mathbf{p} + ij\ m) \quad \text{fermion spin up}$$
$$(kE - ii\ \mathbf{p} + ij\ m) \quad \text{fermion spin down}$$
$$(-kE + ii\ \mathbf{p} + ij\ m) \quad \text{antifermion spin up}$$
$$(-kE - ii\ \mathbf{p} + ij\ m) \quad \text{antifermion spin down}\ .$$

The antifermion then takes up the corresponding column (or row) vector:

$$(-kE + ii\ \mathbf{p} + ij\ m)$$
$$(-kE - ii\ \mathbf{p} + ij\ m)$$
$$(kE + ii\ \mathbf{p} + ij\ m)$$
$$(kE - ii\ \mathbf{p} + ij\ m)\ ,$$

while the spin 1 boson produced by their combination is simply the scalar product:

$$(kE + ii\ \mathbf{p} + ij\ m) \quad (-kE + ii\ \mathbf{p} + ij\ m)$$
$$(kE - ii\ \mathbf{p} + ij\ m) \quad (-kE - ii\ \mathbf{p} + ij\ m)$$
$$(-kE + ii\ \mathbf{p} + ij\ m) \quad (kE + ii\ \mathbf{p} + ij\ m)$$
$$(-kE - ii\ \mathbf{p} + ij\ m) \quad (kE - ii\ \mathbf{p} + ij\ m)\ .$$

The spin 0 boson is obtained by reversing the $\mathbf{p}$ signs in either fermion or antifermion:

$$(kE + ii\ \mathbf{p} + ij\ m) \quad (-kE - ii\ \mathbf{p} + ij\ m)$$
$$(kE - ii\ \mathbf{p} + ij\ m) \quad (-kE + ii\ \mathbf{p} + ij\ m)$$
$$(-kE + ii\ \mathbf{p} + ij\ m) \quad (kE - ii\ \mathbf{p} + ij\ m)$$
$$(-kE - ii\ \mathbf{p} + ij\ m) \quad (kE + ii\ \mathbf{p} + ij\ m)\ .$$

Significantly, massless spin 0 particles (Goldstone bosons) are ruled out on purely algebraic grounds:

$$(kE + ii\ \mathbf{p}) \quad (-kE - ii\ \mathbf{p}) = 0$$
$$(kE - ii\ \mathbf{p}) \quad (-kE + ii\ \mathbf{p}) = 0$$
$$(-kE + ii\ \mathbf{p}) \quad (kE - ii\ \mathbf{p}) \ = 0$$
$$(-kE - ii\ \mathbf{p}) \quad (kE + ii\ \mathbf{p}) \ = 0\ .$$

Massless spin 1 states, however, are allowed, since

$$(kE + ii\ \mathbf{p}) \quad (-kE + ii\ \mathbf{p})$$
$$(kE - ii\ \mathbf{p}) \quad (-kE - ii\ \mathbf{p})$$
$$(-kE + ii\ \mathbf{p}) \quad (kE + ii\ \mathbf{p})$$
$$(-kE - ii\ \mathbf{p}) \quad (kE - ii\ \mathbf{p})$$

has a nonzero scalar sum. Pauli exclusion is also automatic, since:

$$(kE + i\textbf{i}\ \textbf{p} + i\textbf{j}\ m) \quad (kE + i\textbf{i}\ \textbf{p} + i\textbf{j}\ m) \ = 0$$
$$(kE - i\textbf{i}\ \textbf{p} + i\textbf{j}\ m) \quad (kE - i\textbf{i}\ \textbf{p} + i\textbf{j}\ m) \ = 0$$
$$(-kE + i\textbf{i}\ \textbf{p} + i\textbf{j}\ m) \ (-kE + i\textbf{i}\ \textbf{p} + i\textbf{j}\ m) = 0$$
$$(-kE - i\textbf{i}\ \textbf{p} + i\textbf{j}\ m) \ (-kE - i\textbf{i}\ \textbf{p} + i\textbf{j}\ m) = 0 \ .$$

Baryon state vectors may be derived from the fact that we can produce a three-component non-zero structure of the form

$$(kE \pm i\textbf{i}\ p_x + i\textbf{j}\ m)\ (kE \pm i\textbf{i}\ p_y + i\textbf{j}\ m)\ (kE \pm i\textbf{i}\ p_z + i\textbf{j}\ m) \ ,$$

where, for convenience, we show only the first terms of the column and row vectors. We can here imagine **p** as having allowed phases in which only *one* of the three components of momentum, $p_x$, $p_y$, $p_z$, is nonzero and represents the total **p**. The products

$$(kE + i\textbf{j}\ m)\ (kE + i\textbf{j}\ m)\ (kE + i\textbf{i}\ \textbf{p} + i\textbf{j}\ m)$$
$$(kE + i\textbf{j}\ m)\ (kE - i\textbf{i}\ \textbf{p} + i\textbf{j}\ m)\ (kE + i\textbf{j}\ m)$$
$$(kE + i\textbf{i}\ \textbf{p} + i\textbf{j}\ m)\ (kE + i\textbf{j}\ m)\ (kE + i\textbf{j}\ m)$$

then become equivalent to $-p^2(kE + i\textbf{i}\ \textbf{p} + i\textbf{j}\ m)$, while

$$(kE + i\textbf{j}\ m)\ (kE + i\textbf{j}\ m)\ (kE - i\textbf{i}\ \textbf{p} + i\textbf{j}\ m)$$
$$(kE + i\textbf{j}\ m)\ (kE + i\textbf{i}\ \textbf{p} + i\textbf{j}\ m)\ (kE + i\textbf{j}\ m)$$
$$(kE - i\textbf{i}\ \textbf{p} + i\textbf{j}\ m)\ (kE + i\textbf{j}\ m)\ (kE + i\textbf{j}\ m)$$

result in $-p^2(kE - i\textbf{i}\ \textbf{p} + i\textbf{j}\ m)$. Choosing the labels *B*, *G* and *R* to represent the **p** variation within the brackets, with the $+ i\textbf{i}\ \textbf{p}$ phases representing a positive or cyclic combination of the three, and the $- i\textbf{i}\ \textbf{p}$ phases a negative or anticyclic combination, we can represent the total state vector, incorporating all six phases, as

$$\psi \sim (BGR - BRG + GRB - GBR + RBG - RGB) \ ,$$

with the mappings:

$$
\begin{array}{rcl}
BGR & \rightarrow & (kE + i\textbf{j}\ m)\ (kE + i\textbf{j}\ m)\ (kE + i\textbf{i}\ \textbf{p} + i\textbf{j}\ m) \\
- BRG & \rightarrow & (kE + i\textbf{j}\ m)\ (kE - i\textbf{i}\ \textbf{p} + i\textbf{j}\ m)\ (kE + i\textbf{j}\ m) \\
GRB & \rightarrow & (kE + i\textbf{j}\ m)\ (kE + i\textbf{i}\ \textbf{p} + i\textbf{j}\ m)\ (kE + i\textbf{j}\ m) \\
- GBR & \rightarrow & (kE + i\textbf{j}\ m)\ (kE + i\textbf{j}\ m)\ (kE - i\textbf{i}\ \textbf{p} + i\textbf{j}\ m) \\
RBG & \rightarrow & (kE + i\textbf{i}\ \textbf{p} + i\textbf{j}\ m)\ (kE + i\textbf{j}\ m)\ (kE + i\textbf{j}\ m) \\
- RGB & \rightarrow & (kE - i\textbf{i}\ \textbf{p} + i\textbf{j}\ m)\ (kE + i\textbf{j}\ m)\ (kE + i\textbf{j}\ m) \ .
\end{array}
$$

Perfect gauge invariance between these 'three quark' states requires an *SU*(3) symmetry. The same should apply where a bosonic ('quark-antiquark') state can be defined in terms of the same varying directional properties of its **p** operator.

## 4 *CPT* symmetry

*CPT* symmetry is another natural outcome of the nilpotent representation. Here, the *i*, *k*, and *j* operators can be applied to a nilpotent state vector to represent the respective *P*, *T*, and *C* transformations:

Parity (*P*):

$$i\,(kE + ii\,\mathbf{p} + ij\,m)\,i \;=\; (kE - ii\,\mathbf{p} + ij\,m)$$
$$i\,(kE - ii\,\mathbf{p} + ij\,m)\,i \;=\; (kE + ii\,\mathbf{p} + ij\,m)$$
$$i\,(-kE + ii\,\mathbf{p} + ij\,m)\,i \;=\; (-kE - ii\,\mathbf{p} + ij\,m)$$
$$i\,(-kE - ii\,\mathbf{p} + ij\,m)\,i \;=\; (-kE + ii\,\mathbf{p} + ij\,m)$$

Time reversal (*T*):

$$k\,(kE + ii\,\mathbf{p} + ij\,m)\,k \;=\; (-kE + ii\,\mathbf{p} + ij\,m)$$
$$k\,(kE - ii\,\mathbf{p} + ij\,m)\,k \;=\; (-kE - ii\,\mathbf{p} + ij\,m)$$
$$k\,(-kE + ii\,\mathbf{p} + ij\,m)\,k \;=\; (kE + ii\,\mathbf{p} + ij\,m)$$
$$k\,(-kE - ii\,\mathbf{p} + ij\,m)\,k \;=\; (kE - ii\,\mathbf{p} + ij\,m)$$

Charge conjugation (*C*):

$$-j\,(kE + ii\,\mathbf{p} + ij\,m)\,j \;=\; (-kE - ii\,\mathbf{p} + ij\,m)$$
$$-j\,(kE - ii\,\mathbf{p} + ij\,m)\,j \;=\; (-kE + ii\,\mathbf{p} + ij\,m)$$
$$-j\,(-kE + ii\,\mathbf{p} + ij\,m)\,j \;=\; (kE - ii\,\mathbf{p} + ij\,m)$$
$$-j\,(-kE - ii\,\mathbf{p} + ij\,m)\,j \;=\; (kE + ii\,\mathbf{p} + ij\,m)$$

From this, we see immediately that :

$CP = T$: $\quad -j\,(i\,(kE + ii\,\mathbf{p} + ij\,m)\,i)\,j = k\,(kE + ii\,\mathbf{p} + ij\,m)\,k = (-kE + ii\,\mathbf{p} + ij\,m)$

$PT = C$: $\quad i\,(k\,(kE + ii\,\mathbf{p} + ij\,m)\,k)\,i = -j\,(kE + ii\,\mathbf{p} + ij\,m)\,j = (-kE - ii\,\mathbf{p} + ij\,m)$

$TC = P$: $\quad k\,(-j\,(kE + ii\,\mathbf{p} + ij\,m)\,j)\,k = i\,(kE + ii\,\mathbf{p} + ij\,m)\,i = (kE - ii\,\mathbf{p} + ij\,m)$

and that $TCP \equiv$ identity, because:

$$k\,(-j\,(i\,(kE + ii\,\mathbf{p} + ij\,m)\,i)\,j)\,k = -kji\,(kE + ii\,\mathbf{p} + ij\,m)\,ijk = (kE + ii\,\mathbf{p} + ij\,m)\,.$$

Using this formalism for the transformations, the correct intrinsic parities of ground-state baryons and bosons are easily recovered.

It is of interest, in connection with the violation of symmetries that occurs in the weak interaction, that no fundamental process can tell, in principle, whether the symmetry violated, along with charge conjugation, is *P* or *T*. We can only tell whether the violation is of one or two of these symmetries. By convention, and because it is easier to measure, we assume that the first symmetry violated is $P$, but the result does not depend on any fundamental justification.

## 5 The vacuum operator

Since the vacuum plays such a significant part in many fundamental processes, the formulation of a vacuum operator (or vacuum operators) will be necessary to a full theory. Assuming an appropriate normalization, we may constructed a nilpotent vacuum operator as a diagonal matrix, which may be premultiplied by a 4-component quaternion row state vector or postmultiplied by a 4-component column quaternion state vector, representing a fermion state. In the first case, we write:

$$((kE + ii\mathbf{p} + ijm)\ (kE - ii\mathbf{p} + ijm)\ (-kE + ii\mathbf{p} + ijm)\ (-kE - ii\mathbf{p} + ijm)) \times$$

$$k \begin{pmatrix} kE + ii\mathbf{p} + ijm & 0 & 0 & 0 \\ 0 & kE - ii\mathbf{p} + ijm & 0 & 0 \\ 0 & 0 & -kE + ii\mathbf{p} + ijm & 0 \\ 0 & 0 & 0 & -kE - ii\mathbf{p} + ijm \end{pmatrix} e^{-i(Et\, \cdot\, \mathbf{p.r})}$$

$$= ((kE + ii\mathbf{p} + ijm)\ (kE - ii\mathbf{p} + ijm)\ (-kE + ii\mathbf{p} + ijm)\ (-kE - ii\mathbf{p} + ijm))\, e^{-i(Et\, \cdot\, \mathbf{p.r})}\ .$$

The vacuum operator, here, clearly leaves the original fermion state unchanged. The individual creation operators, or individual components of the row vector, ($\pm kE \pm ii\mathbf{p} + ijm$), which specify the complete fermion system, can be considered as being postmultiplied by $k$ ($\pm kE \pm ii\mathbf{p} + ijm$) to return to their original state, after normalization. The process can be continued indefinitely, with the fermion acting continually on the vacuum to reproduce itself:

$$(\pm kE \pm ii\mathbf{p} + ijm)\, k\, (\pm kE \pm ii\mathbf{p} + ijm)\, k\, (\pm kE \pm ii\mathbf{p} + ijm)\, k\, (\pm kE \pm ii\mathbf{p} + ijm) \ldots$$

However, $k$ ($\pm kE \pm ii\mathbf{p} + ijm$) $k$ is the same as the antistate to ($\pm kE \pm ii\mathbf{p} + ijm$), or ($\mp kE \pm ii\mathbf{p} + ijm$), making this equivalent to

$$(\pm kE \pm ii\mathbf{p} + ijm)\, (\mp kE \pm ii\mathbf{p} + ijm)\, (\pm kE \pm ii\mathbf{p} + ijm)\, (\mp kE \pm ii\mathbf{p} + ijm) \ldots$$

Physically, the fermion can be considered to see in the vacuum its 'image' or virtual antistate, producing a kind of virtual bosonic combination, and leading to an infinite alternating series of virtual fermions and bosons. Each real fermion state creates a virtual antifermion mirror image of itself in the vacuum, while each real antifermion state creates a virtual fermion mirror image of itself. The combined real and virtual particle creates a virtual boson state. Real fermions and real antifermions, of course, provide *real* mirror images of each other. The bosons, here, are assumed to be spin 1, created from a fermion-antifermion pair, with the same spin, but opposite helicities, like all known gauge bosons, but we could also imagine a vacuum of the form $\boldsymbol{j}$ ($\pm \boldsymbol{k}E \pm i i \mathbf{p} + i j m$), or $-\boldsymbol{j}$ ($\pm \boldsymbol{k}E \pm i i \mathbf{p} + i j m$), in which the bosonic state would be spin 0. (For reasons which will become apparent in section 8, premultiplication by $\boldsymbol{k}$ could be said to produce a 'weak' vacuum while premultiplication by $\boldsymbol{j}$ produces an 'electric' one.)

We can also consider the possible wavefunction for a spin 2 object, for example, a glueball:

$$(\pm \boldsymbol{k}E \pm i i \mathbf{p} + i j m)\,(\mp \boldsymbol{k}E \pm i i \mathbf{p} + i j m)\,(\pm \boldsymbol{k}E \pm i i \mathbf{p} + i j m)\,(\mp \boldsymbol{k}E \pm i i \mathbf{p} + i j m)\,.$$

A spin 0 glueball would be represented by:

$$(\pm \boldsymbol{k}E \pm i i \mathbf{p} + i j m)\,(\mp \boldsymbol{k}E \mp i i \mathbf{p} + i j m)\,(\pm \boldsymbol{k}E \pm i i \mathbf{p} + i j m)\,(\mp \boldsymbol{k}E \mp i i \mathbf{p} + i j m)\,,$$
or
$$(\pm \boldsymbol{k}E \mp i i \mathbf{p} + i j m)\,(\mp \boldsymbol{k}E \pm i i \mathbf{p} + i j m)\,(\pm \boldsymbol{k}E \mp i i \mathbf{p} + i j m)\,(\mp \boldsymbol{k}E \pm i i \mathbf{p} + i j m)\,,$$

which, significantly, cannot be massless. A gluon of spin 1 would take the form:

$$(\pm \boldsymbol{k}E \pm i i \mathbf{p})\,(\mp \boldsymbol{k}E \pm i i \mathbf{p})$$

which easily transforms to:

$$(\pm \boldsymbol{k}E \pm i i \mathbf{p})\,(\pm \boldsymbol{k}E \mp i i \mathbf{p}) \quad \text{or} \quad (\pm \boldsymbol{k}E \mp i i \mathbf{p})\,(\pm \boldsymbol{k}E \pm i i \mathbf{p})\,,$$

implying a 'strong, vacuum $\boldsymbol{i}$ ($\pm \boldsymbol{k}E \pm i i \mathbf{p}$), with

$$(\pm \boldsymbol{k}E \pm i i \mathbf{p})\,\boldsymbol{i}\,(\pm \boldsymbol{k}E \pm i i \mathbf{p})\,\boldsymbol{i}\,(\pm \boldsymbol{k}E \pm i i \mathbf{p})\,\boldsymbol{i}\,(\pm \boldsymbol{k}E \pm i i \mathbf{p})\,\boldsymbol{i}\,(\pm \boldsymbol{k}E \pm i i \mathbf{p}) \ldots$$

or $\quad (\pm \boldsymbol{k}E \pm i i \mathbf{p})\,(\pm \boldsymbol{k}E \mp i i \mathbf{p})\,(\pm \boldsymbol{k}E \pm i i \mathbf{p})\,(\pm \boldsymbol{k}E \mp i i \mathbf{p})\,(\pm \boldsymbol{k}E \pm i i \mathbf{p})\,(\pm \boldsymbol{k}E \mp i i \mathbf{p}) \ldots$

producing the interactions of an (assumed) massless quark with the gluon sea.

The fermion and antifermion state vectors are not only quantum field operators (removing the need for representation by quantum field integrals), but also supersymmetric operators, equivalent to $Q$ and $Q^{\dagger}$, respectively converting boson to fermion and fermion to boson, and each being the Hermitian conjugate (i.e. vacuum 'image') of the other. With this conception of vacuum, we can imagine a renormalization process, involving an infinite succession of boson and fermion loops cancelling each other out, without needing to generate a new set of extra supersymmetric partners or encountering a hierarchy problem. The formalism also produces a perturbation expansion for a first-order QED coupling with a state vector of the form:

$$\Psi_1 = -e \sum [kE + i\boldsymbol{i}\,\boldsymbol{\sigma}.(\mathbf{p} + \mathbf{k}) + i\boldsymbol{j}m]^{-1} \ (ik\phi \ - \boldsymbol{i}\,\boldsymbol{\sigma}.\mathbf{A}) \ (kE + i\boldsymbol{i}\boldsymbol{\sigma}.\mathbf{p} + i\boldsymbol{j}m) \ e^{-i(Et - (\mathbf{p} + \mathbf{k}).\mathbf{r})} \ ,$$

which automatically becomes 0 for a self-interacting electron, and similar cases.[8] Pure vacuum interactions in this formalism require no renormalization, although charge values vary with the strength of real interactions in the usual way, while a fermion propagator of the form

$$S_F(p) = \frac{1}{(kE + i\boldsymbol{i}\boldsymbol{\sigma}.\mathbf{p} + i\boldsymbol{j}m)} \ ,$$

eliminates any infrared divergence by having a denominator which conjugates to a non-zero scalar using its vacuum 'image'.

Of course, in some cases, the fermion (or antifermion) produces its 'image' in a real antifermionic (or fermionic) state.[5,9] This is the origin of the Berry phase, Jahn-Teller effect, Aharanov-Bohm effect, quantum Hall effect, and many other similar phenomena. A Bose-Einstein condensate in He$^4$ or Cooper pairing in a normal superconductor is a slightly different way of producing a bosonic-type state, as it is composed of a fermion-fermion pairing with opposite spins (total spin 0), as in ($\pm kE \pm i\boldsymbol{i}\mathbf{p} + i\boldsymbol{j}m$) ($\pm kE \mp i\boldsymbol{i}\mathbf{p} + i\boldsymbol{j}m$). The 'vacuum' equivalent for this would be $\boldsymbol{i}$ ($\pm kE \pm i\boldsymbol{i}\mathbf{p} + i\boldsymbol{j}m$) or $\boldsymbol{i}$ ($\pm kE \mp i\boldsymbol{i}\mathbf{p} + i\boldsymbol{j}m$). Another effect which is observed is the creation by a real boson of a fermionic equivalent.

Vacuum fermions and vacuum antifermions have a similar relationship to real fermions and real antifermions, although both states, in this case, are virtual. The mirror image states of all possible fermion states constitute the zero point energy of the vacuum. Each possible state provides a virtual vacuum energy of $\hbar\omega$ / 2, like the ground state of a harmonic oscillator (which, of course, it is). To create a real fermion state, we excite a virtual vacuum state of $-\hbar\omega$ / 2 up to the level $\hbar\omega$ / 2, using a total energy quantum of $\hbar\omega$. Counting real and virtual particles, we have the same number of fermions and antifermions in the universe, but, in a universe with a non-symmetric ground state (such as we will demonstrate must exist), fermions will be predominantly real and antifermions predominantly virtual; and, counting real and virtual particles, and assigning $+E$ to fermions and $-E$ to fermions, we obtain a total energy of zero.

| Real | Fermion | Antifermion |
|---|---|---|
| Vacuum | Antifermion | Fermion |

The existence of mirror image vacuum states for all fermionic particles accounts for the structure of the Dirac quaternion state vector. We incorporate both real and virtual components (interpreting the *zitterbewegung* as a switching between them). The four creation operators create both the real particle and its set of dual vacuum images. All fermion wavefunctions are, in this sense, single-valued, producing an effective combination analogous to a simultaneous consideration of the two sides of Newton's third law of motion or a virial doubling of the kinetic energy in a potential energy term. Fermion and antifermion state vectors thus have identical components;

only the *order* privileges either $+E$ or $-E$ states as the 'real' ones. (A similar principle applies to the spin states.)

The vacuum is really an expression of the continuous or noncountable nature of mass-energy ('mass', as the source of gravity). Continuity, as we will see, automatically makes mass-energy unidimensional and unipolar. Since it is also real, it is therefore restricted to a single mathematical sign, which is usually taken as positive. We can interpret this as implying a non-symmetric ground state or a filled vacuum. The filled vacuum for the ground state is that of negative energy or antifermions. In physical terms, it manifests itself in the Higgs field, which breaks charge conjugation symmetry for the weak interaction, and gives rest masses to the fermions and weak gauge bosons. (The reaction half of the system, in this case, is equivalent to what Newton called the 'impressed force' or the inertia.) It is also responsible for quantum mechanical nonlocality and the instantaneous transmission of the static gravitational force – though not the acceleration-dependent inertial or GTR component, or the inertial reaction force that we actually measure in systems with localised mass (and with which gravity is often confused).[8,10] Significantly, gravitational potential energy is often represented as negative.

## 6 The origin of the nilpotent formalism

The nilpotent formalism is so intrinsically powerful that it seems to imply a foundational status. It is, in fact, possible to relate this to a fundamental principle of duality which acts to preserve a zero totality in our description of nature.[11] That is, the process of setting up fundamental dualities in physics and mathematics can be used to *generate* the nilpotent form of the Dirac state vector. The full argument generates discrete mathematics from an undefined real number system via anticommuting dimensional structures, but it is convenient here to use an abbreviated version, in which the positive integers are taken to be the primary units.

Here, we begin with a unit integer (1) and then imagine finding an infinite series of 'duals' to this unit. We suppose that the dualling process must be carried out with respect to all previous duals, so that the entire set of characters generated becomes the new 'unit', and ensure that the total result is zero at every stage. The first dual then becomes $-1$, generating a new 'unit' consisting of $(1, -1)$. Following this, we have a series of terms to which we can give symbols such as $i_1, j_1$, etc. Each new 'unit' will double the number of terms in the immediately previous unit, so we will have:

order 2      $(1, -1)$
order 4      $(1, -1) \times (1, i_1)$
order 8      $(1, -1) \times (1, i_1) \times (1, j_1)$
order 16      $(1, -1) \times (1, i_1) \times (1, j_1) \times (1, i_2)$
order 32      $(1, -1) \times (1, i_1) \times (1, j_1) \times (1, i_2) \times (1, j_2)$
order 64      $(1, -1) \times (1, i_1) \times (1, j_1) \times (1, i_2) \times (1, j_2) \times (1, i_3)$, etc.

which becomes, when written out in full:

order 2     $\pm\,1$

order 4     $\pm\,1,\pm\,\boldsymbol{i}_1$

order 8     $\pm\,1,\pm\,\boldsymbol{i}_1,\pm\,\boldsymbol{j}_1,\pm\,\boldsymbol{i}\boldsymbol{j}_1$

order 16     $\pm\,1,\pm\,\boldsymbol{i}_1,\pm\,\boldsymbol{j}_1,\pm\,\boldsymbol{i}\boldsymbol{j}_1,\pm\,\boldsymbol{i}_2,\pm\,\boldsymbol{i}_2\boldsymbol{i}_1,\pm\,\boldsymbol{i}_2\boldsymbol{j}_1,\pm\,\boldsymbol{i}_2\boldsymbol{i}\boldsymbol{j}_1$

order 32     $\pm\,1,\pm\,\boldsymbol{i}_1,\pm\,\boldsymbol{j}_1,\pm\,\boldsymbol{i}\boldsymbol{j}_1,\pm\,\boldsymbol{i}_2,\pm\,\boldsymbol{i}_2\boldsymbol{i}_1,\pm\,\boldsymbol{i}_2\boldsymbol{j}_1,\pm\,\boldsymbol{i}_2\boldsymbol{i}\boldsymbol{j}_1,$

             $\pm\,\boldsymbol{j}_2,\pm\,\boldsymbol{j}_2\boldsymbol{i}_1,\pm\,\boldsymbol{j}_2\boldsymbol{j}_1,\pm\,\boldsymbol{j}_2\boldsymbol{i}\boldsymbol{j}_1,\pm\,\boldsymbol{j}_2\boldsymbol{i}_2,\pm\,\boldsymbol{j}_2\boldsymbol{i}_2\boldsymbol{i}_1,\pm\,\boldsymbol{j}_2\boldsymbol{i}_2\boldsymbol{j}_1,\pm\,\boldsymbol{j}_2\boldsymbol{i}_2\boldsymbol{i}\boldsymbol{j}_1$

order 64     $\pm\,1,\pm\,\boldsymbol{i}_1,\pm\,\boldsymbol{j}_1,\pm\,\boldsymbol{i}\boldsymbol{j}_1,\pm\,\boldsymbol{i}_2\boldsymbol{i}_1,\pm\,\boldsymbol{i}_2\boldsymbol{i}_1,\pm\,\boldsymbol{i}_2\boldsymbol{j}_1,\pm\,\boldsymbol{i}_2\boldsymbol{i}\boldsymbol{j}_1,$

             $\pm\,\boldsymbol{j}_2,\pm\,\boldsymbol{j}_2\boldsymbol{i}_1,\pm\,\boldsymbol{j}_2\boldsymbol{j}_1,\pm\,\boldsymbol{j}_2\boldsymbol{i}\boldsymbol{j}_1,\pm\,\boldsymbol{j}_2\boldsymbol{i}_2,\pm\,\boldsymbol{j}_2\boldsymbol{i}_2\boldsymbol{i}_1,\pm\,\boldsymbol{j}_2\boldsymbol{i}_2\boldsymbol{j}_1,\pm\,\boldsymbol{j}_2\boldsymbol{i}_2\boldsymbol{i}\boldsymbol{j}_1$

             $\pm\,\boldsymbol{i}_3,\pm\,\boldsymbol{i}_3\boldsymbol{i}_1,\pm\,\boldsymbol{i}_3\boldsymbol{j}_1,\pm\,\boldsymbol{i}_3\boldsymbol{i}\boldsymbol{j}_1,\pm\,\boldsymbol{i}_3\boldsymbol{i}_2,\pm\,\boldsymbol{i}_3\boldsymbol{i}_2\boldsymbol{i}_1,\pm\,\boldsymbol{i}_3\boldsymbol{i}_2\boldsymbol{j}_1,\pm\,\boldsymbol{i}_3\boldsymbol{i}_2\boldsymbol{i}\boldsymbol{j}_1,$

             $\pm\,\boldsymbol{i}_3\boldsymbol{j}_2,\pm\,\boldsymbol{i}_3\boldsymbol{j}_2\boldsymbol{i}_1,\pm\,\boldsymbol{i}_3\boldsymbol{j}_2\boldsymbol{j}_1,\pm\,\boldsymbol{i}_3\boldsymbol{j}_2\boldsymbol{i}\boldsymbol{j}_1,\pm\,\boldsymbol{i}_3\boldsymbol{j}_2\boldsymbol{i}_2,\pm\,\boldsymbol{i}_3\boldsymbol{j}_2\boldsymbol{i}_2\boldsymbol{i}_1,\pm\,\boldsymbol{i}_3\boldsymbol{j}_2\boldsymbol{i}_2\boldsymbol{j}_1,\pm\,\boldsymbol{i}_3\boldsymbol{j}_2\boldsymbol{i}_2\boldsymbol{i}\boldsymbol{j}_1$

To define these character sets as true 'units', we require that the product of any unit with itself, or with any subunit, generates only the unit. So, for example, at order 8, we will have the products:

$$(\pm\,1)\times(\pm\,1,\pm\,\boldsymbol{i}_1,\pm\,\boldsymbol{j}_1,\pm\,\boldsymbol{i}\boldsymbol{j}_1)=(\pm\,1,\pm\,\boldsymbol{i}_1,\pm\,\boldsymbol{j}_1,\pm\,\boldsymbol{i}\boldsymbol{j}_1)$$

$$(\pm\,\boldsymbol{i}_1)\times(\pm\,1,\pm\,\boldsymbol{i}_1,\pm\,\boldsymbol{j}_1,\pm\,\boldsymbol{i}\boldsymbol{j}_1)=(\pm\,1,\pm\,\boldsymbol{i}_1,\pm\,\boldsymbol{j}_1,\pm\,\boldsymbol{i}\boldsymbol{j}_1)$$

$$(\pm\,\boldsymbol{j}_1)\times(\pm\,1,\pm\,\boldsymbol{i}_1,\pm\,\boldsymbol{j}_1,\pm\,\boldsymbol{i}\boldsymbol{j}_1)=(\pm\,1,\pm\,\boldsymbol{i}_1,\pm\,\boldsymbol{j}_1,\pm\,\boldsymbol{i}\boldsymbol{j}_1)$$

$$(\pm\,\boldsymbol{i}\boldsymbol{j}_1)\times(\pm\,1,\pm\,\boldsymbol{i}_1,\pm\,\boldsymbol{j}_1,\pm\,\boldsymbol{i}\boldsymbol{j}_1)=(\pm\,1,\pm\,\boldsymbol{i}_1,\pm\,\boldsymbol{j}_1,\pm\,\boldsymbol{i}\boldsymbol{j}_1)$$

$$(\pm\,1,\pm\,\boldsymbol{i}_1)\times(\pm\,1,\pm\,\boldsymbol{i}_1,\pm\,\boldsymbol{j}_1,\pm\,\boldsymbol{i}\boldsymbol{j}_1)=(\pm\,1,\pm\,\boldsymbol{i}_1,\pm\,\boldsymbol{j}_1,\pm\,\boldsymbol{i}\boldsymbol{j}_1)$$

$$(\pm\,1,\pm\,\boldsymbol{j}_1)\times(\pm\,1,\pm\,\boldsymbol{i}_1,\pm\,\boldsymbol{j}_1,\pm\,\boldsymbol{i}\boldsymbol{j}_1)=(\pm\,1,\pm\,\boldsymbol{i}_1,\pm\,\boldsymbol{j}_1,\pm\,\boldsymbol{i}\boldsymbol{j}_1)$$

$$(\pm\,1,\pm\,\boldsymbol{i}_1,\pm\,\boldsymbol{j}_1)\times(\pm\,1,\pm\,\boldsymbol{i}_1,\pm\,\boldsymbol{j}_1,\pm\,\boldsymbol{i}\boldsymbol{j}_1)=(\pm\,1,\pm\,\boldsymbol{i}_1,\pm\,\boldsymbol{j}_1,\pm\,\boldsymbol{i}\boldsymbol{j}_1)$$

$$(\pm\,1,\pm\,\boldsymbol{i}_1,\pm\,\boldsymbol{j}_1,\pm\,\boldsymbol{i}\boldsymbol{j}_1)\times(\pm\,1,\pm\,\boldsymbol{i}_1,\pm\,\boldsymbol{j}_1,\pm\,\boldsymbol{i}\boldsymbol{j}_1)=(\pm\,1,\pm\,\boldsymbol{i}_1,\pm\,\boldsymbol{j}_1,\pm\,\boldsymbol{i}\boldsymbol{j}_1),\ \text{etc.}$$

For this to be always true, the terms $\boldsymbol{i}_1,\boldsymbol{j}_1,\boldsymbol{i}_2,\boldsymbol{j}_2,\boldsymbol{i}_3,\boldsymbol{j}_3$, etc. are required to have the properties of imaginary units, or square roots of –1, while the products, such as $\boldsymbol{i}\boldsymbol{j}_1$, must be imaginary or real units, that is, square roots of either –1 or +1. The two possibilities lead to entirely different consequences, for we can generate an unlimited number of complex products which are square roots of 1, but, for any complex number, such as $\boldsymbol{i}_1$, there is *only a single complex product* of the form $\boldsymbol{i}\boldsymbol{j}_1$, which is itself complex. So, if $\boldsymbol{i}\boldsymbol{j}_1$ is complex, then $\boldsymbol{i}_1$, $\boldsymbol{j}_1$, and $\boldsymbol{i}\boldsymbol{j}_1$ form a *closed system* – equivalent to the cyclic quaternion system $\boldsymbol{i},\boldsymbol{j},\boldsymbol{k}$, we have used in generating the Dirac algebra. The choice is arbitrary, but, if we choose the first option as default, we generate an infinite number of identically structured closed systems. In the more general case, where we begin with an undefined $\mathfrak{R}$ rather than an integer unit, and proceed to terms of the form $\mathbb{C}, \mathbb{C}', \mathbb{C}''$, and orders 2, 4, 8, 16, such as

$\mathfrak{R}, -\mathfrak{R}$

$\mathfrak{R}, -\mathfrak{R}, \mathbb{C}, -\mathbb{C}$

$\mathfrak{R}, -\mathfrak{R}, \mathbb{C}, -\mathbb{C}, \mathbb{C}', -\mathbb{C}', \mathbb{C}\mathbb{C}', -\mathbb{C}\mathbb{C}'$

$\mathfrak{R}, -\mathfrak{R}, \mathbb{C}, -\mathbb{C}, \mathbb{C}', -\mathbb{C}', \mathbb{C}\mathbb{C}', -\mathbb{C}\mathbb{C}', \mathbb{C}'', -\mathbb{C}'', \mathbb{C}\mathbb{C}'', -\mathbb{C}\mathbb{C}'', \mathbb{C}'\mathbb{C}'', -\mathbb{C}'\mathbb{C}'', \mathbb{C}\mathbb{C}'\mathbb{C}'', -\mathbb{C}\mathbb{C}'\mathbb{C}'',$

the creation of an infinite series of identical closed systems also effectively creates a system of discrete mathematics from an unspecified 'real' or 'continuous' one.

In generating the quaternion system, through a 'natural' concept of duality, we also generate a 'natural' concept of (3-)dimensionality linked to discreteness. If we take $i_1, j_1$, and $i_1 j_1$ as a quaternion system ($i, j, k$), then any further complexification, to produce, say, $i_2 i_1$, $i_2 j_1$, and $i_2 i_1 j_1$, will produce a system equivalent to the multivariate vectors, complexified quaternions, or Pauli matrices ($\mathbf{i}, \mathbf{j}, \mathbf{k}$), which were the other component in our Dirac algebra. The processes of *complexification* and *dimensionalization*, with their respective open and closed algebras, become simply alternative forms of duality, along with *conjugation*, or the introduction of alternative signs, + and –. The three processes, taken together, and repeated indefinitely, provide the entire structure of mathematical duality required for physical application. Since the repeated application of conjugation makes no change to the structure, the series follows the pattern:

| | | |
|---|---|---|
| order 2 | conjugation | $\times (1, -1)$ |
| order 4 | complexification | $\times (1, i_1)$ |
| order 8 | dimensionalization | $\times (1, j_1)$ |
| order 16 | complexification | $\times (1, i_2)$ |
| order 32 | dimensionalization | $\times (1, j_2)$ |
| order 64 | complexification | $\times (1, i_3)$ |

and the mathematical structures generated in this way become:

| | |
|---|---|
| order 2 | real scalar |
| order 4 | complex scalar (real plus pseudoscalar) |
| order 8 | quaternions |
| order 16 | complex quaternions or multivariate 4-vectors |
| order 32 | double quaternions |
| order 64 | complex double quaternions or multivariate vector quaternions, etc. |

The point at which the extension of the sequence becomes one of repetition is at order 16, and so a complete specification of an interative generating procedure could be made by using the groups of order 2, 4, 8 and 16, which respectively introduce the real scalar, pseudoscalar, quaternion and multivariate vector groups, with units $\pm 1$, $\pm i$, $\pm i$, $\pm j$, $\pm k$, $\pm \mathbf{i}$, $\pm \mathbf{j}$, $\pm \mathbf{k}$. If these are taken as independent entities, then the simplest order group which combines them all is that of order 64, which is the group required by the Dirac nilpotent state vector. In principle, the algebra of the Dirac state vector is constituted as a minimal mathematical 'unit' required to generate the iterative procedure corresponding to an infinite process of dualling. It is significant that the state vector, being nilpotent, is also automatically *self*-dual.

# 7 Duality and the fundamental parameters of physics

If duality were a purely mathematical concept, then it would have no direct relevance to a discussion of the fundamentals of physics. However, duality in mathematics seems also to be reflected in duality in physics, and this principle enables us to find a deeper understanding of the symmetries that lie behind the construction of the Dirac equation and the concepts of particle physics. It has been argued in previous papers that the most fundamental fact in physics is the existence of the four parameters mass, time, charge, and space, which incorporate all the aspects of conservation and nonconservation through which the laws of physics are defined.[12-16] It has also been proposed that the parameters have characteristics which suggest that there is a fundamental group symmetry between them:

| | | | |
|---|---|---|---|
| **space** | nonconserved | real | dimensional / discrete |
| **time** | nonconserved | imaginary | nondimensional / continuous |
| **mass** | conserved | real | nondimensional /continuous |
| **charge** | conserved | imaginary | dimensional / discrete |

Here, charge is used in the generic sense to represent the sources of electric, strong and weak interactions, which form the equivalent of its three 'dimensions'. Ideally, we should expect these to be indistinguishable in type (though, because charge is a conserved quantity, they would remain inconvertible into each other). However, it will be argued that the creation of the Dirac state necessarily breaks the symmetry between them. The dualities here are absolute: properties and 'antiproperties' are completely opposed in every sense. Nonconservation, for example, which includes such effects as gauge invariance and translation and rotation symmetries, means the absolute opposite of conservation, and is equally local in character. Dimensional quantities are also always discrete, just as nondimensional ones are always continuous. That space is discrete (because dimensional) accounts for its representation in terms of the Robinson or Löwenheim-Skolem system of countable reals, rather than the Cantor system, which applies to mass (in the form of energy). Of course, because space is nonconserved, its dimensions and its units would remain unfixed, unlike those of charge. The real / imaginary distinction between mass and charge reflects the distinction in sign for the forces between identical masses and identical charges. The nondimensional or continuous nature of time is manifested in terms of its irreversibility (though the imaginary representation allows changes in its mathematical sign). The same property, applied to mass(-energy), is crucial to the Higgs mechanism, as it implies a continuum of mass-energy, or filled vacuum.

Applying the mathematical dualling processes described in the previous section, we can see that each of the processes has equivalent representation in the parameter group:

| | | | |
|---|---|---|---|
| **space** | nonconjugated | real | dimensional |
| **time** | nonconjugated | complexified | nondimensional |
| **mass** | conjugated | real | nondimensional |
| **charge** | conjugated | complexified | dimensional |

Thus, a conserved quantity could also be taken as 'conjugated', because this would imply that we can only create / destroy positive conserved quantities if we also simultaneously create / destroy negative ones. It can be seen, in addition, that the parameters not only encode the three processes involved in mathematical dualling on an equal basis, but also represent stages in the emergent algebra that it creates:

| | | | |
|---|---|---|---|
| order 2 | real scalar | $1$ | mass |
| order 4 | pseudoscalar | $i$ | time |
| order 8 | quaternions | $i, j, k$ | charge |
| order 16 | multivariate vectors | $\mathbf{i}, \mathbf{j}, \mathbf{k}$ | space |

What becomes immediately apparent here is that, if we put these four mathematical structures together in a single algebra, they constitute the complex double quaternion or multivariate vector quaternion algebra, which occurs at order 64, and which, once again, is the algebra required to represent the Dirac state. The Dirac state, as we will show, is, in fact, the combined state of space, time, mass and charge.

## 8 The creation of the Dirac state

The Dirac algebra is a group of order 64. It is an algebra of 32 parts, each with + and – values, and it is formed from 8 basic units: $1$, $i$, $\mathbf{i}$, $\mathbf{j}$, $\mathbf{k}$, $i, j, k$. The various combinations of these units generate the entire set of 32 parts, which is made up of 2 complex scalars, 6 complex vectors, 6 complex quaternions, and 18 complex vector quaternions. However, 32 parts can also be derived from the binomial combinations of 5 quantities, so we can also generate the entire structure from a *pentad* set, equivalent to the gamma matrices, such as $i\mathbf{k}$; $\mathbf{i}i$; $\mathbf{j}i$; $\mathbf{k}i$; $j$. In effect, generating the algebra from 5 units could be taken as 'simpler' and more efficient than generating it from 8 units, and so the compactified composite set could be taken, in some senses, as more mathematically 'fundamental' than the original basic set.

There are many ways of constructing a pentad to generate the Dirac algebra, but all involve taking the components of one of the two 3-dimensional parameters (space or charge) and superimposing one on the units of each of the 3 other parameters. Physically, of course, if the basic units really do represent those of space, time, mass and charge, the new, composite units must represent entirely new physical parameters, produced by the combinations, and, if we choose to perform the 'compactification' using the units of charge (as being more convenient than using those of space) we will create composite units that incorporate the properties that are characteristic of charge:

conservation and discrete quantization. Diagrammatically, we can represent the process in the following terms:

| *Time* | *Space* | *Mass* | *Charge* |
|--------|---------|--------|----------|
| $i$ | **i  j  k** | 1 | *i  j  k* |

*Superimposing the units of charge*

| | | | |
|--------|---------|--------|----------|
| *k* | *i* | *j* | |

*produces*

| | | | |
|--------|---------|--------|----------|
| *ik* | *i***i**  *i***j**  *i***k** | *j* | |

*and creates*

| | | | |
|--------|---------|--------|----------|
| *E* | **p** | *m* | |

| *Dirac Energy* | *Dirac Momentum* | *Dirac Rest Mass* |
|----------------|------------------|-------------------|

If we multiply by an extra *i* for operational convenience, we obtain

| | | |
|--------|---------|--------|
| *k* | *ii***i**  *ii***j**  *ii***k** | *ij* |

The new composite quantities produced by the application of the conserved and quantized units of charge to the parameters time, space and mass naturally combine the characteristics of their parent quantities. The charge input makes them all conserved and quantized, the act of imposing charge's three-dimensional structure onto the original time, space and mass being identical to the act of quantization; but the Dirac energy (*E*), the Dirac momentum (**p**) and the Dirac rest mass (*m*) also retain the respective pseudoscalar, multivariate vector, and real scalar properties of time, space and mass. (Another conserved and quantized quantity, the Dirac angular momentum, relates to the directional properties of the vector term, and, in some sense, to the Dirac state as a whole.) The combination, however, has another important physical consequence, as the quaternion units, *i*, *j*, *k*, are changed from being symmetrical and indistinguishable representations of independent charges into composite units whose symmetry is broken; and, from the composition of *ik*, the combined (*i***i**, *i***j**, *i***k**), and *j*, it is possible to derive the respective $SU(2)$, $SU(3)$ and $U(1)$ symmetries associated with the weak, strong and electric charges.

Significantly, the three components *E*, **p**, and *m*, of the Dirac state, which we represent in the form ($\pm$ *k**E* $\pm$ *ii***p** + *ij**m*) or ($\pm$ *ik**E* $\pm$ *i***p** + *j**m*), are, from the fundamental properties of their parent-parameters time, space, and mass(-energy), specified by unrestricted real number values (though space's are countable in the Löwenheim-Skolem sense). Thus, it is possible, using the anticommuting properties of the quaternion and vector operators, and the presence of at least one complex term, to find values of the state, which square to a *zero numerical solution*. These, in turn, become the units of an infinite higher algebra (Hilbert space), which provides the basic parameterisation that we describe as physics. In terms of the individual Dirac state, the process of creating a conserved state is paralleled by a description in terms of the equivalent process of *nonconservation*. This is the meaning of the Dirac equation, where the most absolute way of specifying nonconservation, in the idealised free state, is via the expression $e^{-i(Et - \mathbf{p.r.})}$, to which we then apply a nonconservation or *differential* operator of the form $(\pm \, i k \partial / \partial t \pm i \nabla + jm)$ such that the *eigenvalue* or result becomes identical to the expression for the Dirac state $(\pm \, i k E \pm i \mathbf{p} + jm)$. The exponential term is a mathematical represention of maximal variation (or 'nonconservation') for space and time coordinates in the free particle state. For a non-free (or interacting) state, the functional expression would, of course, be different but the eigenvalue format would be the same, so that

$$(\pm \, i k E \pm i \mathbf{p} + jm) \, (\pm \, i k E \pm i \mathbf{p} + j \, m) = E^2 - p^2 - m^2 = 0$$

would always be true.

It is of deep significance here, however, that the application of quaternion operators in an expression such as $(\pm \, i k E \pm i \mathbf{p} + j \, m)$ does not in itself create the Dirac state – the same algebraic expression could have been used in a purely mathematical factorization of the classical special relativistic energy-momentum expression. It is the act of *equating of these operators to the three fundamental charge units*, with their properties of quantization and conservation, that creates the Dirac state by restructuring the meaning of the terms to which they are applied as quantized and conserved ones. The same act also establishes direct and inverse numerical relationships between the units $E$ and $\mathbf{p}$, and between those of $t$ and $\mathbf{r}$, leading to the introduction of the constants $\hbar$ and $c$, and the equations of special relativity. A third constant, $G$, is required when we involve $m$. These constants, as has long been known, have no intrinsic meaning; they are simply the inevitable consequence of creating a composite state.

The connection between the quaternionic operators applied to charge and the (hidden) ones used in the Dirac 4-spinor now gives us a new understanding of the physical meaning of charge as a vacuum generator: $i, j$, and $k$ are, simultaneously, the respective operators applied to strong, electric and weak charges, and also the creators of the strong, electric and weak vacuum images of a real fermion (which itself may be presumed to be 'generated' by the 'mass' operator, 1). Charge is, in effect, a kind of vacuum state, linked to the quantum field nature of the state vector.

The Dirac equation in this quantum field form now becomes the most fundamental equation in physics, incorporating in compactified form all the conservation and nonconservation principles which make up classical and quantum physics. All other physical principles are in some sense defined in relation to it, and can be discovered through exploring its many consequences. Through the equation, for example, $E$ and $t$, and $\mathbf{p}$ and $\mathbf{r}$, become conjugate variables, that is, ones which exchange statements about conservation into equivalent statements about nonconservation, and vice versa. In addition, the existence of four solutions becomes an obvious consequence of the derivation of the Dirac terms $E$, $\mathbf{p}$ and $m$ from the original parameters time, mass and space. The two signs for the $E$ term derive from the two signs for the imaginary time parameter, while the two signs for $\mathbf{p}$ result from its dimensionality and countability. However, this makes the negative versions of $E$ fundamentally different in character from the negative versions of $\mathbf{p}$, being 'mathematical' rather than 'physical' in origin. Thus, negative $\mathbf{p}$ terms are of equal status to positive ones, but negative $E$ terms, like negative time, do not exist in the ground state of the universe. The $m$ term, in addition, remains positive in all cases because the parent parameter, mass, is unipolar. These distinctions are significant in understanding such fundamental physical processes as the Higgs mechanism.

## 9 $SU$(3)

The vector nature of the $\mathbf{p}$ term in the Dirac nilpotent state vector produces a natural $SU$(3) symmetry for the strong interaction, as is evident from the possible phases of the baryon state:

$$(\mathbf{k}E \pm \mathbf{ii}\, p_x + \mathbf{ij}\, m)\, (\mathbf{k}E \pm \mathbf{ii}\, p_y + \mathbf{ij}\, m)\, (\mathbf{k}E \pm \mathbf{ii}\, p_z + \mathbf{ij}\, m)\,.$$

The $SU$(3) symmetry then becomes simply a straightforward expression of perfect gauge invariance between all the possible phases. Gauge invariance is really an expression of the nonconservation codified within a differential operator, and the conventional way of defining this is via a covariant derivative, which, for an $SU$(3) symmetry, takes the form:

$$\partial_\mu \to \partial_\mu + i g_s \frac{\lambda^\alpha}{2} A^{\alpha\mu}(x)\,,$$

or, in terms of the component coordinates:

$$ip_1 = \partial_1 \to \partial_1 + i g_s \frac{\lambda^\alpha}{2} A^{\alpha 1}(x)$$

$$ip_2 = \partial_2 \to \partial_2 + i g_s \frac{\lambda^\alpha}{2} A^{\alpha 2}(x)$$

$$ip_3 = \partial_3 \to \partial_3 + i g_s \frac{\lambda^\alpha}{2} A^{\alpha 3}(x)$$

$$E = i\partial_0 \to i\partial_0 - g_s \frac{\lambda^\alpha}{2} A^{\alpha 0}(x)\,.$$

For an $SU$(3) structure, we require a field with eight generators.

If we insert the coordinate expressions into the differential form of the baryon state vector, we obtain:

$$\left( \boldsymbol{k} \left( E - g_s \frac{\lambda^\alpha}{2} A^{\alpha 0} \right) \pm \boldsymbol{i} \left( \partial_1 + i g_s \frac{\lambda^\alpha}{2} A^{\alpha 1} \right) + \boldsymbol{ij}\, m \right)$$

$$\left( \boldsymbol{k} \left( E - g_s \frac{\lambda^\alpha}{2} A^{\alpha 0} \right) \pm \boldsymbol{i} \left( \partial_2 + i g_s \frac{\lambda^\alpha}{2} A^{\alpha 2} \right) + \boldsymbol{ij}\, m \right)$$

$$\left( \boldsymbol{k} \left( E - g_s \frac{\lambda^\alpha}{2} A^{\alpha 0} \right) \pm \boldsymbol{i} \left( \partial_3 + i g_s \frac{\lambda^\alpha}{2} A^{\alpha 3} \right) + \boldsymbol{ij}\, m \right).$$

The possible phases then become:

$$\left( \boldsymbol{k} \left( E - g_s \frac{\lambda^\alpha}{2} A^{\alpha 0} \right) \pm \boldsymbol{i} \left( \partial_1 + i g_s \frac{\lambda^\alpha}{2} \mathbf{A}^\alpha \right) + \boldsymbol{ij}\, m \right) \left( \boldsymbol{k} \left( E - g_s \frac{\lambda^\alpha}{2} A^{\alpha 0} \right) + \boldsymbol{ij}\, m \right) \left( \boldsymbol{k} \left( E - g_s \frac{\lambda^\alpha}{2} A^{\alpha 0} \right) + \boldsymbol{ij}\, m \right)$$

$$\left( \boldsymbol{k} \left( E - g_s \frac{\lambda^\alpha}{2} A^{\alpha 0} \right) + \boldsymbol{ij}\, m \right) \left( \boldsymbol{k} \left( E - g_s \frac{\lambda^\alpha}{2} A^{\alpha 0} \right) \pm \boldsymbol{i} \left( \partial_1 + i g_s \frac{\lambda^\alpha}{2} \mathbf{A}^\alpha \right) + \boldsymbol{ij}\, m \right) \left( \boldsymbol{k} \left( E - g_s \frac{\lambda^\alpha}{2} A^{\alpha 0} \right) + \boldsymbol{ij}\, m \right)$$

$$\left( \boldsymbol{k} \left( E - g_s \frac{\lambda^\alpha}{2} A^{\alpha 0} \right) + \boldsymbol{ij}\, m \right) \left( \boldsymbol{k} \left( E - g_s \frac{\lambda^\alpha}{2} A^{\alpha 0} \right) + \boldsymbol{ij}\, m \right) \left( \boldsymbol{k} \left( E - g_s \frac{\lambda^\alpha}{2} A^{\alpha 0} \right) \pm \boldsymbol{i} \left( \partial_1 + i g_s \frac{\lambda^\alpha}{2} \mathbf{A}^\alpha \right) + \boldsymbol{ij}\, m \right)$$

in parallel to the six forms incorporated in

$$\psi \sim (BGR - BRG + GRB - GBR + RBG - RGB)\, .$$

Conventionally, we describe three quark 'colours' ($R$, $G$, $B$), which are as inseparable as the three dimensions of space. Though all 'phases' of the interaction are, of course, equally probable, and present at the same time, we can imagine arbitrarily isolating one phase as the carrier of the 'colour' component of the interaction ($i g_s\, \lambda^\alpha\, \mathbf{A}^\alpha / 2$), or, alternatively, the strong charge ($s$); and then picture this as being 'transferred', at a constant rate, to create the next phase, along with the spin or **p** term. The 'current' effecting the 'transfer' of strong charge or 'colour' field will then be carried by the eight generators of the strong field, or 'gluons'; and the 'transfer' will be, simultaneously, an expression of the conservation of the directional aspect of angular momentum. Deriving entirely from the nilpotent structure of the baryon state vector, the interaction will necessarily be nonlocal, and the constant rate of momentum 'transfer', will be equivalent to a force which does not depend on the physical separation of the components. Such a force, requires, in mathematical terms, a potential which is linear with distance, though, as we will see, an additional Coulomb component is needed for spherical symmetry. In addition, exactly the same structure of 'colour' phases and interaction should apply even when the bound state is a bosonic state composed of quark and antiquark, that is, a meson, rather than a three-quark baryon.

## 10 An analytical derivation of the quark-antiquark and three-quark interactions

With the assumption of a linear potential for the strong interaction, we can immediately use the nilpotent form of the Dirac equation to obtain an analytical solution for both the quark-antiquark and three-quark potentials, which predicts both infrared slavery and asymptotic freedom. Beginning with the idea of a linear potential for the strong interaction, let us suppose that the quark-antiquark potential in the bound meson state of the form:

$$V = \sigma r + D \ ,$$

where $D$ is a function of $r$, yet to be determined. With a strong or colour charge for the quark of strength $q$ ($= \sqrt{\alpha_s}$), this is equivalent to a potential energy

$$W = -q\sigma r - qD$$

for the quark-antiquark interaction. Let us assume that, if $D$ contains any constant term ($C$), its effect will be merely to shift the value of $E$ to $E' = E - qC$. It will be convenient to refer to this, simply, as $E$.

To determine the effect of this potential, we now construct the appropriate form of the Dirac equation. Assuming spherical symmetry, it is convenient, to choose a form of the $\mathbf{p}$ operator in which helicity ($\boldsymbol{\sigma}.\mathbf{p}$) is explicit and $\nabla$ becomes an ordinary vector; $\boldsymbol{\sigma}.\nabla$ can then be expressed as a function of $r$ in polar coordinates, with the explicit addition of the angular momentum term which would be required using a multivariate form of $\nabla$. That is,

$$\boldsymbol{\sigma}.\nabla = \left( \frac{\partial}{\partial r} + \frac{1}{r} \right) \pm i \frac{j + \frac{1}{2}}{r}$$

Assuming constant total energy, we can use this expression to construct a nilpotent differential operator of the form:

$$\pm \boldsymbol{k}(E - q\sigma r - qD) \pm \boldsymbol{i}\left( \frac{\partial}{\partial r} + \frac{1}{r} \pm i \frac{j + \frac{1}{2}}{r} \right) + \boldsymbol{ij}m \ .$$

We now need to identify the functional term to which this operator applies. We suppose (on the basis of parallel calculations for the Coulomb potential) that it is of the form:

$$\psi = \exp\left(-ar - br^2\right) r^\gamma \sum_{\nu = 0} a_\nu r^\nu \ ,$$

and consider the ground state (with $\nu = 0$) over the four Dirac solutions. The four-part nilpotent state vector defines the condition:

$$4(E - q\sigma r - qD)^2 = -2\left( \frac{\partial}{\partial r} + \frac{1}{r} + i \frac{j + \frac{1}{2}}{r} \right)^2 - 2\left( \frac{\partial}{\partial r} + \frac{1}{r} - i \frac{j + \frac{1}{2}}{r} \right)^2 + 4m^2$$

for all solutions, from which it becomes clear that – $qD$ must be a scalar phase or Coulomb term of the form $qA \, / \, r$, as would be expected from spherical symmetry or equality in all directions.

Incorporating this term, applying $\psi$ and expanding, we obtain:

$$E^2 - 2q^2A\sigma + \frac{q^2A^2}{r^2} + q^2\sigma^2 r^2 + \frac{2qA}{r}E - 2q\sigma Er \ = m^2 -$$

$$\left( a^2 + \frac{(\gamma + \nu \ldots + 1)^2}{r^2} - \frac{(j + \tfrac{1}{2})^2}{r^2} + 4b^2r^2 + 4abr - 4b(\gamma + \nu \ldots + 1) - \frac{2a}{r}(\gamma + \nu \ldots + 1) \right).$$

With the positive and negative $i(j + \tfrac{1}{2})$ terms cancelling out over the four solutions, and, assuming a termination in the power series, we can equate:

(1) coefficients of $r^2$:
$$q^2\sigma^2 = -4b^2$$

(2) coefficients of $r$:
$$-2q\sigma E = -4ab$$

(3) coefficients of $1 \, / \, r$:
$$2qAE = 2a\,(\gamma + \nu + 1)$$

(4) coefficients of $1 \, / \, r^2$:
$$q^2A^2 = -\,(\gamma + \nu + 1)^2 + (j + \tfrac{1}{2})^2$$

(5) constant terms:
$$E^2 - 2q^2A\sigma = -a^2 + 4b\,(\gamma + \nu + 1) + m^2$$

The first three equations immediately lead to:

$$b = \pm \frac{iq\sigma}{2}$$

$$a = \mp\, iE$$

$$\gamma + \nu + 1 = \pm\, iqA \quad.$$

The case where $\nu = 0$ then requires a state vector with functional component

$$\psi = \ \exp\,(\mp\, iEr \pm iq\sigma\, r^2/2)\ r^{\pm\, iqA - 1}\quad.$$

The imaginary exponential terms in $\psi$ can be seen as representing asymptotic freedom, the $\exp\,(\mp\, iEr)$ being typical for a free fermion. The complex $r^{\gamma-1}$ term can be written as a phase, $\phi\,(r) = \exp\,(\pm\, iqA \ln\,(r))$, which varies less rapidly with $r$ than the rest of $\psi$. We can therefore write $\psi$ as

$$\psi = \frac{\exp\left(kr + \phi\left(r\right)\right)}{r},$$

where

$$k = \left(\mp iE \pm iq\sigma\, r/2\right).$$

At high energies, where $r$ is small, the first term dominates, approximating to a free fermion solution, which can be interpreted as asymptotic freedom. At low energies, when $r$ is large, the second term dominates, with its confining potential $\sigma$, and this can be interpreted as infrared slavery. Significantly, the Coulomb term, which is required to maintain spherical symmetry, is the component which here defines the strong interaction phase, $\phi\left(r\right)$, and this can be related to the directional status of $\mathbf{p}$ in the state vector.

Reducing the quark-quark potential to the Coulomb term, which is what we suppose might happen effectively at short distances, produces a hydrogen-like spectral series. Here, we have

$$4\left(E + q\frac{A}{r}\right)^2 = -2\left(\frac{\partial}{\partial r} + \frac{1}{r} + i\frac{j + \frac{1}{2}}{r}\right)^2 - 2\left(\frac{\partial}{\partial r} + \frac{1}{r} - i\frac{j + \frac{1}{2}}{r}\right)^2 + 4m^2,$$

where the functional part of the state vector has the form

$$\psi = \exp\left(-ar\right) r^\gamma \sum_{\nu = 0} a_\nu r^\nu.$$

Applying this over the four Dirac solutions, and expanding (for the ground state), we obtain:

$$E^2 + \frac{q^2 A^2}{r^2} + \frac{2qA}{r} E$$

$$= -\left(a^2 + \frac{(\gamma + \nu + 1)^2}{r^2} - \frac{(j + \frac{1}{2})^2}{r^2} - \frac{2a}{r}(\gamma + \nu + 1)\right) + m^2.$$

Equating coefficients of $1/r$, coefficients of $1/r^2$, and constant terms, we obtain:

$$2qAE = 2a(\gamma + \nu + 1)$$

$$q^2 A^2 = -(\gamma + \nu + 1)^2 + (j + \frac{1}{2})^2$$

$$E^2 = -a^2 + m^2,$$

leading to:

$$a = \frac{qAE}{(\gamma + \nu + 1)}$$

$$(\gamma + \nu + 1) = \pm\sqrt{(j + \frac{1}{2})^2 - q^2 A^2}$$

$$m^2 = E^2\left(1 + \frac{q^2 A^2}{(\gamma + \nu + 1)^2}\right).$$

According to this equation, there will be a certain value of $E$, below which $a$ is real, suggesting a confined solution, with equations which are identical in form to those for the Coulomb potential defined for atomic states, but with $qA$ replacing $Ze^2$. We assume a state vector, with functional component:

$$\psi = \exp\left(-\sqrt{m^2 - E^2}\right) r^\gamma \sum_{\nu = 0} a_\nu r^\nu,$$

and, allowing the power series to terminate at $\nu = n'$, we obtain the characteristic Coulomb-type solution:

$$\frac{E}{m} = \left(1 + \frac{q^2 A^2}{(\gamma + 1 + n')^2}\right)^{-1/2},$$

or

$$\frac{E}{m} = \left(1 + \frac{q^2 A^2}{(\sqrt{(j + \frac{1}{2})^2 - q^2 A^2} + n')^2}\right)^{-1/2}.$$

The condition resulting from $E^2 > m^2$ is that of *asymptotic* freedom, rather than escape, because of the continued presence (though reduced effect) of the confining linear potential. Combining the full and Coulomb-like solutions, we can make an approximate numerical calculation of the distance at which infrared slavery becomes effective. From the full solution, we let

$$k = (\mp iE \pm iq\sigma r/2) = \frac{2\pi(r)}{\lambda},$$

and take $\lambda = \infty$ at zero energy, or infrared slavery. Then

$$q\sigma r = 2E$$

and

$$r = \frac{2E}{q\sigma}.$$

From the Coulomb-like solution, we take $E$ as the mass or reduced mass of the $c$ quark, in the case of charmonium ($\approx 1.5$ GeV). Taking $\sigma \approx 1$ GeV fm$^{-1}$ and $q \approx 0.4$, we find $r \approx 4$ fm.

Virtually identical arguments can be applied to the three-quark or baryon system, where the potential may be assumed to have the form:[17]

$$V_{3Q} = -A_{3Q} \sum_{i < j} \frac{1}{|\mathbf{r}_i - \mathbf{r}_j|} + \sigma_{3Q} L_{min} + C_{3Q},$$

with $L_{min}$ taken as the minimal total length of the colour flux tubes linking three quarks, arranged in a triangle with sides, $a$, $b$, $c$, is given by

$$L_{min} = \left[\frac{1}{2}(a^2 + b^2 + c^2) + \frac{\sqrt{3}}{2}\sqrt{(a + b + c)(-a + b + c)(a - b + c)(a + b - c)}\right]^{1/2}.$$

and $C_{3Q}$ a constant term which can be absorbed, as $qC_{3Q}$, into the overall energy $E$. For perfect spherical symmetry, $a = b = c$, and $L_{min}$ becomes a multiple of the distance $r$ of any quark from the centre of the flux tubes, while

$$\sum_{i < j} \frac{1}{|\mathbf{r}_i - \mathbf{r}_j|}$$

becomes a multiple of $1 / r$. The three-quark potential $V_{3Q}$ then takes the same form as the quark-antiquark potential, and the same solutions will apply, with variations in the values of $A$, $\sigma$ and $E$.

The linear potential is one of only two polynomial distance-dependent potentials which give a special solution to the Dirac equation for a bound state. The other is for the Coulomb or inverse-distance potential (typically, for the hydrogen atom), which is, in effect, given here for a special case of the strong interaction. All other polynomial potentials (for example, the Lennard-Jones 6-12 potential) result in a harmonic oscillator solution (with a complex Coulomb phase) (see section 18). It is interesting that the two special-case potentials are exactly the same as are required, in classical physics, to produce a factor of 2 between potential and kinetic energies in the virial theorem, and are, in effect, characteristic consequences of the existence of duality and 3-dimensional space.

## 11 Angular momentum

According to the mechanism outlined in the previous sections, angular momentum carries the information relevant to the process of strong charge 'transfer', or gauge invariance, which defines the meaning of the strong interaction. The three-dimensionality of the (angular) momentum operator not only allows for the creation of a three-part (i.e. three-quark) fermionic nilpotent for the baryon, but also generates an $SU(3)$ structure for the strong interaction. Spin, as a consequence, becomes a property of the baryon as a whole, not of the component quarks. A theory, by Brodsky et al, equating baryon spin to the orbital angular momentum of the quarks, is possible in this context.[18]

Angular momentum is also important, however, to the descriptions of weak and electric interactions. The reasons for this lie deep in the foundations of physics. Here, we require Noether's theorem, which relates conserved quantities to symmetries or invariance under particular transformations. The theorem is, in fact, an example of the principle of duality in action. Invariance under transformation is really only another way of describing nonconservation, and duality requires every nonconserved quantity to be paralleled by an equivalent conserved one. The conjugate variables provide the most obvious examples. So, the translation symmetry of space (or non-identifiability of its elements) becomes equivalent to the conservation of the conjugate linear momentum, while the translation symmetry of time (or non-identitifiability of its elements) is equivalent to the conservation of the conjugate energy. At the same time, the rotation symmetry of space (or non-identity of directions in space) requires the conservation of a new conjugate parameter, angular momentum.

Since energy is related to mass by the equation $E = mc^2$, then the translation symmetry of time also becomes an expression of the conservation of mass, a result which could have been predicted directly from the duality of the elements of the parameter group: nonconservation of time implies conservation of mass. It may also be supposed, therefore, that the same kind of reasoning applies also to the dimensional parameters, space and charge. For example, we could expect the translation symmetry of space, or the conservation of linear momentum, to imply the conservation of the value of charge – of any type. In the case of conservation of electric charge, this is observed in its invariance under transformation of the electrostatic potential by a constant representing changes of phase, the phase changes being of the kind involved in the conservation of linear momentum. In a conservative system, of course, electrostatic potential varies only with the spatial coordinates, so we have, in effect, a statement of the principle that the quantity of electric charge is conserved because the spatial coordinates are not, exactly as we would expect from the duality incorporated in the parameter group, though we could also extend it to weak and strong charges.

However, an even more significant result can be predicted from the general symmetry of the parameter group. This is a relationship between the rotation symmetry of space, or the conservation of angular momentum, and the conservation of *type* of charge. Charge, as a conserved quantity, has units which should be conserved in type as well as number, So, if we consider charge units to be arranged along axes, separately representing the electric, strong and weak charges, then duality suggests that these axes, unlike those of space, should be fundamentally irrotational, so that one type of charge (say, electric) can never be converted into another (say, weak or strong). This is, of course, the basis of the laws of lepton and baryon conservation. It is also the reason why baryon decay has never been detected, and it can be seen as the fundamental basis for defining interactions, in which each type of charge acts in such a way that it is oblivious to the presence or absence of charge of a different type. If the property of conservation of charge type is fundamental, it ought to be linked directly to the conservation of angular momentum, according to the following scheme for an extended interpretation of Noether's theorem:

| symmetry | conserved quantity | linked conservation |
|---|---|---|
| space translation | linear momentum | value of charge |
| time translation | energy | value of mass |
| space rotation | angular momentum | type of charge |

A special case of the relationship is evident in the connection between spin and statistics, which requires fermions, with nonzero weak charge, and bosons, with zero weak charge, to have different values of spin angular momentum. (The actual spins are easily calculated using formal procedures, but the relationship between the spin ½ of the fermion and its status as ½ of a dual state will also be immediately evident.) However, it is also possible to show that the conservation of angular momentum requires the separate conservation of weak, strong and electric charges, in a much more fundamental way, through the conservation of the separate properties of orientation (with respect to the linear momentum), direction, and magnitude. It is precisely because of this connection that the electric and weak charges are found to be directly linked in the $SU(2)_L \times U(1)$ symmetry.

## 12 The weak vacuum

The conservation properties of the weak and electromagnetic charges are certainly determined by those of the angular momentum operator, and, in the case of a quark-type arrangement, might be expected to operate the same system of 'privileging' one charge in three during the complete phase cycle (with only one component of angular momentum well-defined). (This is because the 'quarks' are effectively only a way of identifying separate phases for a particular interaction with vector properties.) These charges, however, are not directly attached to the **p** operator, like the strong charge, and so their 'privileged' phases will not necessarily coincide with that of the strong charge or with each other. The weak charge ($w$) is, in fact, attached to $E$ and the electric charge ($e$) to $m$, in the Dirac state, and it is their *combination* which affects **p**. It is because of this that we tend to think of the electric and weak forces as being in some way combined, but the two charges are actually governed by quite separate symmetries.

Just as the character of the strong force and its relationship with the conservation of angular momentum is determined by its association with a vector operator, so we can expect that the pseudoscalar nature of $iE$ and the scalar nature of $m$ will determine the respective character and angular momentum relation of the weak and electric forces. The weak charge ($w$), which is the one associated with the quaternion label $\boldsymbol{k}$, produces two sign options for $iE$, because the algebra demands complexification of $E$, as it does of the parent-parameter time, and, consequently, two mathematical solutions. The sign option, in effect, determines the helicity state, or handedness with respect to the direction of motion, and it is this aspect of angular momentum conservation which is linked to the weak interaction.

One of the special properties of the weak interaction is its confinement to a single helicity state for fermions, with the opposite state reserved for antifermions. This is entirely a result of the fundamental group duality requiring mass-energy to be a continuum, and the consequent generation of a filled vacuum state. Essentially, there is no *physical* state corresponding to $-E$, although the use of a complex operator requires that $-iE$ has the same *mathematical* status as $iE$. Charge conjugation, however, or reversal of the signs of quaternion labels, *is* permitted physically. So $-ikE$ states are interpreted as antifermion or charge-conjugated states; and the mass-energy continuum becomes a filled vacuum for the ground state of the universe, in which such states would not exist.

A filled vacuum of this type was invoked by Dirac in the process of deriving the antiparticle concept, but the filled vacuum is now specifically a *k* or *weak* vacuum. Its manifestation is a violation of charge conjugation symmetry for the weak interaction, with consequent violation of either time reversal symmetry or parity to maintain the invariance of *CPT*. In principle, though the weak interaction can tell the difference between particle and antiparticle, it cannot distinguish between + and – signs of weak charge, and making the transition in the sign of the *k* operator (equivalent to *T*), because it is now interpreted as a charge conjugation (*C*), comes at the price of switching the sign of the *i* operator (*P*) as well.

This is also connected with the existence of just four solutions to the Dirac equation. Antifermions represent conjugate charge states to fermions, and we should, ideally, have eight independent combinations of $\pm \boldsymbol{k}w \pm \boldsymbol{i}s \pm \boldsymbol{j}e$. However, the Dirac equation allows only four solutions, and the charge parameters (*w-s-e*), like the $E$-**p**-*m* terms, require mapping onto a quaternion or 4-vector 4-space. The $E$-**p**-*m* terms and charge parameters, however, present us with alternative problems, for, while $E$-**p**-*m* offers too few solutions, *w-s-e* requires too many. For $E$-**p**-*m*, we have only the + and – states of **p** (which, as a vector, automatically provides these alternatives), while $E$ and *m* are both, strictly speaking, confined to positive values. In the case of the charges, the eight possible combinations of $\pm \boldsymbol{k}w \pm \boldsymbol{i}s \pm \boldsymbol{j}e$ have to be reduced to four. To overcome the problem, each effectively 'borrows' aspects of the other – $E$-**p**-*m* uses charge; *w-s-e* uses mass.

Unphysical $-E$ states appear in the Dirac equation to create the extra alternatives for $E$-**p**-*m*, which are explained physically by Dirac's assumption of a filled vacuum for antifermions in the ground state, with a natural preponderance of matter over antimatter. We then require one other charge (here, assumed to be *e*) to adopt + and – signs within matter, producing what is called 'isospin', as the equivalent of the spin variation produced by the two signs of **p**. In the case of *w*-s-*e*, we apply the category of antifermions using the negative sign of *s*. The filled vacuum then gives us the opportunity to remove the unwanted degree of freedom in the sign of *w*, by making the *effective* signs of *w* for matter and antimatter linked to those of *s*. Where *s* charges are present, the effective sign of *w* is determined by that of *s*, reducing the degrees of freedom in the charge structures from the eight of $\pm w \pm s \pm e$ to the four of $\pm (w + s) \pm e$, because of the linking of the signs of two of the quaternion operators.

It is, therefore, *w*, in effect, that determines the status of matter and antimatter, rather than *s*, though the sign of *s* is linked with the effective sign of *w*. As a result of this, both quarks and free fermions become mixed states, containing +*w*, and suppressed –*w*, states, and involving alternative violations of parity and time reversal symmetry. The removal of a degree of freedom from the charges $\pm w \pm s \pm e$ thus coincides exactly with the acquisition of a degree of freedom by $E \pm$ **p**, when it increases from the physical two to the mathematical four of $\pm E \pm \mathbf{p}$; in each case the sign of the $\mathbf{k}$ operator determines that the Dirac state has the four solutions which result from its quaternionic structure and its 4-D space-time. In principle, the unique $2^{n/2} \times 2^{n/2}$ matrix representation of the Clifford algebra, where $n = 2$, permits an exact quaternionic structure only because it is situated within a universe which has a filled $\mathbf{k}$ vacuum, and a single sign for the term $\mathbf{j}m$ in either fermion or antifermion states; and, ultimately, this is possible only because of the 4-dimensionality of the space-time signature which we have applied to the equation. (The 'coincidence' which makes $2^{n/2} \times 2^{n/2}$ into 4 because $2n = 4$, is, in fact, an expression of the fact that dimensionalization, to create a 3-space, and complexification to convert this to a space-time, are 'dual' processes of exactly equivalent status.) The process is outlined in Figure 1.

**Figure 1**

Exactly four Dirac solutions required:

| $E$ | $\mathbf{p}$ | $m$ |
|---|---|---|
| + | + | + |
| − * | − | |

| | |
|---|---|
| *only with | spin up |
| filled fermion | spin down |
| vacuum | |

$\rightarrow$ antifermions
$\rightarrow$ ground state has fermions only

Exactly four charge accommodation solutions required:

| $w$ | $s$ | $e$ |
|---|---|---|
| + * | + | + |
| − * | − | − |
| +* fermions | fermions | isospin up |
| −*antifermions | antifermions | isospin down |

*effective because of
filled weak (fermion)
vacuum

## 13 The origin of the Higgs mechanism

Just as the problem with $E$-**p**-$m$ was solved by invoking charge, so the problem with $w$-$s$-$e$ is solved by invoking mass. Physically, the loss of a degree of freedom for $w$ means that both quarks and free fermions become mixed states, containing both +$w$, and suppressed –$w$, states, and involving respective violations of parity and time reversal symmetry for the latter. A violation of parity or time reversal symmetry, consequent upon the violation of charge conjugation, as we have said, also means that only one state of helicity or **σ.p** exists for the pure weak interaction for fermions, with the opposite helicity applying to antifermions. Because (according to the Dirac equation) **σ** = –**1**, the fermionic state acquires negative helicity or left-handedness. The Dirac formalism, however, requires the creation of alternative states of positive helicity or right-handedness, through the existence of –**p**. If we wish to create these states, then the only remaining mechanism is through the introduction of rest mass in the term $jm$. The nilpotent version of the Dirac state thus associates the mass with the $j$ quaternion label, which defines what we call the electric charge; and the presence of $m$ simultaneously mixes $E$ and **p** terms, right-handed and left-handed components, and the effects of $e$ and $w$ charges.

In fact, for a particle with any other kind of charge as well as weak, the charge conjugation violation is not absolute and the alternative state of helicity is allowed. However, a finite probability of alternative helicity requires a nonzero rest mass (because the speed must be < $c$), the amount being determined by the probability of the state and the strength of the interaction involved ($e$, $s$). Since it is the 'filled' weak vacuum (that is, one with a nonzero expectation value, or 'Higgs field'), that gives rise to the nonzero rest mass of the fermions involved, then the mass of a particle must be determined by the strength of its coupling to this field; and the strength of the coupling will depend ultimately on the degree of symmetry-breaking which the creation of that particle requires. The process is shown in Figure 2.

**Figure 2**

Filled weak vacuum

$\rightarrow$ Violation of weak charge conjugation symmetry (+ *P* / *T*)

$\rightarrow$ Fermion with only weak charge in a single state of helicity

$\rightarrow$ For fermion with other charge, violation not absolute

$\rightarrow$ Finite probability of alternative helicity

$\rightarrow$ Nonzero rest mass (speed < *c*)

$\rightarrow$ Amount of mass depends on probability of state (e.g. number of zero charges)

and strength of the interaction involved (*e*, *s*)

Filled weak vacuum (with nonzero expectation value) = 'Higgs field'

$\rightarrow$ nonzero rest mass of fermions involved

$\rightarrow$ mass of fermion determined by the strength of coupling to this field

$\rightarrow$ strength of the coupling depends on degree of symmetry-breaking
in creation of fermion (e.g. number of missing or zero charges)

## 14 $SU(2)_L \times U(1)$

Just as the character of the strong force and its relationship with the conservation of angular momentum is determined by its association with a vector operator, so we can expect that the pseudoscalar nature of *iE* and the scalar nature of *m* will determine the respective character and angular momentum relation of the weak and electric forces. In the case of the weak force, we have two sign options for *iE*, because we are using complex algebra, and there are necessarily two mathematical solutions. The sign option, in effect, determines the helicity state, and it is this aspect of angular momentum conservation which is linked to the weak interaction.

The separate conservation laws for *w*, *s*, and *e* charges, which are axiomatic in this theory, determine that each type of charge must be independent of the other. It is particularly essential to the *characterization* of the weak interaction to express its independence from the presence or absence of electric charges, for it is precisely this independence that creates the characteristic $SU(2)_L$ 'isospin' pattern associated with the interaction. If the mixing of $E$ and **p** terms, or right-handed and left-handed components, is also equivalent to the mixing of $e$ and $w$ charges, then it is important to establish that this mixing *does not affect the weak interaction* as such. Otherwise, the whole idea of defining the weak interaction through charge-conjugation violation would be compromised. The weak interaction must be simultaneously left-handed for fermion states and indifferent to the presence or absence of the electric charge, which introduces the right-handed element.

There are two possible $SU(2)_L$ states, with electric charge or without electric charge; these are the two states of weak isospin, and the weak interaction must behave in such a way that they are indistinguishable. Mathematically, these two $SU(2)_L$ states are described by a quantum number, $t_3$ (the third component of weak isospin), whose value is such that $(t_3)^2 = (\frac{1}{2})^2$ in half the total number of possible states, that is, in the left-handed ones. For the electric force, in the case of free fermions, the relevant quantum number ($Q$) is determined by the absence or presence of the electric charge, and takes the values 0 and –1, equivalent to the charges 0 and –$e$, the – sign being purely historical in origin, with the + sign reserved for antistates. So $Q^2 = 1$ in half the total number of possible states (though a different half – including the right-handed ones), and 0 in the others. By a standard argument,[19-20] it can be shown that, if the weak and electric interactions are described by some grand unifying gauge group, irrespective of its particular structure, then, to satisfy orthogonality and normalisation conditions, the parameter which describes the mixing ratio, $\sin^2\theta_W$, is precisely determined by $\Sigma (t_3)^2 / \Sigma Q^2$, which in this case must be 0.25.

However, the ratio cannot apply only to free fermions. The weak interaction is also required to be indifferent to the presence or absence of the strong charge, that is, to the directional state of the angular momentum operator, and so the same mixing proportion, as observed in free fermion states, should exist also for quark states, and separately for each 'colour' phase, so that none is preferred, and colour is not directly detected through $w$. Applying this to quarks, we can create the same weak isospin states for one lepton-like 'colour', that is, we have one quark state with alternative $Q$ values of –1 and 0, or charge values of –$e$ and 0. We now find that the only corresponding isospin states for the other colours that retain both the accepted value of $\sin^2\theta_W$ and the variation of only one 'privileged' quark phase 'instantaneously' in three, are 1 and 0 (or $e$ and 0). In effect, the variation 0 0 –$e$ must be taken against either an empty background or 'vacuum' (0 0 0) or a full background ($e\ e\ e$), so that the two states of weak isospin in the three colours become:

$$
\begin{array}{ccc}
e & e & 0 \\
0 & 0 & -e
\end{array}.
$$

A filled 'electromagnetic' vacuum might be considered to generate an antifermion 'image' of the form ***j*** ($\pm$ ***k****E* $\pm$ *ii***p** + *ij*m) for a fermion with state vector ($\pm$ ***k****E* $\pm$ *ii***p** + *ij*m), and, significantly, the bosonic form generated would have spin 0.

If the weak interaction is characterized by $SU(2)$, the electromagnetic interaction takes on the required $U(1)$ structure for a pure scalar magnitude by introducing a required phase. Conventionally, if $SU(2)$ breaks parity, the only way of maintaining a group structure, and the only way of ensuring that $SU(2)$ remains renormalizable, is to incorporate $U(1)$. This becomes significant in defining a Higgs ground state which is nonsymmetric and parity violating through finding the one such state that $SU(2)$ and $U(1)$ have in common.

## 15 The weak interaction and the Dirac formalism

The argument here suggests that the pattern of $SU(3) \times SU(2)_L \times U(1)$ for the strong, weak and electric interactions between fermions can be established from first principles, and that the reasoning applied to the state vectors for $SU(3)$ can also be applied to those used for $SU(2)_L \times U(1)$, together with the formalisms relating to these symmetries for the derivation of Lagrangians, generators, covariant derivatives, and so forth. To apply the Dirac formalism to the weak interaction, we observe, first, that, experimentally, weak interactions all follow a pattern, which is determined by the $SU(2)_L$ symmetry. In the case of leptons, it is

$$e + v \rightarrow e + v \,.$$

For quarks, it is

$$u + d \rightarrow u + d \,,$$

and, for weak interactions involving both leptons and quarks (for example, $\beta$ decay):

$$d + v \rightarrow e + u \,.$$

These can all be seen to involve the same two-isospin state structure, as should apply irrespective of the presence or absence of strong charges.

Considering the lepton case as exemplar, we find that there are four possible vertices (assuming left-handed components only).

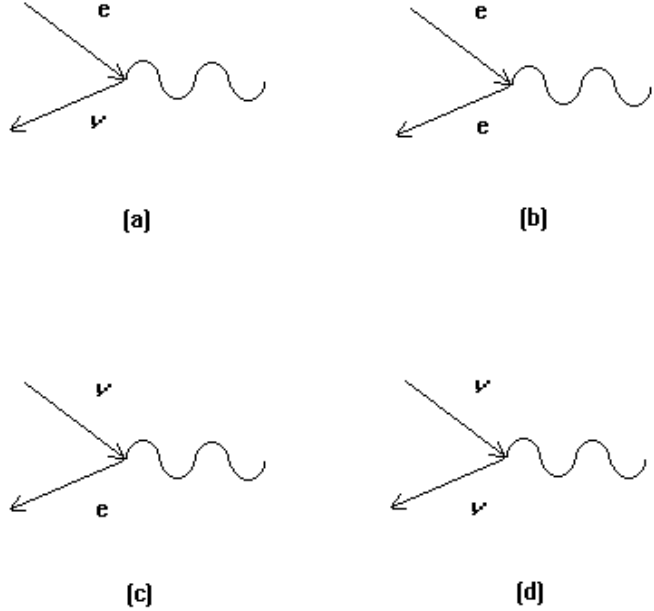

[a]          [b]

[c]          [d]

All the vertices occur at once, and so the interaction can be described as a mixing or superposition of the four possibilities. However, vertex (b), and this one alone, also represents a possible electromagnetic interaction, giving us a 1 to 4 ratio for the occurrence of the electromagnetic to weak interaction at the energy which the vertices characteristically represent (phenomenologically, that of the *W* / *Z* bosons). This results from the fact that particle charge structures at this energy are such that the electroweak mixing ratio becomes

$$\sin^2\theta_W = \frac{e^2}{w^2} = \frac{\Sigma\, t_3^{\,2}}{\Sigma\, Q^2} = 0.25 \ .$$

Since, we are concerned here only with the mass generated via the $SU(2)_L$ electroweak mechanism (weak isospin), and not with any mass associated with the mixing of generations (or direct violation of weak charge conjugation symmetry), whether quark or lepton, we will take the unmixed 'pure' weak state to be massless in this respect, and consider the interactions of a 'massless' or near-massless pure state with a massive mixed one (as is effectively the case with $v_e$ and *e*). Taking the quaternion state vectors for the fermionic components of the four vertices, we obtain, for the case where the spins of the interacting fermions are assumed parallel (total 0 for fermion-antifermion combination):

(a)    $(kE - i i\mathbf{p} + i j m) \ldots (-kE + i i\mathbf{p}) \ldots = 4m^2$ ;

(b)    $(kE - i i\mathbf{p} + i j m) \ldots (-kE + i i\mathbf{p} + i j m) \ldots = 4m^2$ ;

(c)    $(kE - i i\mathbf{p}) \ldots (-kE + i i\mathbf{p} + i j m) \ldots = 4m^2$ ;

(d)    $(kE - i i\mathbf{p}) \ldots (-kE + i i\mathbf{p}) \ldots = 4m^2$ .

where ($kE - ii\mathbf{p} + ijm$) … represents a column or row vector with the terms:

$$(kE - ii\mathbf{p} + ijm); (kE + ii\mathbf{p} + ijm); (-kE + ii\mathbf{p} + ijm); (-kE - ii\mathbf{p} + ijm) ,$$

and so on. Using a standard normalisation, these sums become $m^2 / E^2$, implying that, without an $m$ term, all four vertices would become 0. The $m$ term arises, as we have seen, from the fact that $\mathbf{p}$ is not purely composed of left-handed helicity states (with $-\mathbf{p}$ right-handed), but incorporates a right-handed component, which itself cannot contribute to the weak interaction because of charge-conjugation violation and the presence of a weak filled vacuum. So the right-handed component *can only arise from the presence of the electromagnetic interaction*. The weak interaction, therefore, cannot exist as a pure left-handed interaction, without a mixing with the electromagnetic interaction to produce the necessary non-zero mass through the introduction of right-handed states.

If we now put into the $E$ and $\mathbf{p}$ terms of the state vector the covariant derivatives for an $SU(2)_L \times U(1)$ electroweak interaction, the scalar part goes with $E$ and the vector part with $\mathbf{p}$. Mass is produced by the mixing of $E$ with $\mathbf{p}$ via the relativistic connection between these terms. It is, similarly, produced by the mixing of the gauge field $B^0$ with $W^+$, $W^0$, and $W^-$, which we may now identify with the four vertices (d), (a), (b), and (c). Choosing the single, well-defined direction of spin or angular momentum ($\mathbf{p}$) to be, in principle, the one where the total value for the interacting fermion-antifermion combination is 0, we can ensure that the mixing is specifically between the neutral components, $B^0$ and $W^0$, and create one massless *combination* to represent the carrier of the pure electromagnetic interaction ($\gamma$), with the other being the massive neutral weak carrier $Z^0$. If the mixing must be such as to define the ratio of the two interactions, $\sin^2\theta_W$, at 0.25, the other two vertices, $W^+$ and $W^-$, then fulfil the requirements for the existence of states corresponding to total spin values of $+1$ and $-1$.

For left-handed leptons, we have the covariant derivatives:

$$\partial_\mu \to \partial_\mu + ig \frac{\boldsymbol{\tau}.W^\mu}{2} - ig' \frac{B^\mu}{2} ,$$

and, for right-handed:

$$\partial_\mu \to \partial_\mu - ig' \frac{B^\mu}{2} .$$

The energy operator and the single well-defined component of spin angular momentum give us:

$$E = i\partial_0 \to i\partial_0 + g' \frac{B^0}{2} + ig' \frac{B^3}{2}$$

and

$$ip_3 = \partial_3 \to \partial_3 + ig \frac{\boldsymbol{\tau}.W^3}{2} + ig \frac{\boldsymbol{\tau}.W^0}{2} .$$

So, the state vector for the (d) vertex can be written in the form:

$$(kE - i\textbf{i}\textbf{p}) \dots (-kE + i\textbf{i}\textbf{p}) \dots = \left(k\left(\partial_0 + g'\frac{B^0}{2} + g'\frac{B^3}{2}\right) - i\left(\partial_3 + ig\frac{\boldsymbol{\tau}.\textbf{W}^3}{2} + ig\frac{\boldsymbol{\tau}.\textbf{W}^0}{2}\right)\right) \times$$

$$\left(-k\left(\partial_0 + g'\frac{B^0}{2} + g'\frac{B^3}{2}\right) + i\left(\partial_3 + ig\frac{\boldsymbol{\tau}.\textbf{W}^3}{2} + ig\frac{\boldsymbol{\tau}.\textbf{W}^0}{2}\right)\right)$$

and the state vector for the (b) vertex in the form:

$$(kE - i\textbf{i}\textbf{p} + i\textbf{j}m) \dots (-kE + i\textbf{i}\textbf{p} + i\textbf{j}m) \dots =$$

$$\left(k\left(\partial_0 + g'\frac{B^0}{2} + g'\frac{B^3}{2}\right) - i\left(\partial_3 + ig\frac{\boldsymbol{\tau}.\textbf{W}^3}{2} + ig\frac{\boldsymbol{\tau}.\textbf{W}^0}{2}\right) + i\textbf{j}m\right) \times$$

$$\left(-k\left(\partial_0 + g'\frac{B^0}{2} + g'\frac{B^3}{2}\right) + i\left(\partial_3 + ig\frac{\boldsymbol{\tau}.\textbf{W}^3}{2} + ig\frac{\boldsymbol{\tau}.\textbf{W}^0}{2}\right) + i\textbf{j}m\right).$$

With $m$ determined from the combination of $E$ and $\textbf{p}$, we can, by appropriate choice of the value of $m$, make these compatible by additionally defining a combination of the coupling constants related to the $SU(2)_L$ and $U(1)$ symmetries, $g'$ and $g$, which removes $B^3$ from $E$ and $W^0$ from $\textbf{p}$. It is, of course, significant here that it is $B^\mu$ which is characteristic of right-handed lepton states, and therefore associated with the production of mass. Writing these combinations as $\gamma^0$ and $Z^3$, and those of $g'$ and $g$, as $e$ and $w$ ($= g$), we obtain:

$$(kE - i\textbf{i}\textbf{p} + i\textbf{j}m) \dots (-kE + i\textbf{i}\textbf{p} + i\textbf{j}m) \dots =$$

$$\left(k\left(\partial_0 + e\frac{\gamma^0}{2}\right) - i\left(\partial_3 + iw\frac{\boldsymbol{\tau}.\textbf{Z}^3}{2}\right) + i\textbf{j}m\right)\left(-k\left(\partial_0 + e\frac{\gamma^0}{2}\right) + i\left(\partial_3 + iw\frac{\boldsymbol{\tau}.\textbf{Z}^3}{2}\right) + i\textbf{j}m\right).$$

Here, $\gamma^0 / 2$ becomes the same as the electrostatic potential $\phi$. So, we can write this in the form:

$$(kE - i\textbf{i}\textbf{p} + i\textbf{j}m) \dots (-kE + i\textbf{i}\textbf{p} + i\textbf{j}m) \dots =$$

$$\left(k(\partial_0 + e\phi) - i\left(\partial_3 + iw\frac{\boldsymbol{\tau}.\textbf{Z}^3}{2}\right) + i\textbf{j}m\right)\left(-k(\partial_0 + e\phi) + i\left(\partial_3 + iw\frac{\boldsymbol{\tau}.\textbf{Z}^3}{2}\right) + i\textbf{j}m\right).$$

With $e$ and $w$, the combinations of $g'$ and $g$, now representing the pure electromagnetic and weak coupling constants, we must necessarily obtain the ratio $e^2 / w^2 = 0.25$, and both quarks and leptons must be structured to observe this.

Significantly, the *exchange* of electromagnetic charge, through $W^+$ or $W^-$, is not itself an electromagnetic interaction, but rather an indication of the weak interaction's indifference to the presence of the electromagnetic charge. A 'weak interaction', in principle, is a statement that all states of a particle with the same weak charge are equally probable, given the appropriate energy conditions, and that gauge invariance is maintained with respect to them. Weak bosons are massive because they act as carriers of the electromagnetic charge, whereas electromagnetic bosons (or photons) are massless because they do not. The quantitative value of the mass must be determined from the coupling of the weak charge to the asymmetric vacuum state which produces the violation of charge conjugation in the weak interaction. The weak interaction is also indifferent to the presence of the strong charge, and so cannot distinguish between quarks and leptons (hence, the intrinsic identity of purely lepton weak interactions with quark-lepton or quark-quark ones) and, in the case of quarks, it cannot tell the difference between a filled 'electromagnetic vacuum' (up quark) and an empty one (down quark). The weak interaction, in addition, is also indifferent to the sign of the weak charge, and responds (via the vacuum) only to the status of fermion or antifermion – hence, the Cabibbo-Kobayashi-Maskawa mixing.

## 16 The Higgs mechanism for $U(1)$ and $SU(2)_L$

It is now possible to relate the work in the preceding sections to the conventional treatment of the Higgs mechanism.[21] It is usual to first illustrate the procedure using a $U(1)$ symmetry group. We take the Lagrangian for a complex scalar field $\phi = (\phi_1 + i\phi_2) / \sqrt{2}$,

$$L = (\partial_\mu \phi)^* (\partial^\mu \phi) - V(\phi^*, \phi) = (\partial_\mu \phi)^* (\partial^\mu \phi) - \mu^2 \phi^* \phi - \lambda (\phi^* \phi)^2 ,$$

where $\lambda(\phi^*\phi)^2$ is a self-interaction, and make this invariant under a $U(1)$ local gauge transformation,

$$\phi \rightarrow e^{i\alpha(x)} \phi ,$$

by replacing $\partial_\mu$ with the covariant derivative,

$$D_\mu = \partial_\mu - ieA_\mu ,$$

with the gauge field transforming as

$$A_\mu \rightarrow A_\mu + \frac{1}{e} \partial_\mu \alpha .$$

The gauge invariant Lagrangian then becomes of the same form as the QED Lagrangian for a charged scalar particle of mass $\mu$ without the self-interaction term:

$$L = (\partial^\mu + ieA_\mu)\phi * (\partial_\mu - ieA_\mu)\phi - \mu^2 \phi^* \phi - \lambda(\phi^*\phi)^2 - \frac{1}{4} F_{\mu\nu} F^{\mu\nu} .$$

We now need to introduce the specific changes that will produce a filled vacuum state or spontaneous symmetry-breaking. To spontaneously break the symmetry, we take $\mu^2 < 0$, $\lambda > 0$. The potential $V$ now has a local maximum at $\phi = 0$, and a minimum at

$$\phi_1{}^2 + \phi_2{}^2 = v^2 = -\frac{\mu^2}{2\lambda} .$$

Without loss of generality, we are free to choose one of the degenerate vacua represented by this equation as the physical one. So we choose $\phi_1 = v$ and $\phi_2 = 0$, the so-called physical or unitary gauge. Now, expanding the Lagrangian about this vacuum by defining fields, $\eta(x)$ and $\xi(x)$, so that

$$\phi(x) = \sqrt{\frac{1}{2}} \left( v + \eta(x) + i\xi(x) \right) .$$

and substituting $\phi(x)$ into the Lagrangian, we obtain

$$L = \frac{1}{2} (\partial_\mu \xi)^2 + \frac{1}{2} (\partial_\mu \eta)^2 - v^2 \lambda \eta^2 + \frac{1}{2} e^2 v^2 A_\mu A^\mu - e v A_\mu \partial^\mu \xi - \frac{1}{4} F_{\mu\nu} F^{\mu\nu} + \text{interaction terms}$$

The $\eta$-field mass now becomes $\sqrt{2\lambda v^2}$, and the $A$-field mass is $ev$, but the $\xi$-field has only a kinetic energy term and no mass. This Goldstone boson is a massless spin 0 scalar, which is denied physical existence by the nilpotent algebra, but we can choose a gauge to eliminate it, in which, using polar coordinates,

$$\phi(x) \rightarrow \sqrt{\frac{1}{2}} \left( v + h(x) \right) e^{i\theta(x)/v} .$$

and

$$A_\mu \rightarrow A_\mu + \frac{1}{ev} \partial_\mu \theta .$$

With $h$ real, we now obtain

$$L = \frac{1}{2} (\partial_\mu h)^2 - v^2 \lambda h^2 + \frac{1}{2} e^2 v^2 A_\mu^2 - \lambda v h^3 - \frac{1}{4} \lambda h^4 + \frac{1}{2} e^2 A_\mu^2 h^2 + v e^2 A_\mu^2 h - \frac{1}{4} F_{\mu\nu} F^{\mu\nu} .$$

The Lagrangian now includes only two massive particles, the vector gauge boson $A_\mu$ and the massive spin 0 scalar (Higgs boson) $h$.

The application to $U(1)$ was simply an illustration of the mechanism, but, for $SU(2)$, where we believe the symmetry is truly broken, we will need to justify our assumptions on a fundamental basis. In the case of an $SU(2)$ local gauge symmetry, we apply an $SU(2)$ doublet of complex scalar fields,

$$\phi = \sqrt{\frac{1}{2}} \begin{pmatrix} \phi_1 + i\phi_2 \\ \phi_3 + i\phi_4 \end{pmatrix} ,$$

to what is essentially the same Lagrangian:

$$L = (\partial_\mu \phi)^\dagger (\partial^\mu \phi) - V(\phi^\dagger, \phi) = (\partial_\mu \phi)^\dagger (\partial^\mu \phi) - \mu^2 \phi^\dagger \phi - \lambda (\phi^\dagger \phi)^2 .$$

The complex doublet is chosen here because it is the simplest that will produce the gauge fields that we already know must exist for a spontaneously-broken $SU(2)$.

Replacing $\partial_\mu$ with the covariant derivative,

$$D_\mu = \partial_\mu - i g \frac{1}{2} \tau_a W_\mu^a ,$$

with $W_\mu{}^a$ representing the three new gauge fields where $a = 1, 2, 3$, and $\tau_a W_\mu{}^a$ can be written $\boldsymbol{\tau} . \mathbf{W}_\mu$. The gauge invariant Lagrangian then becomes

$$L = (\partial_\mu \phi + ig \tfrac{1}{2}\boldsymbol{\tau} . \mathbf{W}_\mu)^\dagger \, (\partial^\mu \phi + ig \tfrac{1}{2}\boldsymbol{\tau} . \mathbf{W}_\mu) - \mu^2 \phi^\dagger \phi - \lambda(\phi^\dagger \phi)^2 - \tfrac{1}{4}\mathbf{W}_{\mu\nu}\mathbf{W}^{\mu\nu} \ ,$$

with the last term representing the additional kinetic energy of the gauge fields. Again assuming $\mu^2 < 0$ , $\lambda > 0$, to spontaneously break the symmetry, and produce the filled vacuum state that we already know must exist for the weak interaction, we find the minimum potential occurs when

$$\phi^\dagger \phi = \frac{1}{2}\,(\phi_1{}^2 + \phi_2{}^2 + \phi_3{}^2 + \phi_4{}^2) = -\frac{\mu^2}{2\lambda} \ .$$

Choosing $\phi_1{}^2 = \phi_2{}^2 = \phi_4{}^2 = 0$, and $\phi_3{}^2 = v^2 = -\mu^2 / 2\lambda$, as our gauge (as we are free to do), and expanding $\phi(x)$ about the vacuum,

$$\phi_0 = \sqrt{\frac{1}{2}} \begin{pmatrix} 0 \\ v \end{pmatrix} \ ,$$

we substitute

$$\phi(x) = \sqrt{\frac{1}{2}} \begin{pmatrix} 0 \\ v + h(x) \end{pmatrix}$$

into the Lagrangian to gauge away the unphysical Goldstone bosons, which, according to our nilpotent formalism, simply cannot exist. The fluctuations from the vacuum can be parameterized in terms of the four real fields $\boldsymbol{\theta}$ and $h$, using

$$\phi(x) = \sqrt{\frac{1}{2}} \begin{pmatrix} 0 \\ v + h(x) \end{pmatrix} e^{i\boldsymbol{\tau} . \boldsymbol{\theta}(x)/v} \ .$$

Substituting $\phi_0$ into the Lagrangian, we obtain an expression containing the term

$$(ig \tfrac{1}{2}\boldsymbol{\tau} . \mathbf{W}_\mu)^\dagger \, (ig \tfrac{1}{2}\boldsymbol{\tau} . \mathbf{W}_\mu)$$

$$= \frac{g^2}{8}\left(\begin{pmatrix} W_\mu{}^3 & W_\mu{}^1 - iW_\mu{}^2 \\ W_\mu{}^1 + iW_\mu{}^2 & W_\mu{}^3 \end{pmatrix}\begin{pmatrix} 0 \\ v \end{pmatrix}\right)^\dagger \left(\begin{pmatrix} W_\mu{}^3 & W_\mu{}^1 - iW_\mu{}^2 \\ W_\mu{}^1 + iW_\mu{}^2 & W_\mu{}^3 \end{pmatrix}\begin{pmatrix} 0 \\ v \end{pmatrix}\right)$$

$$= \frac{g^2 v^2}{8}\,[(W_\mu{}^1)^2 + (W_\mu{}^2)^2 + (W_\mu{}^3)^2] \ . \tag{3}$$

The Lagrangian now describes three massive vector gauge fields and one massive scalar $h$. The first three are required to explain the weak interaction, as we already understand it physically; the last is the Higgs boson, a spin 0 particle, which must necessarily be massive in the nilpotent formalism.

## 17 The renormalizability of the electroweak interaction

Of course, as is well known, the weak interaction theory is not renormalizable taken on its own, but only when combined with the $U(1)$ electromagnetic theory, which provides a necessary scalar phase term. It is interesting that the renormalizability of the combined electroweak interaction is related to the very mechanism which gives masses to the fermions and gauge bosons. For obvious reasons, a quantum field integral taken over all values of $p$ will only be finite, as the nilpotent algebra demands, if the index of $p$ in the integrand (or divergence $D$) is less than 0. Now, the propagator for the combined electroweak gauge bosons is[22]

$$\Delta_{\mu\nu} = \frac{1}{p^2 - m^2}\left(-g_{\mu\nu} + (1 - \xi)\left(\frac{p_\mu p_\nu}{p^2 - \xi m^2}\right)\right),$$

where $\xi$ is the 't Hooft gauge term, which appears in the gauge fixing term in the Lagrangian for the interaction:

$$-\frac{1}{2\xi}\left(\partial_\mu A^\mu + \xi m \phi_2\right)^2 .$$

This term removes the unphysical (massless scalar) Goldstone boson $\phi_2$, which arises from the spontaneous symmetry breaking produced by the filled weak vacuum used to eliminate negative energy states. If $\xi$ is finite, then as $p_\mu \to \infty$, $\Delta_{\mu\nu} \to p^{-2}$, like the pure photon propagator, which, in the absence of any gauge choice, becomes:

$$\Delta_{\mu\nu} = \frac{1}{q^2}\left(-g_{\mu\nu} + (1 - \xi)\left(\frac{q^\mu q^\nu}{q^2}\right)\right).$$

However, for $\xi \to \infty$, we have the propagator for a massive vector boson theory without massless component,

$$\Delta_{\mu\nu} = \frac{1}{p^2 - m^2}\left(-g_{\mu\nu} + \left(\frac{p_\mu p_\nu}{m^2}\right)\right),$$

which becomes a constant when $p_\mu \to \infty$, leading to infinite sums in the diagrams equivalent to those in QED.

One of the most convenient choices of gauge is $\xi = 1$ (Feynman gauge), which leads to an electroweak boson propagator,

$$\Delta_{\mu\nu} = \frac{-g_{\mu\nu}}{p^2 - m^2} ,$$

entirely analogous to that for the photon in the same gauge, and similarly linked by the factor ($k E + i i \mathbf{p} + i j m$) to the fermion propagator, as in QED. (It may be possible, here, to link the existence of massive weak bosons to the creation, in the nilpotent representation, of massive bosonic states via the interactions of fermions with the vacuum.)

## 18 The spherical harmonic oscillator

To gain a deeper understanding of the weak interaction, we need to explore further the solutions of the Dirac equation for spherically symmetric potentials. The spherical harmonic oscillator provides a particularly significant case. Here, we have a potential energy of the form $\frac{1}{2}\,cr^2$. We may also suppose that spherical symmetry requires a Coulomb 'phase' term, say $A / r$, the exact form and significance of which will become apparent in the calculation. (The nilpotent method shows that spherically symmetric solutions are impossible without such a term.) The covariant form of the differential operator will then produce a Dirac equation of the form:

$$\left( \pm \boldsymbol{k} \left( E + \frac{1}{2}\,cr^2 + \frac{A}{r} \right) \pm \boldsymbol{i} \left( \frac{\partial}{\partial r} + \frac{1}{r} \pm i\,\frac{j+\frac{1}{2}}{r} \right) + \boldsymbol{i}jm \right) \Psi = 0 \ .$$

As usual, the solution of the equation will require finding $F$, the variable part of $\Psi$ which will make the eigenvalue nilpotent. Polynomial potential terms which are multiples of $r^n$ require the incorporation into the exponential of terms which are multiples of $r^{n+1}$. So, extending our work on the strong interaction and the Coulomb field, we may suppose that the solution is of the form:

$$F = \exp\,(-ar + br^3)\ r^\gamma \sum_{\nu\,=\,0} a_\nu r^\nu \ .$$

So

$$\frac{\partial F}{\partial r} = \left( -a + 3br^2 + \frac{\gamma}{r} + \frac{\nu}{r} + \dots \right) F \ ,$$

and the eigenvalue produced by the differential operator then becomes:

$$\left( \pm \boldsymbol{k} \left( E + \frac{1}{2}\,cr^2 + \frac{A}{r} \right) \pm \boldsymbol{i} \left( -a + 3br^2 + \frac{\gamma}{r} + \frac{\nu}{r} + \dots + \frac{1}{r} \pm i\,\frac{j+\frac{1}{2}}{r} \right) + \boldsymbol{i}jm \right).$$

Assuming that this is nilpotent, and that the power series terminates, we obtain:

$$4 \left( E + \frac{1}{2}\,cr^2 + \frac{A}{r} \right)^2 = -2 \left( -a + 3br^2 + \frac{\gamma}{r} + \frac{\nu}{r} + \frac{1}{r} + i\,\frac{j+\frac{1}{2}}{r} \right)^2$$

$$-2 \left( -a + 3br^2 + \frac{\gamma}{r} + \frac{\nu}{r} + \frac{1}{r} - i\,\frac{j+\frac{1}{2}}{r} \right)^2 + 4m^2 \ .$$

Equating constant terms, we find

$$E^2 = -a^2 + m^2 \ ,$$

$$a = \sqrt{m^2 - E^2} \tag{4}$$

Equating terms in $r^4$, with $\nu = 0$, we obtain:

$$\frac{1}{4}\,c^2 = -9c^2 \ ,$$

from which

$$b = \pm \frac{ic}{6} \ .$$

Since $c$ is real, $b$ must be imaginary.

Equating coefficients of $r$, where $\nu = 0$, we find

$$Ac = -6b\,(1 + \gamma)\ ,$$

and

$$(1 + \gamma) = \pm\,iA\ .$$

This means that, if $(1 + \gamma)$ is real, then $A$ must be imaginary.

Equating coefficients of $1 / r^2$ and coefficients of $1 / r$, and assuming the power series terminates in $\nu = n'$, we obtain

$$A^2 = -(1 + \gamma + n')^2 + (j + \tfrac{1}{2})^2 \tag{5}$$

and

$$EA = a\,(1 + \gamma + n')\ . \tag{6}$$

Using (4), (5) and (6), we obtain

$$\left(\frac{m^2 - E^2}{E^2}\right)(1 + \gamma + n')^2\ = -(1 + \gamma + n')^2 + (j + \tfrac{1}{2})^2$$

or

$$E = -\frac{m}{(j + \tfrac{1}{2})}\,(\pm\,iA + n')\ .$$

If we now take $A$ to have a half-unit value ($\pm\,\tfrac{1}{2}\,i$), in line with the value of spin, we obtain a set of energy levels of the form expected in the simple harmonic oscillator:

$$E = -\frac{m}{(j + \tfrac{1}{2})}\,(\tfrac{1}{2} + n')\ .$$

We can now associate the phase term required for spherical symmetry ($A = \pm\,\tfrac{1}{2}\,i$), directly with the random directionality of the spin of the fermion. In the case of the harmonic oscillator, the term, with its unit value and imaginary coefficient, is clearly not a fundamental component of the potential for the interaction, as it is in the case of the strong interaction between quarks, and it is effectively brought in when the spin component is added to the $\boldsymbol{\sigma}.\nabla$ term in transforming from rectilinear to polar coordinates (although it is, of course, present implicitly where $\boldsymbol{\sigma}$ and $\nabla$ are taken as multivariate vectors). It is also significant that, although the nilpotent produces the same three equations as were required to generate the energy level series in the case of the pure Coulomb interaction, this solution cannot be applied to the case of the harmonic oscillator as it would result in a series of imaginary or complex energy levels.

The same method applied can be applied to any case where the potential can be expressed as a polynomial function of the radial distance. The Lennard-Jones potential,

$$V = \frac{B}{r^6} - \frac{C}{r^{12}}$$

provides a characteristic instance. Again incorporating a Coulomb or phase term for spherical symmetry, the Dirac equation becomes:

$$\left( \pm k \left( E + \frac{A}{r} + \frac{B}{r^6} - \frac{C}{r^{12}} \right) \pm i \left( \frac{\partial}{\partial r} + \frac{1}{r} \pm i \frac{j + \frac{1}{2}}{r} \right) + ijm \right) \Psi = 0 \; ,$$

suggesting a solution of the form:

$$F = \exp \left( - ar - br^{-5}/5 - br^{-11}/11 \right) r^\gamma \sum_{\nu = 0} a_\nu r^\nu \; .$$

The nilpotent now becomes:

$$\pm k \left( E + \frac{A}{r} + \frac{B}{r^6} - \frac{C}{r^{12}} \right) \pm i \left( - a + \frac{1}{r} + \frac{\gamma}{r} + \frac{\nu}{r} + \ldots + \frac{1}{r} \pm i \frac{j + \frac{1}{2}}{r} \right) + ijm \; .$$

Equating constant terms, as usual,

$$E^2 = - a^2 + m^2 \; ,$$

$$a = \sqrt{m^2 - E^2} \; .$$

Equating respective coefficients of $r^{-24}$ and $r^{-18}$, we find that $C^2 = - c^2$ and $B^2 = - b^2$, from which we obtain $c = \pm iC$ and $b = \mp iB$, with the two coefficients having opposite signs. Equating coefficients of $r^{-7}$, for the case when $\nu = 0$, leads to

$$AB = - b \left( 1 + \gamma \right) \; ,$$

with

$$\left( 1 + \gamma \right) = \pm iA \; ,$$

which is the same result as for the harmonic oscillator. Finally, equating respective coefficients of $r^{-2}$ and $r^{-1}$, for a series terminating in $\nu = n'$, produces the identical relations:

$$A^2 = - \left( 1 + \gamma + n' \right)^2 + \left( j + \frac{1}{2} \right)^2 \; ,$$

$$EA = a \left( 1 + \gamma + n' \right) \; ,$$

and

$$E = - \frac{m}{(j + \frac{1}{2})} \left( \frac{1}{2} + n' \right) \; ,$$

which demonstrate that the energy levels are again those of the harmonic oscillator.

This result is not sensitive to the particular terms used in the polynomial form of the potential. For any spherically symmetric polynomial potential with terms of the form $Ar^n$, where $|n| \geq 2$, the solution will be that of a harmonic oscillator. Particularly important examples are the dipolar case, where $V \propto r^{-3}$, and the multipolar case, with $V \propto r^{-n}$, where $n > 3$. Only in the special cases of $n = 1$ (the strong interaction potential) and $n = -1$ (the pure electrostatic or gravitational Coulomb potential), do we expect particular bound solutions, and the special nature of these solutions is a result of the symmetry of 3-dimensional space, just as it is in the analogous case of classical physics, where only constant and inverse-square forces produce a virial relation between the potential and kinetic energies of exactly 2.

## 19 The weak interaction as a harmonic oscillator

It would appear that there are three solutions of the Dirac equation for spherically-symmetric distance-dependent potentials. This becomes particularly significant in that specifying the spherical symmetry of space is an equivalent way of expressing the conservation of angular momentum. So, the Dirac equation effectively specifies three types of interacting potential under which angular momentum conservation is preserved, and we may imagine that these are also equivalent to specifications of the types of *charge* states that can be conserved.

Now, $V$ proportional to $1 / r$ gives the Coulomb solution for the electric force. $V$ proportional to $r$ gives the quark-confinement solution for the strong force. Any other polynomial-type $r$ dependence gives a harmonic oscillator. Let us suppose that this is *the solution for the weak force*. It doesn't actually matter what the shape of the function is; the solution will be a harmonic oscillator as long as it is not proportional to $r$ or $1 / r$. In fact, there is a good reason why it will not be either of these for the weak interaction. Essentially the weak interaction is *always dipolar*. It always involves one fermion-antifermion combination becoming another. This is not true of the electric interaction where the interacting fermions remain unchanged. We can therefore think of the weak charge as only manifesting itself when it is part of a dipole-dipole interaction. If this is true of the weak charge, then the weak *field* will behave in essentially the same way.

In addition, the weak charge should have a dipole moment because of its left-handedness for fermion and right-handedness for antifermion states. Searches for electric and strong dipole moments have produced negative results to many orders of magnitude, implying the indistinguishability of left- and right-handed states, but weak dipole moments have not even been conceived. Yet the single-handedness of the weak charge necessarily implies that a weak dipole moment should exist; and if weak interactions always involve a weak dipole, then a weak dipole moment, as the direct expression of single-handedness, should, in some sense, be the very *manifestation* of the weak interaction.

From this it would seem that any interaction between a weak dipole and a weak field will be manifested as a dipole-dipole or dipole-multipole interaction. In this case, the weak potential $V$ will be proportional to $r^{-n}$, where $n$ is 3 or greater, or to a polynomial incorporating terms of this kind; and, for a single weak charge (fermion or antifermion), taken separately, the action of the field will require a similar potential with $n = 2$ or greater. In either case, the solution of the Dirac equation will be a harmonic oscillator.

Now, the harmonic oscillator is a classic way of representing the production of spin ½ fermion (and antifermion) states emerging from, or disappearing into, the vacuum using the appropriate creation or annihilation operators; and these operators are, of course, essentially the same in principle as those used in Quantum Field Theory. So we can represent the action of the weak dipole moment as the *cause* of the production of fermion-antifermion combinations from the vacuum or of the reverse process of mutual annihilation. It is, of course, the presence of a single unit of positive or negative weak charge which distinguishes fermions from bosons or characterizes spin ½ states. In the case of neutrinos and antineutrinos, it is their *only* charge-related characteristic. It is the weak interaction, therefore, that is particularly significant in the production of fermion and antifermion states; and the fact that it must be single-handed produces, via the weak dipole moment, the driving mechanism for the creation (or annihilation) of spin ½ states.

The handedness is, of course, ultimately a vacuum property because it stems from the existence of a filled weak vacuum, with a zero point energy, with ½ quantum values, corresponding to the ½ quantum value for intrinsic fermion spin, that could be taken to be the physical manifestation of the weak dipole moment. What this means in effect is that the weak vacuum is in a continual state of proclaiming its filled state, by creating weak dipoles which have a dipole moment or specific handedness. So it is not surprising that 'fluctuations in this vacuum' are the same thing as the production or annihilation of a weak dipolar fermion-antifermion pair, each of spin ½, via a harmonic oscillator creation-annihilation mechanism. The same fluctuations are also responsible for the Casimir or Van der Waals force, which in the simplest case produces a $V$ proportional to $r^{-3}$ from the zero point energy, corresponding to the potential for a fluctuating dipole-dipole interaction. Their existence is proclaimed by the three 'vacuum' terms in the Dirac 4-spinor, one corresponding to each charge type (through the connection with the hidden $\boldsymbol{i}, \boldsymbol{j}, \boldsymbol{k}$ column operators), in addition to each Dirac solution, and, physically, by the *zitterbewegung*. It is the dipole moment that privileges the real fermion (or antifermion) over its accompanying vacuum states.

Significantly, all three interactions require a Coulombic phase term ($V$ proportional to $1 / r$) in a nilpotent version of the Dirac equation. In the case of the electric field, of course, this produces the entire interaction, but in all cases it represents spherical symmetry – or arbitrary direction of spin in the absence of a field external to the system. The harmonic oscillator case, however, uses a complex phase term, presumably relating to the complex nature of $E$ in the nilpotent state vector on the site corresponding to the position of the weak charge. It is the complex nature of this term which associates the two states of handedness with the respective concepts of fermion and antifermion, and which allows the possibility of pair creation or annihilation.

The complexity of the term, in this sense, drives the creation of a dipolar field, the algebra requiring simultaneous positive and negative solutions and privileging neither. The nilpotent nature of the Dirac operator also demands that one of the terms is complex; so a spin ½ or nilpotent state is, in principle, impossible without a complex aspect. It is significant that the three terms involved in the Dirac nilpotent operator – $iE$, $\mathbf{p}$ and $m$ – which we have associated with the respective weak, strong and electric charge operators $\boldsymbol{k}$, $\boldsymbol{i}$, $\boldsymbol{j}$, are different algebraic objects, being, respectively, pseudoscalar, vector and real scalar. While the pseudoscalar suggests a dipolar mechanism (with $V \propto r^{-n}$), one can imagine a gauge invariant vector term leading to a mechanism of constant rate of change of vector momentum with respect to distance (i.e. constant force, or $V \propto r$), while a gauge invariant real scalar structure produces a simple $U(1)$ symmetry (requiring $V \propto 1 / r$). These will be the appropriate variations with respect to $r$ applied to find the covariant form of the time differential.

The dipolar nature of the weak force is thus intrinsically connected with the complex associations of the weak charge. Complex numbers are not privileged as to sign, and the weak charge tends to behave in such a way as to make its sign irrelevant; fermion and antifermion are distinguishable, but $+w$ and $–w$ are not. Complex equations necessarily have dual solutions, and we can consider the weak charge as carrying with it its alternative sign as a vacuum image, the dipole moment thus created setting up the handedness which ensures chirality.

In principle, the space variations for the strong, electric and weak potentials in the covariant time derivative are definitions of all those states that are equivalent under the conditions of conservation of charge and energy. We avoid using the space derivative (with vector potential terms and time variation) by making our frame of reference the 'static' one, while the mass term in the differential operator is necessarily a constant. This allows us to use spherical spatial symmetry (or no variation with respect to spatial direction) to convert the problem to one conceived in terms of angular momentum, and by implication pure charge, conservation. In principle, then, the three interaction potentials are describing three separate requirements for conserving angular momentum – its behaviour with respect to space (the vector or directional term), its behaviour with respect to time (the pseudoscalar or helicity term), and its behaviour with respect to mass (the scalar or pure magnitude term), although all the terms necessarily have a scalar or magnitude component. (Ultimately, this means that it is not necessary, *in principle*, for a fermion state to have mass.)

Converting from the differential operator formalism to the state vector eigenvalue links the respective directional, helicity and magnitude terms with $\mathbf{p}$, $iE$ and $m$; while associating the potentials via their $r$-dependence with the respective ones produced by strong, weak and electric charges enables us to specify the aspects of angular momentum conservation which are preserved by the separate conservation properties of each charge type. It is significant that each charge type requires a separate statement of angular momentum conservation in the Dirac equation; so the presence or absence of each charge can be defined by a separate angular momentum operator. To define a unified fermion state it is convenient to assign angular momentum operators which are arbitrarily variable, but orthogonal, components of a single angular momentum pseudovector. Different fermionic states arise according to whether these are or are not aligned to each other, or to a defined projection of the actual angular momentum state of the fermion. All fermion states have weak charges because all necessarily have a helicity state, but, where the arbitrarily defined components are aligned with each other, separate information on charge conservation becomes unnecessary and the charge structure of the fermion will reflect this.

## 20 The charge structures of quarks and leptons

Fermions (that is, quarks and leptons) are states characterized by the presence of the weak charge, and to specify all possible fermion states we simply have to enumerate all particle states *which are indistinguishable from each other in terms of the weak interaction*. Essentially, the weak interaction cannot tell whether a strong or electric action interaction is also operating, and so fermions, for example, with strong charges (that is, quarks) ought to be indistinguishable by this interaction from particles without strong charges (that is, leptons). In terms of the weak interaction, quarks ought to be *lepton-like*. For quarks, also, the weak interaction cannot tell the difference between a filled 'electromagnetic vacuum' (or weak isospin up state) and an empty one (or weak isospin down state). The weak interaction, in addition, is also indifferent to the sign of the weak charge, and responds (via the vacuum) only to the status of fermion or antifermion; this results in mixing between the respective fermion *generations*, defined with $+w$, with $-w$ and $P$ violation, and with $-w$ and $T$ violation. From the sets of equally probable states thus specified (excluding energy considerations), we define all the possible distinctions between fermion / antifermion; quark / lepton; isospin up / isospin down; and the three quark-lepton generations. The distinctions are made in terms of the strong and electric charges, and of mass.

The process can be represented in terms of conservation of angular momentum, which we have already associated with the conservation of each of the charges. Taking $\boldsymbol{\sigma}.\hat{\mathbf{p}}$ (or $-\boldsymbol{\sigma}.\hat{\mathbf{p}}$, using the historically-established sign conventions for charges) as equivalent in unit charge terms to an expression in which $\hat{\mathbf{p}}$ becomes the unit vector components $\hat{\mathbf{p}}_1$, $\hat{\mathbf{p}}_2$, $\hat{\mathbf{p}}_3$, in successive phases of the strong interaction, and applying this to the strong charge quaternion operator $\boldsymbol{i}$, the units of strong charge will become $0\boldsymbol{i}$ or $1\boldsymbol{i}$, depending on the supposed instantaneous direction of the angular momentum vector. Only one component of a baryon will have this unit at any instant. In reality, of course, gauge invariance ensures that all possible phases exist at once, so spin becomes a property of the entire system and not of the component quarks.

The same angular momentum term ($\boldsymbol{\sigma}.\hat{\mathbf{p}}$) carries the information concerning the conservation of the other two charge terms; the three charges are, as we have seen, separately conserved because they represent three different aspects of the angular momentum conservation process. In the case of the weak charges, the random unit vector components $\hat{\mathbf{p}}_1$, $\hat{\mathbf{p}}_2$, $\hat{\mathbf{p}}_3$, are associated respectively with the sign of the angular momentum state, and, in the case of the electric interaction, they are associated with the magnitude. This occurs through the connections of $\mathbf{p}$ with $E$ and $\mathbf{p}$ with $m$. We can, thus, generalise the procedure by applying $\boldsymbol{\sigma}.\hat{\mathbf{p}}_1$, $\boldsymbol{\sigma}.\hat{\mathbf{p}}_2$, $\boldsymbol{\sigma}.\hat{\mathbf{p}}_3$ to the quaternion operators ($\boldsymbol{k}$ and $\boldsymbol{j}$) specifying $w$ and $e$, but with the sequence of unit vectors determined separately in each case. The various alignments between the sequences of unit vectors or *phases* applied to *s*, *w* and *e* then determine the nature of the fermion state produced.

If we align the unit vectors applied to *w* and *e*, we are effectively aligning the *E* and *m* phases with each other, and so necessarily with the **p** phase (by alignment of the magnitudes), which means that the system has a single phase and so cannot be baryonic. The **p** phase is defined with *E* and *m*, and there is no strong charge. We thus define a free fermion or lepton. In a baryon system, with strong charges present, the vectors assigned to the weak and electric charges, and hence to *E* and *m*, will not be aligned, and, consequently, the **p** phase is not fixed with respect to them.

To complete the representation of all possible fermions, we need to incorporate the effects of weak isospin, and the parity- and time-reversal-violations which will create second and third 'generations'. Reversal of isospin can be accomplished by replacing a term such as $-j\hat{\mathbf{p}}_1$ with $-j(\hat{\mathbf{p}}_1 - \mathbf{1})$, the $j\mathbf{1}$ representing the filled 'electric vacuum' state. Charge conjugation violation may be represented by the non-algebraic symbols $z_P$ and $z_T$, depending on whether it is accompanied by *P* or *T* violation. In using these symbols, we are merely saying that we are treating the *–w* of the second and third generations as though it were positive in the same way as the *w* of the first generation. We can now express quark structures in the following form:

| | |
|---|---|
| down | $-\boldsymbol{\sigma}.\,(-j\hat{\mathbf{p}}_a + i\hat{\mathbf{p}}_b + k\hat{\mathbf{p}}_c)$ |
| up | $-\boldsymbol{\sigma}.\,(-j(\hat{\mathbf{p}}_a - \mathbf{1}) + i\hat{\mathbf{p}}_b + k\hat{\mathbf{p}}_c)$ |
| strange | $-\boldsymbol{\sigma}.\,(-j\hat{\mathbf{p}}_a + i\hat{\mathbf{p}}_b - z_P k\hat{\mathbf{p}}_c)$ |
| charmed | $-\boldsymbol{\sigma}.\,(-j(\hat{\mathbf{p}}_a - \mathbf{1}) + i\hat{\mathbf{p}}_b - z_P k\hat{\mathbf{p}}_c)$ |
| bottom | $-\boldsymbol{\sigma}.\,(-j\hat{\mathbf{p}}_a + i\hat{\mathbf{p}}_b - z_T k\hat{\mathbf{p}}_c)$ |
| top | $-\boldsymbol{\sigma}.\,(-j(\hat{\mathbf{p}}_a - \mathbf{1}) + i\hat{\mathbf{p}}_b - z_T k\hat{\mathbf{p}}_c)$ |

In this representation, $-j$ stands for electric charge (which is conventionally negative), $i$ for strong, $k$ for weak. *a*, *b*, *c* are *each* randomly 1, 2, 3, except that $b \neq c$. Both $-z_P k$ and $-z_T k$ become equivalent to $k$, for the purposes of the weak interaction. For the corresponding leptons, the components are all in phase $(\hat{\mathbf{p}}_a)$, and there is no directional component:

| | |
|---|---|
| electron | $-\boldsymbol{\sigma}.\,(-j\hat{\mathbf{p}}_a + k\hat{\mathbf{p}}_a)$ |
| *e* neutrino | $-\boldsymbol{\sigma}.\,(-j(\hat{\mathbf{p}}_a - \mathbf{1}) + k\hat{\mathbf{p}}_a)$ |
| muon | $-\boldsymbol{\sigma}.\,(-j\hat{\mathbf{p}}_a - z_P k\hat{\mathbf{p}}_a)$ |
| $\mu$ neutrino | $-\boldsymbol{\sigma}.\,(-j(\hat{\mathbf{p}}_a - \mathbf{1}) - z_P k\hat{\mathbf{p}}_a)$ |
| tau | $-\boldsymbol{\sigma}.\,(-j\hat{\mathbf{p}}_a - z_T k\hat{\mathbf{p}}_a)$ |
| $\tau$ neutrino | $-\boldsymbol{\sigma}.\,(-j(\hat{\mathbf{p}}_a - \mathbf{1}) - z_T k\hat{\mathbf{p}}_a)$ |

Both antiquarks and antileptons simply replace $-\boldsymbol{\sigma}$ with $\boldsymbol{\sigma}$.

# 21 A unified representation for quarks / leptons

It is possible to incorporate all the information outlined in the previous section into a single unified representation for the entire set of charge structures for quarks and leptons (and their antistates):

$$\boldsymbol{\sigma}_z.(\boldsymbol{i}\,\hat{\mathbf{p}}_a\,(\delta_{bc}-1)+\boldsymbol{j}\,(\hat{\mathbf{p}}_b-\mathbf{1}\delta_{0m})+\boldsymbol{k}\,\hat{\mathbf{p}}_c\,(-1)^{\delta_{1g}}\,g)$$

As previously, the quaternion operators $\boldsymbol{i}$, $\boldsymbol{j}$, $\boldsymbol{k}$ are respectively strong, electric and weak charge units; $\boldsymbol{\sigma}_z$ is the spin pseudovector component defined in the $z$ direction (here used as a reference); $\hat{\mathbf{p}}_a$, $\hat{\mathbf{p}}_b$, $\hat{\mathbf{p}}_c$ are each units of quantized angular momentum, selected *randomly* and *independently* from the three orthogonal components $\hat{\mathbf{p}}_x$, $\hat{\mathbf{p}}_y$, $\hat{\mathbf{p}}_z$. $\boldsymbol{\sigma}_z$ and the remaining terms are logical operators representing existence conditions, and defining four fundamental divisions in fermionic states. Each of the operators creates one of these fundamental divisions – fermion / antifermion; quark / lepton (colour); weak up isospin / weak down isospin; and the three generations – which are identified, respectively, by the weak, strong, electromagnetic and gravitational interactions.

(1) $\boldsymbol{\sigma}_z = -\mathbf{1}$ defines left-handed states; $\boldsymbol{\sigma}_z = \mathbf{1}$ defines right-handed. For a filled weak vacuum, left-handed states are predominantly fermionic, right-handed states become antifermionic 'holes' in the vacuum.

(2) $b = c$ produces leptons; $b \neq c$ produces quarks. If $b \neq c$ we are obliged to take into account the three directions of $\mathbf{p}$ at once. If $b = c$, we can define a single direction. Taking into account all three directions at once, we define baryons composed of three quarks (and mesons composed of quark and antiquark), in which each of $a$, $b$, $c$ cycle through the directions $x$, $y$, $z$.

(3) $m$ is an electromagnetic mass unit, which selects the state of weak isospin. It becomes 1 when present and 0 when absent. So $m = 1$ is the weak isospin up state; and $m = 0$ weak isospin down. The unit condition can be taken as an empty electromagnetic vacuum; the zero condition a filled one.

(4) $g$ represents a conjugation of weak charge units, with $g = -1$ representing maximal conjugation. If conjugation fails maximally, then $g = 1$. $g$ can also be thought of as a composite term, containing a parity element ($P$) and a time-reversal element ($T$). So, there are two ways in which the conjugated $PT$ may remain at the unconjugated value (1). $g = -1$ produces the generation $u$, $d$, $\nu_e$, $e$; $g = 1$, with $P$ responsible, produces $c$, $s$, $\nu_\mu$, $\mu$; $g = 1$, and, with $T$ responsible, produces $t$, $b$, $\nu_\tau$, $\tau$.

The weak interaction can only identify (1). This occupies the $\boldsymbol{ik}E$ site in the anticommuting Dirac pentad ($\boldsymbol{ik}E + \boldsymbol{i}\mathbf{p} + \boldsymbol{j}m$), with the $\boldsymbol{i}$ term being responsible for the fermion / antifermion distinction. Because it is attached to a complex operator, the sign of $\boldsymbol{k}$ has two possible values even when those of $\boldsymbol{i}$ and $\boldsymbol{j}$ are fixed; the sign of the weak charge associated with $\boldsymbol{k}$ can therefore only be determined physically by the sign of $\boldsymbol{\sigma}_z$. The filled weak vacuum is an expression of the fact that the 'ground state of the universe' can be specified in terms of positive, but not negative, energy ($E$), because, physically, this term represents a continuum state.

The strong interaction identifies (2). This occupies the $i\mathbf{p}$ (or $i\boldsymbol{\sigma}.\mathbf{p}$) site and it is the three-dimensional aspect of the $\mathbf{p}$ (or $\boldsymbol{\sigma}.\mathbf{p}$) term which is responsible for the three-dimensionality of quark 'colour'. A separate 'colour' cannot be identified any more successfully than a separate dimension, and the quarks become part of a system, the three parts of which have $\hat{\mathbf{p}}_a$ values taking on one each of the orthogonal components $\hat{\mathbf{p}}_x$, $\hat{\mathbf{p}}_y$, $\hat{\mathbf{p}}_z$. Meson states have corresponding values of $\hat{\mathbf{p}}_a$, $\hat{\mathbf{p}}_b$ and $\hat{\mathbf{p}}_c$ in the fermion and antifermion components, although the logical operators $\delta_{0m}$ and $(-1)^{\delta_{1g}} g$ may take up different values for the fermion and the antifermion, and the respective signs of $\boldsymbol{\sigma}_z$ are opposite.

The electromagnetic interaction identifies (3). This occupies the $j\mathbf{m}$ site in the Dirac pentad. Respectively, the three interactions ensure that the orientation, direction and magnitude of angular momentum are separately conserved. Gravity (mass), finally, identifies (4).

The charge conjugation from $-w$ to $w$, in the second and third generations, which is represented in the previous section by $z_P$ or $z_T$, is brought about, as we have said, by the filled weak vacuum needed to avoid negative energy states. The two weak isospin states are associated with this idea in (3), the $\mathbf{1}$ in $(\hat{\mathbf{p}}_b - \mathbf{1}\delta_{0m})$ being a 'filled' state, with its absence an unfilled state, and the weak interaction acts by annihilating and creating $e$, either filling the vacuum or emptying it – which is why, unlike the strong interaction, it always involves the equivalent of particle + antiparticle = particle + antiparticle, and involves a massive intermediate boson. We thus create two possible vacuum states to allow variation of the sign of electric charge by weak isospin, and this variation is linked to the filling of the vacuum which occurs in the weak interaction, and could be connected with a mass-related 'bosonic' spin 0 linking of the two isospin states (in addition to the spin 1 gauge bosons involved in the interaction). The weak and electric interactions are linked by this filled vacuum in the $SU(2)_L \times U(1)$ model, as they are in our description of weak isospin, and we can regard these as alternative formalisms for representing the same physical truth. It is significant that the Higgs mechanism for generating masses of intermediate weak bosons and fermions requires the same Higgs vacuum field both for $SU(2)_L$ and for $U(1)$. In addition, the combination of scalar and pseudoscalar phases in the mathematical description of the combined electric and weak interactions clearly relates to the use of a complex scalar field in the conventional derivation of the Higgs mechanism.

## 22 Phase diagrams for charge conservation

Phase diagrams provide a useful way of picturing lepton, baryon and meson charge structures. In the case of the strong interaction, only one component of angular momentum is well-defined at any moment, and the strong charge appears to act in such a way that the well-defined direction manifests itself by 'privileging' one out of three independent phases making up the complete phase cycle. In a truly gauge invariant system, this can only be accomplished in relative terms. If the weak and electric charges are also related to angular momentum, then the same must apply to them, and the relative 'privileging' of phase can only be defined between the different interactions. We have, here, two options. If the 'privileged' or 'active' phases of $E$ and $m$ (or $w$ and $e$) coincide with each other, then this also determines the 'privileged' phase of $\mathbf{p}$; the result is no 'privileged' relative phase. Since the strong charge is defined only through the directional variation of $\mathbf{p}$, via a 'privileged' relative phase, a system in which the phases coincide cannot be strongly bound. If, however, they are different, then this information can only be carried through $\mathbf{p}$ (or $s$), and the strong interaction must be present.

We can imagine the arrangements diagrammatically using a rotating vector to represent the 'privileged' direction states for the charges. Each charge has only one 'active' phase out of three at any one time to fix the angular momentum direction; the symbols $e$, $s$, and $w$, here refer to these states, not the actual charges. The vectors may be thought of as rotating over a complete spherical surface. In the case of the quark-based states – baryons and mesons – the total information about the angular momentum state is split between three axes, whereas the lepton states carry all the information on a single axis

The axes in Figure 3 represent both charge states and angular momentum states for leptons, mesons and baryons. As previously stated, each type of charge carries a different aspect of angular momentum (or helicity) conservation; $s$ carries the directional information (linked to $\mathbf{p}$); $w$ carries the sign information (+ or – helicity) (linked to $iE$); $e$ carries information about magnitude (linked to $m$). Another way of looking at this is to associate these properties, respectively, with the symmetries of rotation, inversion, and translation.

**Figure 3**

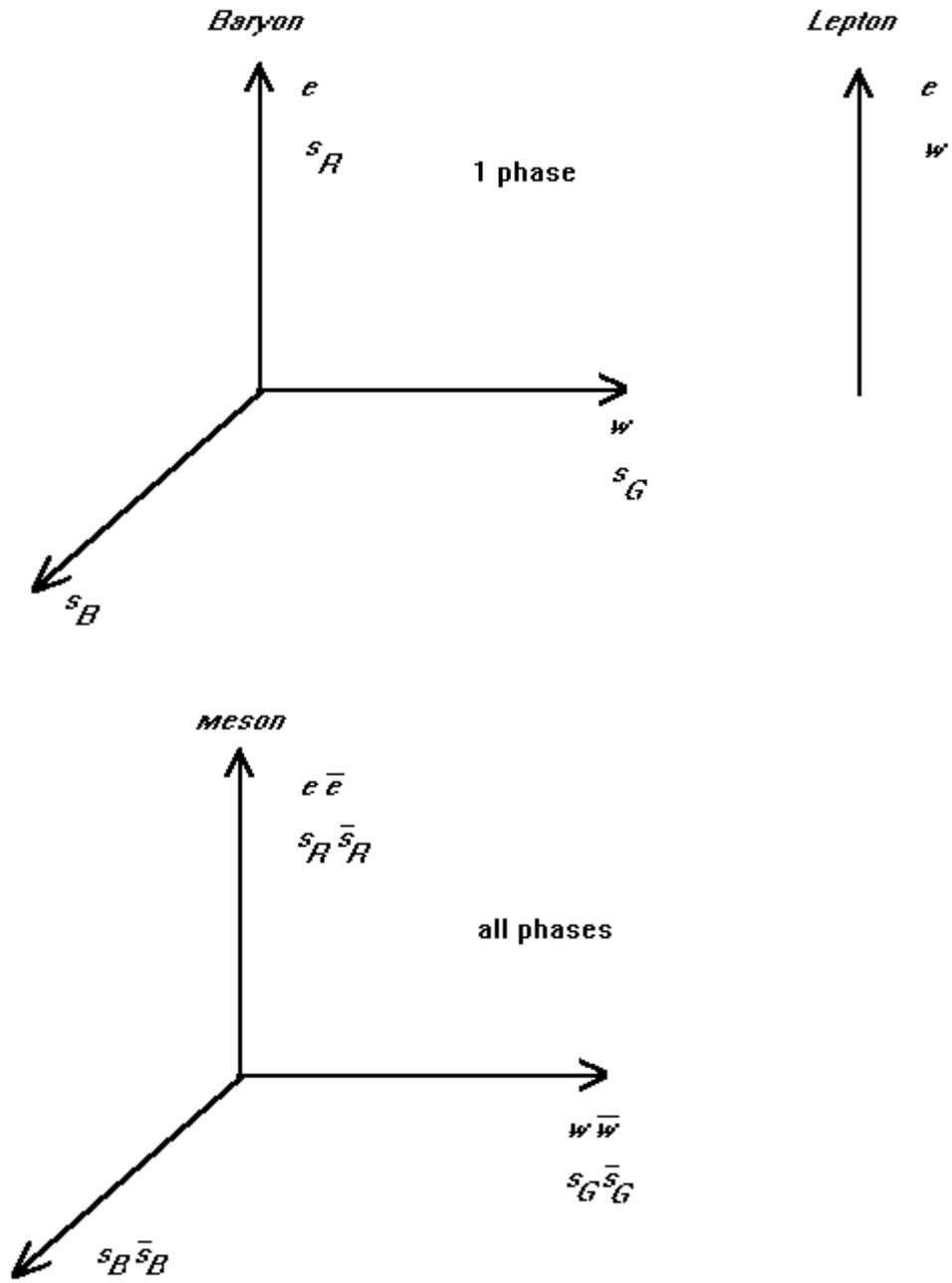

## 23 Quark and lepton charge structures

From both the separate formulae in section 20 and the unified representation in section 21, the 0 and 1 charge structures of the fundamental fermions may be expressed in terms of a set of three 'quark' tables, A-C, with an extra table L for the left-handed leptons and antileptons:[13,16]

### A

|     |        | B        | G   | R   |
|-----|--------|----------|-----|-----|
| $u$ | $+e$   | 1$j$     | 1$j$ | 0$i$ |
|     | $+s$   | 1$i$     | 0$k$ | 0$j$ |
|     | $+w$   | 1$k$     | 0$i$ | 0$k$ |
|     |        |          |     |     |
| $d$ | $-e$   | 0$j$     | 0$k$ | 1$j$ |
|     | $+s$   | 1$i$     | 0$i$ | 0$k$ |
|     | $+w$   | 1$k$     | 0$j$ | 0$i$ |
|     |        |          |     |     |
| $c$ | $+e$   | 1$j$     | 1$j$ | 0$i$ |
|     | $+s$   | 1$i$     | 0$k$ | 0$j$ |
|     | $-w$   | $z_P k$  | 0$i$ | 0$k$ |
|     |        |          |     |     |
| $s$ | $-e$   | 0$j$     | 0$k$ | 1$j$ |
|     | $+s$   | 1$i$     | 0$i$ | 0$k$ |
|     | $-w$   | $z_P k$  | 0$j$ | 0$i$ |
|     |        |          |     |     |
| $t$ | $+e$   | 1$j$     | 1$j$ | 0$i$ |
|     | $+s$   | 1$i$     | 0$k$ | 0$j$ |
|     | $-w$   | $z_T k$  | 0$i$ | 0$k$ |
|     |        |          |     |     |
| $b$ | $-e$   | 0$j$     | 0$k$ | 1$j$ |
|     | $+s$   | 1$i$     | 0$i$ | 0$k$ |
|     | $-w$   | $z_T k$  | 0$j$ | 0$i$ |
|     |        |          |     |     |

### B

|     |        | B        | G   | R   |
|-----|--------|----------|-----|-----|
| $u$ | $+e$   | 1$j$     | 1$j$ | 0$k$ |
|     | $+s$   | 0$i$     | 0$k$ | 1$i$ |
|     | $+w$   | 1$k$     | 0$i$ | 0$j$ |
|     |        |          |     |     |
| $d$ | $-e$   | 0$i$     | 0$k$ | 1$j$ |
|     | $+s$   | 0$j$     | 0$i$ | 1$i$ |
|     | $+w$   | 1$k$     | 0$j$ | 0$k$ |
|     |        |          |     |     |
| $c$ | $+e$   | 1$j$     | 1$j$ | 0$k$ |
|     | $+s$   | 0$i$     | 0$k$ | 1$i$ |
|     | $-w$   | $z_P k$  | 0$i$ | 0$j$ |
|     |        |          |     |     |
| $s$ | $-e$   | 0$i$     | 0$k$ | 1$j$ |
|     | $+s$   | 0$j$     | 0$i$ | 1$i$ |
|     | $-w$   | $z_P k$  | 0$j$ | 0$k$ |
|     |        |          |     |     |
| $t$ | $+e$   | 1$j$     | 1$j$ | 0$k$ |
|     | $+s$   | 0$i$     | 0$k$ | 1$i$ |
|     | $-w$   | $z_T k$  | 0$i$ | 0$j$ |
|     |        |          |     |     |
| $b$ | $-e$   | 0$i$     | 0$k$ | 1$j$ |
|     | $+s$   | 0$j$     | 0$i$ | 1$i$ |
|     | $-w$   | $z_T k$  | 0$j$ | 0$k$ |
|     |        |          |     |     |

| | | **B** | **G** | **R** |
|---|---|---|---|---|
| $u$ | $+e$ | $1j$ | $1j$ | $0k$ |
| | $+s$ | $0i$ | $1i$ | $0j$ |
| | $+w$ | $1k$ | $0k$ | $0i$ |
| | | | | |
| $d$ | $-e$ | $0j$ | $0k$ | $1j$ |
| | $+s$ | $0i$ | $1i$ | $0k$ |
| | $+w$ | $1k$ | $0j$ | $0i$ |
| | | | | |
| $c$ | $+e$ | $1j$ | $1j$ | $0k$ |
| | $+s$ | $0i$ | $1i$ | $0j$ |
| | $-w$ | $z_P k$ | $0k$ | $0i$ |
| | | | | |
| $s$ | $-e$ | $0j$ | $0k$ | $1j$ |
| | $+s$ | $0i$ | $1i$ | $0k$ |
| | $-w$ | $z_P k$ | $0j$ | $0i$ |
| | | | | |
| $t$ | $+e$ | $1j$ | $1j$ | $0k$ |
| | $+s$ | $0i$ | $1i$ | $0j$ |
| | $-w$ | $z_T k$ | $0k$ | $0i$ |
| | | | | |
| $b$ | $-e$ | $0j$ | $0k$ | $1j$ |
| | $+s$ | $0i$ | $1i$ | $0k$ |
| | $-w$ | $z_T k$ | $0j$ | $0i$ |
| | | | | |

| | | $\overline{e}$ | $\overline{e}$ | $\nu_e$ |
|---|---|---|---|---|
| | $+e$ | $1j$ | $1j$ | $0j$ |
| | $+s$ | $0k$ | $0i$ | $0i$ |
| | $+w$ | $0i$ | $0k$ | $1k$ |
| | | | | $e$ |
| | $-e$ | $0i$ | $0k$ | $1j$ |
| | $+s$ | $0j$ | $0i$ | $0i$ |
| | $+w$ | $0k$ | $0j$ | $1k$ |
| | | $\overline{\mu}$ | $\overline{\mu}$ | $\nu_u$ |
| | $+e$ | $1j$ | $1j$ | $0j$ |
| | $+s$ | $0k$ | $0i$ | $0i$ |
| | $-w$ | $0i$ | $0k$ | $z_P k$ |
| | | | | $\mu$ |
| | $-e$ | $0i$ | $0k$ | $1j$ |
| | $+s$ | $0j$ | $0i$ | $0i$ |
| | $-w$ | $0k$ | $0j$ | $z_P k$ |
| | | $\overline{\tau}$ | $\overline{\tau}$ | $\nu_\tau$ |
| | $+e$ | $1j$ | $1j$ | $0j$ |
| | $+s$ | $0k$ | $0i$ | $0i$ |
| | $-w$ | $0i$ | $0k$ | $z_T k$ |
| | | | | $\tau$ |
| | $-e$ | $0i$ | $0k$ | $1j$ |
| | $+s$ | $0j$ | $0i$ | $0i$ |
| | $-w$ | $0k$ | $0j$ | $z_T k$ |
| | | | | |

Applying these to the known fermions, A-C would appear to have all the properties of the coloured quark system, with $s$ (or the $A^{con}$ term in the covariant derivative) pictured as being 'exchanged' between the three states (although in reality, of course, all the states exist simultaneously), in the same way as the operator **p** in the nilpotent baryon wavefunction. We can see symmetry-breaking, in general, as a consequence of the setting up of the algebraic model for charges. When we map time, space and mass onto the charges $w$-$s$-$e$, to create the anticommuting Dirac pentad, only one charge ($s$) has the full range of vector options. 'Fixing' one of the others (say $e$) for $s$ to vary against, gives us only 2 remaining options for $w$, unit on the same colour as $e$ or unit on a different one. Putting both $w$ and $e$ on the same colour denies the necessary three degrees of freedom in the direction of angular momentum, so this is forbidden in a quark system.

The tables explain many facts related to particle physics, and also make some new predictions. For example, if we derive baryon and meson charge structures from those of the component quarks and antiquarks, we find that all baryons have a weak charge structure of *w*, while all mesons have a weak charge structure 0, with the exception of states like the *K* mesons, and the other meson states combining a fermion and antifermion of two different generations. For these states, we find alternative weak charge structures of 0 or $\pm$ (1 + *z*)*w*, where *z* may be $z_P$ or $z_T$. (Mesons combining the second and third generations will have structures combining $z_P$ and $z_T$.) The alternatives depend on the particular colour-anticolour combination, and particular quark representation, we choose. Clearly, 0 and $\pm$ (1 + *z*)*w* have to be indistinguishable in weak terms, but we have already broken a symmetry (either *P* or *T*) in creating *z*, which means that we are also obliged to break another to maintain overall *CPT* invariance. In the case of a bosonic state, charge conjugation must be preserved, so we are obliged to break *CP* (or *T*) as well as *P*, or *CT* (or *P*) as well as *T*. A prediction may be made that such an additional violation will be found in all states of this kind. It has already been observed in $K^o$ and $\overline{K}^o$ mixed states, and is now known to occur in $K^o$ and $\overline{K}^o$, taken separately, as well as being extended to incorporate mesons combining first and third generation components. However, it should also be observable in $K^+$ and $K^-$, and in the equivalent states in other generations. It should also be observed in the weak decays of Bose-Einstein condensates, which again involve a (1 + *z*)*w*, weak charge structure.

## 24 Lepton-like quarks

Both the formulae and the tables suggest that quarks are fundamentally lepton-like objects, with similar fundamental charge structures. In such a theory, the fractional electric charges observed (indirectly) for quarks (2*e* / 3 for *u*, *c*, *t*, or −*e* / 3 for *d*, *s*, *b*), in experiments such as the ratio of hadron / muon production in electron-positron annihilation events, and the rate of decay of neutral pions to two photons, are attributable to the absolutely unbroken gauge invariance of the strong interaction, which means that individual phases of the interaction will never be observed.

Lepton-like quarks have, of course, a long history. They appeared in the original paper by Han and Nambu which introduced the concept of colour to explain the strong interaction.[23] The Han and Nambu quarks had electric charge assignments of the form:

|        | Blue | Green | Red  |
|--------|------|-------|------|
| up     | *e*  | *e*   | 0    |
| down   | 0    | 0     | −*e* |

by contrast with those assumed in the first version of the quark theory proposed in the previous year by Gell-Mann and Zweig:

|        | Blue    | Green   | Red     |
|--------|---------|---------|---------|
| up     | 2*e* / 3 | 2*e* / 3 | 2*e* / 3 |
| down   | −*e* / 3 | −*e* / 3 | −*e* / 3 |

The theory had the advantage of making the 'colour' differences a natural result of the existence of different charge structures rather than an arbitrarily added extra property, though 'colour' was later added also to the Gell-Mann-Zweig version. The original implication of Han and Nambu's theory may have been that, at some sufficiently high energy, the integral nature of the charges would become manifest and the colours directly revealed. However, under conditions of perfect gauge invariance or perfect infrared slavery, this transition would never occur, and the Han-Nambu model would provide a precise way of predicting the observation of fractional charges. Close has expressed it in the following way: 'Imagine what would happen if the colour nonsinglets were pushed up to infinite masses. Clearly only colour **1** [singlets] would exist as physically observable states and quarks would in consequence be permanently confined. At any finite energy we would only see the 'average' quark changes and phenomenonologically we could not distinguish this from the Gell-Mann model where the quarks form three identical triplets.'[24] In this picture, the observed fractional charges are not even 'averages', but exact values, because they reflect an effectively infinite rate of 'rotation' between the coloured states or phases. They are QED or electroweak eigenstates.

As it happens, an almost exactly parallel phenomenon has been observed in condensed matter, in the fractional quantum Hall effect. Here, ensembles of particles with only exact units of $e$ acquire the characteristics of perfect (odd) fractions of this unit, by becoming associated with the odd number of magnetic flux lines needed to create an overall boson state. Thus, if an electron becomes attached to 3 flux lines, its charge is divided between them in units of $e / 3$. Laughlin explained this, with reference to the work of Anderson, as 'a low-energy collective effect of huge numbers of particles that cannot be deduced from the microscopic equations of motion in a rigorous way and that disappears completely when the system is taken apart'.[25,26] In his 1998 Nobel Lecture, he even suggested the connection with particle physics: 'The fractional quantum Hall effect is fascinating for a long list of reasons, but it is important in my view primarily for one: It establishes experimentally that both particles carrying an exact fraction of the electron charge $e$ and powerful gauge forces between these particles, two central postulates of the standard model of elementary particles, can arise spontaneously as emergent phenomena. Other important aspects of the standard model, such as free fermions, relativity, renormalizability, spontaneous symmetry breaking, and the Higgs mechanism, already have apt solid-state analogues and in some cases were even modeled after them (Peskin, 1995), but fractional quantum numbers and gauge fields were thought to be fundamental, meaning that one had to postulate them. This is evidently not true.'[27,28]

There are fundamental problems with the assumption that the charges of the Gell-Mann-Zweig formalism are intrinsic, rather than 'emergent'. Thus, either charge is not properly quantized, or quarks and leptons are not truly fundamental, because their charges do not come in fundamental units. We can, of course, choose to redefine the fundamental unit of charge as $e / 3$, rather than $e$, but this would make a particle with two units of charge, like *u*, less fundamental in some respect than one with a single unit, such as *d*; and electrons, which give no evidence of being composite, would need three units of this new fundamental charge. Again, lepton-like quarks would seem to be required by grand unified theories which propose a single overall unification scheme for quarks and leptons, with possible quark-lepton transitions. Different fundamental units of charge, however, would certainly make this difficult to accomplish.

The Han-Nambu proposal has often been seen as an 'alternative' to the Gell-Mann-Zweig fractional theory, which, although never experimentally refuted, has gradually fallen into disuse because of its less direct relationship to QED phenomenology, but we should really see the two models as being different representations of a more fundamental underlying theory. They are, thus, not alternative theories, but different parts of the same one. The Gell-Mann model is really a representation of the strong interaction between quarks, which forces their electric charges to be phenomenologically fractional. The Han-Nambu model seems to suggest the underlying group structure which enables us to propose a scheme of grand unification, and makes sense of the fundamental nature of electric charge. In fact, as we will show, the group theoretical aspects of this representation allow us to propose an *exact* grand unification of the weak, strong and electric charges at the Planck mass, $(\hbar c / G)^{1/2}$, the energy at which quantum gravity becomes significant. In addition, they resolve an anomaly in the application of the Higgs mechanism, which cannot be resolved in a group representation based on fractional charges.

## 25 The grand unification group

The proposal in the previous section suggests that, although electric charge-related *phenomenology* is determined by the fractional charges generated by the perfect gauge invariance of the strong interaction, the *gauge relations between interactions* must reflect the more fundamental underlying lepton-like quark structures producing the observed effects. Grand unification of the three non-gravitational forces is currently believed to occur at an energy of order $10^{15}$ GeV, which is about four orders below the Planck mass ($M_P$). Minimal $SU(5)$, however, the main model used to make the prediction, which is based on the simplest available group incorporating $SU(3) \times SU(2) \times U(1)$, is known to be seriously flawed. In the first place, it fails to predict an exact convergence of the three interactions. In addition, the assumed electroweak mixing parameter, $\sin^2\theta_W = 0.375$, is very different from the experimental value of 0.231, and has different values for quarks and leptons. Although a 'renormalization' procedure can be adopted to reduce the predicted value (assumed to be that for grand unification) to about 0.21 at the *Z* boson mass ($M_Z$), a reapplication of the renormalized value to the equations for the coupling constants leads to a completely contradictory grand unification value of 0.6! Again, the supposed grand unification is imperfect: the weak and strong coupling constants are assumed to be exactly unified, but the electric coupling constant occurs only in a mixed state with the weak one via an assumed group structure.

Here, it is proposed that the true group structure can be found through the direct connection between the charge states and the Dirac nilpotent state vector in which they are incorporated.[4,7,29] The five charge units ($e$, $s_G$, $s_B$, $s_R$, $w$, taking into account the vector nature of $s$) map directly onto the five Dirac operators ($i\mathbf{k}$; $i\mathbf{i}$; $j\mathbf{i}$; $k\mathbf{i}$; $j$), and the five quantities ($m$, $p_x$, $p_y$, $p_z$, $E$) involved in the Dirac equation, and generate both an overall $SU(5)$ and its breakdown to $SU(3) \times SU(2) \times U(1)$. The 24 $SU(5)$ generators can be represented in terms of any of these units. For example:

|       | $\overline{s_G}$ | $\overline{s_B}$ | $\overline{s_R}$ | $\overline{w}$ | $\overline{e}$ |
|-------|------------------|------------------|------------------|----------------|----------------|
| $s_G$ |                  |                  |                  |                |                |
| $s_B$ |                  | Gluons           |                  | $Y$            | $X$            |
| $s_R$ |                  |                  |                  |                |                |
| $w$   |                  | $Y$              |                  | $Z^0, \gamma$  | $W^-$          |
| $e$   |                  | $X$              |                  | $W^+$          | $Z^0, \gamma$  |

or:

|       | $\overline{p_x}$ | $\overline{p_y}$ | $\overline{p_z}$ | $\overline{E}$ | $\overline{m}$ |
|-------|------------------|------------------|------------------|----------------|----------------|
| $p_x$ |                  |                  |                  |                |                |
| $p_y$ |                  | Gluons           |                  | $Y$            | $X$            |
| $p_z$ |                  |                  |                  |                |                |
| $E$   |                  | $Y$              |                  | $Z^0, \gamma$  | $W^-$          |
| $m$   |                  | $X$              |                  | $W^+$          | $Z^0, \gamma$  |

The only unobserved generators here are $X$ and $Y$, which earlier $SU(5)$ schemes have taken to imply direct proton decay. However, such decay would be forbidden by separate charge conservation rules, as it involves the complete elimination of a strong charge unit. Here, the $X$ and $Y$ generators remain linked to the particle + antiparticle mechanism of the ordinary weak interaction, and it may be that, below grand unification energies, they are connected with nothing more exotic than the ordinary process of beta decay, which links strong and electroweak interactions in the manner required.

$SU(5)$, however, is not the full story. If we had a 25[th] generator (which the Standard Model disregards on the grounds that it is not observed) the group would become $U(5)$, and all the generators would be entirely equivalent to scalar phases. Such a particle, if it existed, would couple to all matter in proportion to the amount, and, as a colour singlet, would be ubiquitous. Gravity suggests itself here (possibly through a spin 1 generator for the inertial reaction, rather than a spin 2, or other hypothetical, generator for gravity itself), and this links up with the demonstration that grand unification occurs at the Planck mass. The gravity generator would then link the 2 colourless gluons with the $Z^0$ and $\gamma$, along the diagonal of the group table, suggesting a link between all four interactions. Reduction of the generators to scalar phases would mean that, at grand unification, all interactions would be identical in effect, and all non-Coulombic structure would disappear. The unification would be exact.

## 26 A Dirac equation for charge

The analogy between the components of the Dirac equation $E$-$\mathbf{p}$-$m$ and the charge structures of $w$-$s$-$e$ is so close that we can represent energy conservation and charge conservation by equations of the same type. Here, it is most convenient to begin with one of the more standard forms of the Dirac equation

$$(\boldsymbol{\alpha}.\mathbf{p} + \beta m - E)\,\psi = 0\ ,$$

which we expand, using a $4 \times 4$ matrix, to

$$(\boldsymbol{\alpha}.\mathbf{p} + \beta m - E)\,\psi = \begin{pmatrix} -E & 0 & im & -ip \\ 0 & -E & ip & im \\ -im & -ip & -E & 0 \\ ip & -im & 0 & -E \end{pmatrix} \begin{pmatrix} \psi_1 \\ \psi_2 \\ \psi_3 \\ \psi_4 \end{pmatrix} = 0\ .$$

The column vector, here, is the usual 4-component spinor, and the terms $E$ and $\mathbf{p}$ represent the quantum differential operators rather than their eigenvalues.

An expression for conserved charge can then be obtained by taking a product of a $4 \times 4$ matrix and a 4-component column vector in the same way as conserved energy in the Dirac equation:

$$\begin{pmatrix} kw & 0 & -je & -is \\ 0 & kw & -is & je \\ -je & is & -kw & 0 \\ is & je & 0 & -kw \end{pmatrix} \begin{pmatrix} kw+is+je \\ kw+is-je \\ -kw-is+je \\ -kw-is-je \end{pmatrix} \tag{7}$$

The $4 \times 4$ matrix used here is almost identical in form to the matrix for the Dirac differential operator, although the + and − signs are in different places. The $s$ term effectively takes up the vector-type properties of $\mathbf{p}$, and can be represented as a vector with a single well-defined direction. The sign applied to $e$ is that of the charge itself, but $e$ has the added property of isospin, so that the $e$'s on the first and fourth rows of the matrix and on the first and fourth rows of the column vector can be considered as isospin 'up' and the others as isospin 'down'. The opposite states of isospin are not + and − but 1 and 0. So, we should apply to these $e$ terms the matrices:

$$\begin{pmatrix} 1 \\ 0 \end{pmatrix}; \begin{pmatrix} 0 \\ 1 \end{pmatrix}; \begin{pmatrix} -1 \\ 0 \end{pmatrix}; \begin{pmatrix} 0 \\ -1 \end{pmatrix}.$$

The result of this is that all terms involving $e$ disappear on multiplication.

Multiplying out the expression in (7) results in a product consisting of a unit column vector times a scalar factor. If we apply the factor $i$ to the $s$ term in the column vector, we derive $w^2 - s^2$, which becomes 0 when $w = s = \pm 1$. We can write this out in the form:

$$\begin{pmatrix} kw & 0 & -ije\uparrow & -iis \\ 0 & kw & -iis & ije\downarrow \\ -ije\downarrow & iis & -kw & 0 \\ iis & ije\uparrow & 0 & -kw \end{pmatrix} \begin{pmatrix} kw + iis + ije\uparrow \\ kw + iis - ije\downarrow \\ -kw - iis + ije\downarrow \\ -kw - iis - ije\uparrow \end{pmatrix} = 0 \qquad (8)$$

where ↑ represents isospin up and ↓ isospin down. If we now create an exponential term $e^{-i(wt - \mathbf{s}.\mathbf{r})}$, to produce a state vector for charge, and define $i\partial/\partial t = -iw$ and $-i\boldsymbol{\nabla} = i\mathbf{s}$, we obtain (9):

$$\begin{pmatrix} ik\partial/\partial t & 0 & -ije\uparrow & -i\boldsymbol{\nabla} \\ 0 & ik\partial/\partial t & -i\boldsymbol{\nabla} & ije\downarrow \\ -ije\downarrow & i\boldsymbol{\nabla} & -ik\partial/\partial t & 0 \\ i\boldsymbol{\nabla} & ije\uparrow & 0 & -ik\partial/\partial t \end{pmatrix} \begin{pmatrix} kw + iis + ije\uparrow \\ kw + iis - ije\downarrow \\ -kw - iis + ije\downarrow \\ -kw - iis - ije\uparrow \end{pmatrix} e^{-i(wt - \mathbf{s}.\mathbf{r})} = 0 .$$

The weak isospin terms cancel, suggesting why this becomes the phase term. We can therefore write equation (8) as:

$$\begin{pmatrix} kw & 0 & 0 & -iis \\ 0 & kw & -iis & 0 \\ 0 & iis & -kw & 0 \\ iis & 0 & 0 & -kw \end{pmatrix} \begin{pmatrix} kw + iis \\ kw + iis \\ -kw - iis \\ -kw - iis \end{pmatrix} = 0,$$

or

$$\begin{pmatrix} kw & 0 & 0 & -iis \\ 0 & kw & -iis & 0 \\ 0 & iis & -kw & 0 \\ iis & 0 & 0 & -kw \end{pmatrix} \begin{pmatrix} 1 \\ 1 \\ -1 \\ -1 \end{pmatrix} (kw + iis) = 0 .$$

The left-hand side reduces to

$$\begin{pmatrix} \boldsymbol{k}w + i\boldsymbol{i}s \\ \boldsymbol{k}w + i\boldsymbol{i}s \\ \boldsymbol{k}w + i\boldsymbol{i}s \\ \boldsymbol{k}w + i\boldsymbol{i}s \end{pmatrix} (\boldsymbol{k}w + i\boldsymbol{i}s) \ ,$$

in which each row of the column vector becomes

$$- w^2 + s^2 = 0 \ ,$$

as in the parallel case of the Dirac equation.

Without the 'phase' terms, equation (3) becomes:

$$\begin{pmatrix} i\boldsymbol{k}\partial/\partial t & 0 & 0 & -i\boldsymbol{\nabla} \\ 0 & i\boldsymbol{k}\partial/\partial t & -i\boldsymbol{\nabla} & 0 \\ 0 & i\boldsymbol{\nabla} & -ik\partial/\partial t & 0 \\ i\boldsymbol{\nabla} & 0 & 0 & -ik\partial/\partial t \end{pmatrix} \begin{pmatrix} \boldsymbol{k}w + i\boldsymbol{i}s \\ \boldsymbol{k}w + i\boldsymbol{i}s \\ -\boldsymbol{k}w - i\boldsymbol{i}s \\ -\boldsymbol{k}w - i\boldsymbol{i}s \end{pmatrix} e^{-i(wt - \mathbf{s} \cdot \mathbf{r})} = 0 \ .$$

Here, each term of the resultant column vector becomes a pseudo-Dirac or Dirac-type equation for charge:

$$(i\boldsymbol{k}\partial/\partial t + i\boldsymbol{\nabla}) \, (\boldsymbol{k}w + i\boldsymbol{i}s) \, e^{-i(wt - \mathbf{s} \cdot \mathbf{r})} = 0 \ ,$$

in the same way as each term of the resultant column matrix becomes a Dirac equation for the $E$-$\mathbf{p}$-$m$ combination. This equation provides a convenient representation of the parallel between the mathematics for charge allocation, determining particle structures, and that for the Dirac state.

Starting with the real Dirac equation (for $E$-$\mathbf{p}$-$m$), we introduce a filled fermion vacuum to create the two-sign degree of freedom required for $E$. We also define a particular status for antifermions beyond the original requirement that each charge-type has two possible signs. We assume, therefore, that a particular type of charge, say $s$, can only be unit in one of the three 'colours' needed to make up an observed state. This excludes charges of the opposite sign, so we take the concept of antistates from the Dirac equation, and assign $-s$ to the antifermions. We cannot, however, repeat the same procedure for, say, $e$, which must have both signs in both states and antistates. So, we preserve the rule that a charge ($-e$ in this case) can be unit in only one of the three 'colours', but make the 'default' position ($e$, $e$, $e$) as opposed to (0, 0, 0) for $s$, and so produce two signs by creating 'weak isospin', with alternatives ($e$, $e$, 0) and (0, 0, $-e$). Subsequently, we find that using 'weak isospin' actually gives us a suitable zero for the matrix equation for charge. Finally, to accommodate two signs of $w$, we have to refer to the fact that a filled vacuum, with antiparticles nonexistent in the ground state, violates charge conjugation symmetry for the charge ($w$) which specifies the fermion state.

## 27 Grand unification and the Planck mass

It will first be convenient to look at the (not completely successful) grand unification proposed for minimal $SU(5)$. We begin with the formula for the electroweak mixing angle:

$$\sin^2 \theta_W = \frac{\sum t_3{}^2}{\sum Q^2} . \tag{10}$$

Taking a weak component with only left-handed contributions to weak isospin, over 3 colours of $u$, 3 colours of $d$, and the leptons $e$ and $v$, we obtain:

$$\sum t_3{}^2 = \frac{1}{4} \times 8 = 2 \quad .$$

This is, of course, independent of the electric charge structure. For the electromagnetic component, however, the phenomenological and lepton-like structures diverge. Taking the phenomenological values, with both left- and right-handed contributions, would lead to

$$\sum Q^2 = 2 \times \left( \frac{4}{9} \times 3 + \frac{1}{9} \times 3 + 1 + 0 \right) = \frac{16}{3} \quad ,$$

from which

$$\sin^2 \theta_W = 0.375 .$$

For lepton-like quarks, however, we have

$$\sum Q^2 = 2 \times (1 + 1 + 0 + 0 + 0 + 1 + 1 + 0) = 8 \quad ,$$

leading to

$$\sin^2 \theta_W = 0.25 .$$

Weinberg[30] is one of many who have observed that the value 0.375 for $\sin^2 \theta_W$ is in 'gross disagreement' with the experimental value of 0.231. 0.25 is, of course, much closer, and second order corrections could account for the relatively small discrepancy. It is usual, in the standard approaches, to take the equations for the running weak and strong coupling constants, derived from their respective $SU(2)$ and $SU(3)$ structures:

$$\frac{1}{\alpha_2(\mu)} = \frac{1}{\alpha_G} - \frac{5}{6\pi} \ln \frac{M_X{}^2}{\mu^2} \tag{11}$$

and

$$\frac{1}{\alpha_3(\mu)} = \frac{1}{\alpha_G} - \frac{7}{4\pi} \ln \frac{M_X{}^2}{\mu^2} , \tag{12}$$

and assume that a particular grand unified gauge group structure will modify the equivalent $U(1)$ equation for the electromagnetic coupling $(1 / \alpha)$ to one in which it is mixed with the weak value, based on $SU(2) \times U(1)$. So, now we have

$$\frac{1}{\alpha_1(\mu)} = \frac{1}{\alpha_G} + \frac{1}{\pi} \ln \frac{M_X^2}{\mu^2} \ , \tag{13}$$

where

$$\frac{5}{3\alpha_1(\mu)} + \frac{1}{\alpha_2} = \frac{1}{\alpha} \ . \tag{14}$$

From equations (11), (12) and (13), we derive a grand unified mass scale ($M_X$) of order $10^{15}$ GeV, and proceed to apply (11) and

$$\sin^2\theta_W = \frac{\alpha(\mu)}{\alpha_2(\mu)} \ , \tag{15}$$

to give 'renormalized' values of $\sin^2\theta_W$ of order 0.19 to 0.21.

The big disadvantage of this procedure is that it does not achieve a true equalization of the interactions, even at grand unification. The strong and weak interactions achieve exact equalization with each other, but not with the electromagnetic interaction. However, equations (13) and 14) are not well-established results, like (11), (12) and (15). They are not supported by the experimental evidence, and make assumptions about group structure, such as relying on a particular value for the Clebsch-Gordan coefficient, $C^2 = 5 / 3$, that have, as yet, no experimental or theoretical justification. The fit to the data is also poor, for substitution of the calculated grand unification constants into the equations for the individual couplings (11), (12), (14), produces a result which manifestly fails to converge to a single value for the grand unified coupling ($\alpha_G$), leading to ad hoc suggestions that a supersymmetric model may be the only solution.[31] Even worse than this, however, is the fact that, compensating errors in the combination tend to disguise the massive inconsistencies between the separate equations. In particular, recalculation of the value of $\sin^2\theta_W$ at $\mu = 10^{15}$ GeV gives 0.6 rather than the 0.375 which was initially assumed in setting up the equations!

In the present theory, however, we have an *independent* value for $\sin^2\theta_W$ of the right order. We can, therefore, perform a much simpler calculation for $M_X$ without making assumptions about the group structure, and avoiding, in the first instance, the problematic running coupling constant equation for $1 / \alpha_1$. We avoid the speculative equations (13) and (14), and combine the well-established (11), (12) and (15) to give:

$$\sin^2\theta_W (\mu) = \alpha(\mu) \left( \frac{1}{\alpha_3(\mu)} + \frac{11}{6\pi} \ln \frac{M_X}{\mu} \right). \tag{16}$$

Equally significantly, we immediately obtain a remarkable value for the Grand Unified mass scale, $M_X$. Taking typical values for $\mu = M_Z = 91.1867(21)$ GeV, $\alpha(M_Z^2)$ = 1 / 128 (or 1/129), $\alpha_3(M_Z^2) = 0.118$ (or 0.12), and $\sin^2\theta_W = 0.25$, we obtain $2.8 \times 10^{19}$ GeV for $M_X$. This is of the order of the Planck mass ($1.22 \times 10^{19}$), and may well be exactly so, as purely first-order calculations overestimate the value of $M_X$. Assuming that $M_X$ *is* the Planck mass, we obtain $\alpha_G$ (the Grand Unified value for all interactions) = 1 / 52.4, and $\alpha_2(M_Z^2)$ = 1 / 31.5, which is exactly the kind of value we would expect for the weak coupling with $\sin^2\theta_W = 0.25$. We also obtain unit strength for the strong interaction ($\alpha_3 = 1$) at the approximate energy level of baryonic and mesonic structure (that is, in the range $\mu \sim m_\pi$).

Higher order calculations based on the phenomenological quark model, using a two-loop approximation, reduce the value of $M_X$ by a factor of about 0.64, while theoretical plots for $\sin^2\theta_W$ against $\mu^2$ show a distinct dip at $M_W$ - $M_z$, against an overall upward trend, suggesting that the emergence of massive gauge bosons depresses the effective values of $1 / \alpha_2$ and $\sin^2\theta_W$ in the energy range $M_W$ - $M_z$, where they are normally measured.[32] (The actual decrease, from about 0.22 to 0.21, represents a possible decrease up to about 0.02 in what would be otherwise expected.) Similar calculations, applied to the lepton-like quark model may well yield similar results, or perhaps an even better fit to the data.

Use of lepton-like quarks, however, means that the hypercharge numbers for the $U(1)$ electromagnetic running coupling equation will be no longer identical to those for a quark model based purely on QED phenomenology. In the lepton-like model, $\binom{u}{d}_L$ changes from 1 / 6 to 1 / 2, while $(u^c)_L$ goes from $-2$ / 3 to $-1$, $-1$ or 0, depending on the colour, and $(d^c)_L$ from 1 / 3 to 0, 0 or 1. The fermionic contribution to vacuum polarization is, conventionally,

$$\frac{4}{3}\times\frac{1}{2}\times\left(\frac{1}{36}\times 3+\frac{1}{36}\times 3+\frac{1}{9}\times 3+\frac{4}{9}\times 3+\frac{1}{4}\times 1+\frac{1}{4}\times 1+1\right)\frac{n_g}{4\pi}=\frac{5}{3\pi},$$

where $n_g = 3$ is the number of fermion generations; but, modifying this for lepton-like quarks, we obtain:

$$\frac{4}{3}\times\frac{1}{2}\times\left(\frac{1}{4}\times 3+\frac{1}{4}\times 3+1+1+0+0+0+1+\frac{1}{4}\times 1+\frac{1}{4}\times 1+1\right)\frac{n_g}{4\pi}=\frac{3}{\pi}\;.$$

(The result corresponds to the change in Clebsch-Gordan coefficient from $C^2 = 5$ / 3 to $C^2 = 3$, when $\sin^2\theta_W = 1 / (1 + C^2)$ changes from 0.375 to 0.25.)

One of our objections to minimal $SU(5)$ was that the strong, weak, and electric interactions were not unified on an equal basis. This suggests that our grand unification treatment of the electric action should be in terms of the pure electric coupling parameter $\alpha$, and not of a modified, mixed electroweak parameter, $\alpha_1$, normalized to fit an overall gauge group, as assumed in most grand unification schemes. By this understanding, 0.25 is specifically the value of $\sin^2\theta_W$ for a *broken* symmetry, produced by asymmetric values of charge, and is the value that would be expected at the mass scale appropriate to the electroweak coupling, that is at $\mu = M_W$ - $M_z$, the energy scale at which the symmetry-breaking takes place. It should *not* be the value expected at grand unification.

Using the new values we have obtained for the hypercharge numbers, the running coupling of the pure electromagnetic interaction, will be:

$$\frac{1}{\alpha(\mu)} = \frac{1}{\alpha_G} + \frac{3}{\pi} \ln \frac{M_X^2}{\mu^2} \quad . \tag{17}$$

Remarkably, when we substitute in the values $M_X = 1.22 \times 10^{19}$ GeV, $\mu = M_Z = 91.1867$ GeV, and $\alpha_G = 1 / 52.4$, we obtain $1 / \alpha = 128$, which is *exactly the value obtained experimentally* at energies corresponding to $M_Z$. It would appear, therefore, that the unification which occurs at $M_X$ might well involve a direct numerical equalization of the strengths of the three, or even four, physical force manifestations, without reference to the exact unification structure.

Since this unification apparently also occurs at the Planck mass, the exact energy scale relevant to quantum gravity, we may propose, in addition, that the actual symmetry group, incorporating gravity in some form, is $U(5)$, with the additional generator, coupling to all others, representing the gravitational interaction (at least in numerical terms), and that $SU(5)$ occurs as the first stage of the symmetry breakdown. At grand unification, also, we would also have $C^2 = 0$ and $\sin^2\theta_W = 1$, creating an exact symmetry in every respect between weak and electric interactions, as well as between weak and strong. The mixing parameter, $\sin^2\theta_W$, would then be interpretable as the electroweak constant for a specifically broken symmetry, taking the value of 0.25 at the energy range where the symmetry breaking occurs ($M_W$ - $M_z$), and gradually decreasing from the maximum to this value at intermediate energies.

A $U(5)$ grand unification would have the advantage of making all the generators become pure scalar phases, and identical in form, at the grand unification energy. A likely possibility is that the grand unification energy represents a kind of 'event horizon', or unattainable limit, at which separate conservation laws for charges would have no meaning. In fact, the necessity for separate conservation laws would prohibit its attainment, as it already prohibits direct proton decay.

The Planck mass, which is here identified as the grand unification energy, is also the likely candidate for the cut-off energy which ensures the finite summation of self-energies for interacting fermions required by the nilpotent formulation.[8] Through the need for a filled vacuum and the continuous nature of mass-energy, gravity may well be the instantaneous carrier of the state vector correlations involved in nonlocality, and the Planck mass may be taken as the quantum of the (GTR-related) inertial interactions, which have been proposed elsewhere as the result of the effect of gravity on the time-delayed nature of nongravitational interactions.[10] These in turn might produce the inertial masses associated with charged particles, by a coupling to the Higgs field which fills the vacuum state.

A grand unification at the Planck mass would have important consequences for reducing the number of free parameters in the Standard Model. Essentially, the three fundamental constants $G$, $\hbar$ and $c$ have no intrinsic meaning. They are simply numbers which relate the arbitrary units which we choose to assign to space, time and mass, but, since these parameters are fundamental, it is meaningless to look for additional significance in the units themselves. Only the numerical values attached to structures, such as the electron, have this kind of intrinsic meaning. Now, if $M_X$ is the Planck mass, then it, too, becomes a fundamental unit, since it is composed entirely from $G$, $\hbar$ and $c$. The value of $\sin^2\theta_W$ is also, apparently, known from an exact conceptual argument, In setting up the conditions for grand unification, then, we have four equations with just five unknowns at any particular energy ($\mu$), namely $\mu$, $\alpha$, $\alpha_2$, $\alpha_3$, and $\alpha_G$. Of course, these equations, as we write them, are merely first-order approximations, but we could, *in principle*, refine them to any degree of exactness. In effect, given any assumed $\mu$, we could have exact predictions for any of the other four constants, with no other empirical input. To go further, it is quite possible that one of the other constants has a theoretically exact value at some particular specified value of $\mu$. The most likely possibility is that $\alpha_3 = 1$ (that is, $\hbar c$) when $\mu$ is, say, $m_e c^2 / \alpha$, the mass-energy equivalent for a unit charge coupling, if the electron's mass is derived directly from the electromagnetic coupling. At present, agreement is moderately good but not perfect, as $\alpha_3 = 1$ seems to occur for $\mu = 1.5 \, m_e / \alpha$ (the muon mass), but the equations here are very sensitive to the approximations employed, and the true value may really be closer to the one we expect. If this is so, then *not even $\mu$ need be assumed* to derive the four fine structure constants; we have, rather, a fifth fundamental equation to derive $m_e / \alpha$ or $m_e$ itself.

The theory proposed here has the particular merit of being eminently testable by a measurement of any of the three interaction strengths at increasing energies, where there will be divergences from values predicted by other models. The most dramatic changes will occur in $1 / \alpha$, which, on this model would be $1/118$ at 14 TeV (the maximum energy of the LHC under construction at CERN), in comparison with the $1/125$ predicted by minimal $SU(5)$ and the quark model with phenomenological electric charges.

## 28 The generation of mass

According to our understanding of the Higgs mechanism, mass is generated when an element of partial right-handedness is introduced into an intrinsically left-handed system. Thus, anything which alters the signs of the terms in the expression ($i \, \hat{\mathbf{p}}_a \, (\delta_{bc} - 1) + j \, (\hat{\mathbf{p}}_b - 1\delta_{bm}) + k \, \hat{\mathbf{p}}_c \, (-1)^{\delta_{1g}} g$), or reduces any of the terms to zero, should, in principle, be a mass generator, because it is equivalent to introducing the opposite sign of $\boldsymbol{\sigma}_z$ or a partially right-handed state. The three main terms in this expression can be specified as sources for producing mass. They can be described as weak isospin, quark confinement, and weak charge conjugation violation.

The production of mass by the zeroing of charge is a particularly significant process, which seems to be responsible, at least, for the masses of the Higgs boson, $Z^0$, and the composite baryons and mesons. That mass and charge are, in some sense, mutually exclusive components of the vacuum (effectively combining to form an invariant in the same manner as space and time), is implied by standard treatments of the $U(1)$ component of the Weinberg-Salam theory, in addition to being required by the quaternionic form of the Dirac 4-spinor. For example, Aitchison and Hey, writing on the hypercharge value of the Higgs field, state that: 'we do not allow the particle physics vacuum to give an electrically charged field a non-zero value. Thus we require that the component of $\phi$ with non-zero vacuum value has zero charge.'[20] Missing charges can be seen as 'unused' vacuum, and occur where there is a superposition of allowed states.

The two states of weak isospin specified by the term $(\hat{\mathbf{p}}_b - \mathbf{1}\delta_{0m})$ are effectively equivalent to taking an undisturbed system in the form $\boldsymbol{j}\boldsymbol{\sigma}_z\cdot\hat{\mathbf{p}}_b$ and of taking the same system with the added 'right-handed' term $-\boldsymbol{j}\boldsymbol{\sigma}_z.\mathbf{1}$. In the pure lepton or free fermion states, when $b = c \neq z$, and hence the weak component, $\boldsymbol{k}\boldsymbol{\sigma}_z\cdot\hat{\mathbf{p}}_c = 0$, the equation generates residual right-handed electron / muon / tau states, specified by $-\boldsymbol{j}$, with the equivalent left-handed antistates specified by $\boldsymbol{j}$. The right-handed terms may be considered as the intrinsically right-handed or non-weak-interacting parts of the fermions, generated by the presence of nonzero rest mass. As we have seen, the mixing of the left- and right-handed terms illustrates the fact that the electromagnetic interaction cannot identify the presence or absence of a weakly interacting component. The quarks follow the same procedure as leptons in generating the two states of weak isospin, but there are no separate representations of 'right-handed' quarks, as two out of any three quarks in any baryon system will always require $c \neq z$ and $\boldsymbol{k}\boldsymbol{\sigma}_z\cdot\hat{\mathbf{p}}_c = 0$.

Mass is again generated by quark confinement, because each baryonic system requires quarks in which one or more of $\boldsymbol{i}\boldsymbol{\sigma}_z\cdot\hat{\mathbf{p}}_a$, $\boldsymbol{j}\boldsymbol{\sigma}_z\cdot\hat{\mathbf{p}}_b$, or $\boldsymbol{k}\boldsymbol{\sigma}_z\cdot\hat{\mathbf{p}}_c$ is zero. Zero charges represent complete coupling to the Higgs field; nonzero charges represent a reduction of the vacuum state to less vacuum. This mechanism is more likely to be relevant to composite and superposed states, such as mesons and baryons, than to 'pure' ones, such as quarks and leptons. In these cases, the mass equivalent for a zero charge would appear to be that of a fundamental unit $m_f$, defined for unit coupling ($\hbar c$), from which we *derive* the electron mass, via the electromagnetic coupling $\alpha$, as $m_e = \alpha\, m_f$. Hence, $m_f = m_e\, /\, \alpha$. The use of a fundamental mass unit for zero charges irrespective of origin appears to derive from the fact that these 'missing' charges are a result of a perfectly random rotation of the momentum states $\hat{\mathbf{p}}_a$, $\hat{\mathbf{p}}_b$, or $\hat{\mathbf{p}}_c$, in exactly the same manner as applies in the strong interaction to produce its linear potential; $\hat{\mathbf{p}}_a$ is, of course, actually an expression of this interaction, but $\hat{\mathbf{p}}_b$ and $\hat{\mathbf{p}}_c$ follow the identical pattern of variation.

The third mechanism for mass generation arises from the fact that the sign of the intrinsically complex $\boldsymbol{k}$ term is not specified with those of the $\boldsymbol{i}$ and $\boldsymbol{j}$ terms. Physically, however, a filled weak vacuum requires that the weak interaction recognizes only one sign for the $\boldsymbol{k}$ term when the sign of $\boldsymbol{\sigma}_z$ is specified. Hence, negative values of $\boldsymbol{k}\boldsymbol{\sigma}_z\cdot\hat{\mathbf{p}}_c$ must act, in terms of the weak interaction, as though they were positive. Reversal of a sign is equivalent to introducing opposite handedness or mass. So, the two intrinsic signs of the $\boldsymbol{k}\boldsymbol{\sigma}_z\cdot\hat{\mathbf{p}}_c$ term become the source of a mass splitting between a first generation, involving no sign reversal, and a second generation in which the reversal is accomplished by charge conjugation violation. However, since charge conjugation violation may be accomplished in two different ways – either by violating parity or time reversal symmetry – there are actually two further mass generations instead of one. In addition, because the weak interaction cannot distinguish between them, the three generations represented by the quarks *d*, *s* and *b*, are mixed, like the left-handed and right-handed states of *e*, *μ* and *τ*, in some proportion related to the quark masses.

## 29 The Higgs model for fermions

According to a well-known textbook by Halzen and Martin: 'An attractive feature of the standard model is that the same Higgs doublet which generates *W* and *Z* masses is also sufficient to give masses to the leptons and quarks.'[21] After application of this to electrons, the authors state: 'The quark masses are generated in the same way. The only novel feature is that to generate a mass for the upper member of a quark doublet, we must construct a new Higgs doublet from $\phi$.' 'Due to the special properties of $SU(2)$, $\phi_c$ transforms identically to $\phi$, (but has opposite weak hypercharge to $\phi$, namely $Y = -1$). It can therefore be used to construct a gauge invariant contribution to the Lagrangian.' Significantly, the hypercharge of $(u_L, d_L) = -1$ in the lepton-like quark model, when the charge structure actually matches that of the leptons, and $\sigma_z.\hat{\mathbf{p}}_b = -1$; but it becomes 1, when $\sigma_z.\hat{\mathbf{p}}_b = 0$, and the electric charge component is provided purely by the filled electromagnetic vacuum.

However, this is only true for lepton-like quarks. For quarks with phenomenological electric charges, the hypercharge is an invariable 1/3 and there is no negative term: the phenomenological charge values allow only one hypercharge state, though the mechanism requires two. The necessary asymmetry introduced by the lepton-like model is lost. The only way round this problem is by the invention of an arbitrary and unphysical linear combination, relating the Higgs terms to the *u* and *d* quark masses. And, of course, the reason why the hypercharge must be reversed in the lepton-like model is that the transition involves a reversal of the 'electromagnetic vacuum' or background condition, from empty to full. The Higgs mechanism seems to make perfect sense of this procedure, where it is just a mathematical operation that 'works' with appropriate (unexplained) adjustments in the conventional view.

In the lepton-like model, the Higgs Lagrangian for the mass of *e* directly transfers from the usual covariant derivative Lagrangian – it is virtually a direct copy now applied to the Higgs doublet. The fermion mass Lagrangian for *d* when the charge structure matches that of the leptons is then a direct copy of that for *e*, while the fermion mass Lagrangian for *d* when the charge structure is not lepton-like is a direct copy of that with reversed hypercharge. In the phenomenological model, the hypercharge for quark mass is different from the hypercharge for quarks; here it is the same. The use of the Higgs mechanism with lepton-like quarks requires no extra modelling at all, and it stems from a charge 'vacuum' (the absence of charges).

# 30 The masses of bosons and baryons

The Higgs mechanism, in the conventional sense, specifies the manner in which bosons and fermions acquire mass, but is unable to predict specific values for the couplings to the Higgs field, and cannot predict the mass of the Higgs boson. The Higgs field and the mechanism of generating specific masses are clear consequences of the present theory, and it should, in principle, be possible to derive numerical results. Of course, the masses of particles originate in the requirements for energy balance between states subject to various conditions of symmetry. Many of these operate at the same time and the prediction of exact masses is a very complex procedure which has not yet been worked out for any single particle; but, where one particular condition is dominant, it would appear to be possible to predict masses to quite good approximations. Difficulties in physical interpretation, however, mean that such results must remain tentative for the present, and the ones presented are suggestive rather than definitive.

The most probable meaning for the Higgs boson is that it represents complete vacuum or a zeroing of all possible bosonic states, that is, a zeroing of all the charges in the complete range of fermion-antifermion combinations within the A-C and L representations. Assuming 6 flavours, 6 anti-flavours, 3 colours (or equivalent states), 3 charge types for each quark / antiquark, and 2 for each quark-antiquark pairing, over 4 representations, gives a total of 2592 zeros. Assigning a mass-energy of $m_e c^2 / \alpha$ (the unit coupling value), to each zeroed charge gives an approximate total mass of 182 GeV, which is, interestingly, within the range of phenomenological values for $m_H$ (approximately 170-180 GeV) which would exclude the production of explicit supersymmetric particles.[33-35]

Similar procedures may apply to the electroweak bosons, which require only the calculation of the mass of $Z^o$, because the $W$ mass then follows from $M_W = M_Z \cos \theta_W$. The masses of particles are determined by the strength of their coupling to the Higgs field. $Z^o$ is completely coupled, $\gamma$ does not couple. Complete coupling implies full strength of vacuum, i.e. zero charges. If $Z^o$, $\gamma$ is derived from the reduced A/B/C - L representations for the pure electroweak case (A/B/C being indistinguishable from each other, with no $s$, and no electroweak recognition of colour), then complete summation of the zeros over 2 representations produces 91 GeV. If $m_H$ really is 182 GeV then the most favoured decay mode for the Higgs boson would be via the two $Z^o$ or four lepton route.

It is of interest also that the total mass of the twelve known fermions ($\Sigma m$) again appears to be about 182 GeV. Now, according to the standard treatment of the Higgs mechanism,[20] fermion masses $m$ are generated by (harmonic oscillator) couplings $g_f$ to the Higgs field of the form

$$g_f = \frac{e}{\sqrt{2} \, \sin \theta_w} \frac{m}{M_w} .$$

Taking $M_W = M_Z \cos \theta_w$, $\sin \theta_w$ at $M_Z = 0.5$, $\Sigma m = M_H = 2 \, M_Z$, and the weak coupling constant, $g = e / \sin \theta_w$, the total coupling to all the fermion states would be

$$\Sigma g_f = \frac{g}{\sqrt{2}} \frac{2}{\cos \theta_w} = \sqrt{\frac{8}{3}} \, g \quad .$$

The total coupling producing the fermion masses would thus be *directly determined* by the weak coupling, and the fermion masses would be related to $M_H$ in the ratio of the Higgs coupling to the weak coupling:

$$\frac{m}{M_H} = \frac{g_f}{g} \sqrt{\frac{3}{8}} \quad .$$

If the vacuum energy is distributed or partitioned in this way between the possible fermion states, it is noticeable that the three quark *generations* are separated from each other by a factor of the order of $\alpha$, effectively the separating factor between strong, or exactly unit, and electroweak couplings.

The vacuum expectation value for the Higgs field ($f$) is clearly another important parameter which ought to be calculable, in some sense, from the zeroing of charge. Phenomenologically, $f$ can be calculated from the Fermi constant at ~ 246 GeV, which appears to be, at least approximately, 3 $M_W$ (3 × 80.45 ~ 241 GeV).[20] There is, in fact, a fundamental reason for a connection with $M_W$, as, using (3) from section 16 for a complex electrically charged field $W^+$, $f$ is given by $M_W = gf / 2$, with $g$ the weak coupling constant expressed in charge units. Now, since $g = e / \sin \theta_W$, with $e$ the electrical coupling, then $g = 2e$ if $\sin^2 \theta_W = 0.25$. This might indicate that the vacuum field value, in its most idealised form, is determined by that expected for a three-phase system with the charge divided between three phases.

The masses of the low-lying baryon and meson states are more definitely predictable. These composite particles generate their masses through the term $j$ ($\hat{\mathbf{p}}_b -$ $\mathbf{1}\delta_{0m}$) and the strong interaction mechanism, which is, of course, electromagnetic charge independent. Here, we need to consider the global symmetries such as $SU(3)_f$ which group together particles which are indistinguishable by the strong interaction in various isospin multiplets, which effectively represent a single particle in different states of electromagnetic charge. All states of one multiplicity exist simultaneously, and so the zero charge components of all states must be accommodated in determining the mass of the particle. The global $SU(3)_f$ symmetry shows the spin 3/2 baryons as a decuplet, with four $\Delta$ states, three $\Sigma$ states, two $\Xi$ states and one $\Omega$ state. The zero charge components of the $\Delta$ particles are simply those of all four states added together, but the four $\Delta$ states, when excited, have to be averaged between three $\Sigma$ states, so each $\Sigma$ state represents an average of 4/3 states, and the average number of charge components has to be multiplied by 4/3. The four $\Delta$ states are eventually excited to one $\Omega$ state, so each $\Omega$ state represents an average 4 states. If $M_0$ is the highest multiplicity in a particular baryon octet or decuplet, and $n_0$ is the total number of zero charges in the components of a multiplet of multiplicity $M$, then the minimum mass of the components of the multiplet is given by

$$\text{mass} = \frac{n_0 \, M_0}{M} \frac{m_e}{\alpha} \,.$$

For example, the multiplet $\Sigma$ in the spin 3/2 baryon decuplet has $M = 3$, and $M_0 = 4$ (the multiplicity of the $\Delta$ particles), while $n_0$, the total number of zero charges, computed from the quark tables A-C, in the combinations $dds$, $uds$ and $uus$, is 15, 17 or 19. For the ground state,

$$\text{mass of } \Sigma \text{ multiplet} = \frac{15 \times 4}{3} \frac{m_e}{\alpha} = 20 \, \frac{m_e}{\alpha} \,,$$

which may be compared with the experimental value of 19.8 $m_e \,/\, \alpha$ or 1385 MeV. The derivation of the masses for the entire spin 3/2 decuplet may be set out in the following table:

|   | quark structure | $n_0$ | $M_0$ | $M$ | predicted mass | measured mass |
|---|---|---|---|---|---|---|
|   |   |   |   |   |   |   |
| $\Delta$ | $ddd,udd,uud,uuu$ | 20,22,24 | 4 | 4 | 20 $m_e \,/\, \alpha$ | $\approx$ 17.6 - 19.6 $m_e \,/\, \alpha$ |
| $\Sigma$ | $dds,uds,uus$ | 15,17,19 | 4 | 3 | 20 | 19.8 |
| $\Xi$ | $d$ss,$uss$ | 11,13 | 4 | 2 | 22 | 21.9 |
| $\Omega$ | $sss$ | 6 | 4 | 1 | 24 | 23.9 |

The masses are all calculated using the ground state values for $n_0$. The $\Delta$ particle is unusual in showing a large spread of measured mass, because, in this case, the energy width (approximately 120 MeV at half-maximum) makes a significant contribution, in addition to the rest mass; this energy width is much greater than that for any other member of the decuplet and explains the particle's instability and very rapid decay. The rest mass value (17.6 $m_e \,/\, \alpha$) preserves the difference of 2 $m_e \,/\, \alpha$ (140 MeV) between each multiplet which occurs due to successive transitions of one $d$ quark to $s$ with the net loss of two $w$ charges (in line with an $s$ mass of 80-155 MeV). The increasing accuracy of the predictions, from $\Delta$ through to $\Omega$, may be related to the fact that the heavier particles represent fewer alternative states.

It is debatable whether similar principles can be applied to an extended $SU(4)_f$ multiplet including the fourth quark ($c$). The very existence of such multiplets depends on the idea that the quarks in the second and third generation have sufficiently high masses to remove the degeneracy between different sets of three-quark and quark-antiquark states. So the existence of higher $SU(n)_f$ symmetries depends crucially on the assumption that $m_b \gg m_t \gg m_c \gg m_s$. However, if the idea can be extended in this way, we would have a count of 50 zero charges (from 10 base states) for $ccc$, and a mass of 50 $m_e \,/\, \alpha$ = 3.5 GeV, in line with the assumed mass for the $c$ quark of between 1.0 and 1.4 GeV. The $b$ and $t$ quarks may be assumed to be too massive for this mechanism to be the main factor in determining the masses of baryons and mesons incorporating these as components. Extending to $SU(5)_f$, with 20 base states would produce 120 zero charges (8.4 GeV) for $bbb$, while $SU(6)_f$, with 35 base states, would produce 175 zero charges (12.25 GeV) for $ttt$.

A zero-charge analysis may be applied, however, to the spin 1/2 baryon octet, generated from $u$, $d$ and $s$, though here the value for $n_0$ is taken at the ground state for the $N$ multiplet ($n$, $p$), which contains no $s$ quark component, and the other mass values are assumed to be of mixed states determined from within the predicted range by their accommodation within the Gell-Mann-Okubo formula required for $SU(3)_f$:

$$\frac{1}{2}(m_N + m_\Xi) = \frac{3}{4}m_\Lambda + \frac{1}{4}m_\Sigma .$$

|   | quark structure | $n_0$ | $M_0$ | $M$ | predicted mass | measured mass |
|---|---|---|---|---|---|---|
|   |   |   |   |   |   |   |
| $N$ | $udd,uud$ | 9,11,13 | 3 | 2 | 13.5 $m_e$ / $\alpha$ | 13.4 $m_e$ / $\alpha$ |
| $\Lambda$ | $uds$ | 5,7 | 3 | 1 | 15 - 21 | 15.9 |
| $\Sigma$ | $dds,uds,uus$ | 15,17,19 | 3 | 3 | 15 - 19 | 17 |
| $\Xi$ | $dss,uss$ | 11,13 | 3 | 2 | 16.5 - 19.5 | 18.9 |

The meson octets do not represent the regular progression of excited states from the lowest member which we observe in the baryon octet and decuplet, and which ultimately derive from $d \rightarrow s$ quark transitions. The multiplets are, in this sense, independent, with mass determined by $n_0$ $m_e$ / $\alpha$, where $n_0$ is the number of zero charge components in the multiplet. For the pseudoscalar $0^-$ meson octet, the ground state value of $n_0$ is once again chosen for the lowest lying member of the octet ($\pi$) − which again contains no symmetry-breaking s quark component − and the values for $K$ and $\eta$ selected from within the predicted range, again to fit a Gell-Mann-Okubo formula for $SU(3)_f$:

$$m_K{}^2 = \frac{1}{4}m_\pi{}^2 + \frac{3}{4}m_\eta{}^2 .$$

$n_0$ is 2, 6, 8, 10, 12, 14, or 16 for $\pi$, so the ground state value is 2; the predicted mass is therefore 2 $m_e$ / $\alpha$, which is exactly the observed value. For $K$, $n_0$ takes values 3, 5, 7, 9 or 11, and so the predicted mass is between 3 and 11 $m_e$ / $\alpha$, compared with the observed mass of 7.1 $m_e$ / $\alpha$; while $\eta$ (which is additionally mixed with a singlet state) has $n_0$ values of 4, 6, 8, 10, 12, leading to a predicted mass between 4 and 12 $m_e$ / $\alpha$, compared with an observed mass of 7.8 $m_e$ / $\alpha$.

Regge trajectories may provide observational evidence for the use of charge counting in determining the masses of strongly-bound composite particles. If the strong interaction is carried with the angular momentum operator $\mathbf{p}$, the covariant derivative introduces the term $q\sigma r$ or $ig_s\lambda^\alpha \mathbf{A}$ / 2, which incorporates a quantity equivalent to the strong coupling or the strong charge squared. In principle, therefore, increasing the angular momentum value assigned to any particular state should also increase the effective value of the strong charge squared in the same proportion. If the masses of the strongly-bound composite particles are determined on the basis of strong charge equivalents, then a change in angular momentum should produce a proportional change in mass squared. This, of course, would only be true if the rate of change of momentum, or linear energy density ($\kappa$), remained constant, at all distances, as effectively assumed in the semi-classical 'string' or 'gluon flux tube' explanation of the trajectories. In this case, if a quark-antiquark pair are connected by a flux tube of length $2R$, then the total mass-energy of the string becomes $m = \pi\kappa R$ and the angular momentum $J = \pi\kappa R^2 / 2$, leading to the relationship, $J = m^2 / 2\pi\kappa$, with $2\pi\kappa$ determined phenomenologically at $\sim 0.9$ GeV$^2$.

## 31 The masses of fermions

Fermion masses, like those of bosons, may be assumed to be due to a coupling between originally massless fermion fields and the nonzero background Higgs field. For a weak $SU(2)$ transformation, acting only on the left-handed component of a fermion state vector, the free particle equation contains a mass term of the form $m\psi_L + m\psi_R$ and so cannot be locally phase invariant because $\psi_L$ and $\psi_R$ are transformed differently under $SU(2)$. The symmetry is only preserved if the fermions are initially massless and acquire their observed masses by interaction with the Higgs field. The coupling strength, however, varies with the mass of the individual fermion and cannot be predicted independently of the known masses. Some other input is needed, as the present account has suggested.

Though the masses for the composite baryons and mesons give indications for limits on the masses of the heavier quarks, deriving exact masses, by direct methods, for the twelve known fermions is a particularly difficult problem, especially as the concept of quark mass seems to be somewhat ill-defined, with the masses 'running', like the values of the coupling constants, with the energy of interaction. However, the mass of the $t$ quark, at least, seems to be obtainable from first principles, on the assumption that it represents maximal coupling to the Higgs field. The mass of $t$ ($\sim$ 174 GeV) seemingly represents the maximum possible energy for a state $f / \sqrt{2}$, where $f$ is the vacuum expectation value. It may also be possible to make some tentative approaches to calculating some of the other quark masses.

If the fermion masses are generated by the Higgs mechanism, and the ultimate origin of mass is in the introduction of the electric charge, to overcome symmetry violation in the weak interaction, then it is conceivable that the mass scales of the three generations of fermions are related by successive applications of the scaling factor $\alpha$, as noted in the previous section. In a related way, the Cabibbo mixing between the quark generations seems to be determined (as we might expect) by the same factor as the electroweak mixing (0.23 – 0.25), and the additional mixing produced with the third generation involves terms which are the square of this factor ($\approx 0.06$).

From both the Higgs mechanism, and our own representation, the weak isospin up state of the quarks $u$, $c$, $t$ represents a filled electromagnetic vacuum. We may therefore expect the separation of the generation masses to be determined by the electromagnetic factor $\alpha$ (at some suitable energy). (The second and third generations, where this factor might be assumed to apply, notably reverse the mass ordering to isospin 'up' > isospin 'down'.) The electromagnetic connection is also obvious from the origin of this mass in the term $\boldsymbol{j}\,(\hat{\mathbf{p}}_b - \mathbf{1}\boldsymbol{\delta}_{0m})$. So the mass of $c$ is $\alpha$ times that of $t$, and the mass of $u$ is $\alpha$ times that of $c$. Possibly this applies to the quark *generations* ($u + d$, $c + s$, $t + b$), or even quark-lepton generations, rather than the individual particles, giving 179, 1.3, and $9.5 \times 10^{-3}$ GeV. The masses due to the weak isospin 'up' states, as is evident from the general formula for fermions, do not come from the perfectly random rotation, which determines the masses of all other states.

A fundamental fermion mass (probably $m_e$, via $m_f = m_e / \alpha$) seems definitely derivable from the relations between $\alpha$, $\alpha_2$ and $\alpha_3$, *without any empirical input*, but the perturbation calculations are too approximate at this stage to yield the exact value. The value produced for first order calculations using a 'unit' charge ($\alpha_3 = 1$) seems to be about 0.112 GeV (slightly above the muon mass). It is quite possible that a calculation with higher order corrections might lead to the fundamental 'unit mass' ($m_f = m_e / \alpha = 0.070$ GeV) involved in the zero-charge $SU(3)_f$ procedure (or perhaps $m_\pi = 2m_e / \alpha = 0.14$ GeV). The unit nature of the strong fine structure constant at the proposed 'unit mass' would be a natural result of the strong interaction being a completely unbroken symmetry connected with an unvarying principle of 3-D rotation – an expression of 'perfect' randomness.

Other approaches to the fermion masses are more phenomenological. The masses of the $d$, $s$, $b$ quarks certainly run as a result of the QCD coupling of the strong interaction and it is generally believed that they would become identical to the respective masses of $e$, $\mu$, $\tau$ at the energy of grand unification ($M_X$, which we have fixed at the Planck mass, $1.22 \times 10^{19}$ GeV). More specific predictions become highly model-dependent, and none has yet produced a completely self-consistent set of results. They are unlikely, I believe, to produce the *fundamental* explanations for quark masses, though they will be significant in determining their running values. One set of calculations, for instance,[31] suggests that, at some unspecified energy ($\mu$), a relationship of the form

$$\frac{m_b(\mu)}{m_\tau(\mu)} = \alpha_3(\mu)^{12/23}\ \alpha_3(m_t)^{8/161}\ \alpha_3(M_X)^{-4/7} \left(\frac{\alpha(\mu)}{\alpha(M_W)}\right)^{10/41}$$

should hold. If $m_t = 173.8$ GeV, $\mu = 182$ GeV, $M_X = 1.22 \times 10^{19}$ GeV, we obtain $\alpha_3(\mu) = 0.10827$; $\alpha_3(m_t) = 0.1088$; $\alpha_3(M_X) = 0.01908$. Also $1/\alpha(\mu) = 126.40$, $1/\alpha_2(\mu) = 31.846$, $1/\alpha_1(\mu) = 31.517$; $1/\alpha(M_W) = 127.9$, $1/\alpha_2(M_W) = 31.846$; $1/\alpha_1(M_W) = 32.018$. So

$$\left(\frac{\alpha_1(\mu)}{\alpha_1(M_W)}\right)^{10/41} \approx \left(\frac{\alpha(\mu)}{\alpha(M_W)}\right)^{10/41} \approx 1.003\ .$$

From these, we derive

$$\frac{m_b(\mu)}{m_\tau(\mu)} = 2.705 ,$$

and if $m_\tau = 1.770$ GeV, then $m_b = 4.79$ GeV. Adapting this to $m_s(\mu) / m_\mu(\mu)$, with $\alpha_3(m_c)$ replacing $\alpha_3(m_t)$, we obtain $\alpha_3(m_c) = 1/3.64$ if $m_c \approx 1.2$ GeV. Hence,

$$\frac{m_s(\mu)}{m_\mu(\mu)} = 2.832 ,$$

and, for $m_\mu = 0.10566$ GeV, $m_s \approx 0.299$ GeV. The results are reasonable, if slightly high (and interestingly close to what would result from a combination of the bare lepton mass and a contribution from $SU(3)_f$), but any decrease in $\mu$ would make them higher still. Also, for $m_d(\mu) / m_e(\mu)$, the perturbation expansion for $\alpha_3(m_d)$ becomes impossible if $m_d \approx 6 \times 10^{-3}$ GeV, as $\alpha_3$ then increases uncontrollably. A value of $\alpha_3(m_d) \approx 10^{12}$ would be required to generate the approximate ratio 6 / 0.511, which appears to apply.

## 32 The CKM mixing

The Cabibbo-Kobayashi-Maskawa mixing between generations is, of course, a significant aspect of the fermion mass problem, and it is produced by the term $\boldsymbol{k}\ \hat{\boldsymbol{p}}_c$ $(-1)^{\delta_{1g}}\ g$ in the expression for quark-lepton generation. Using the Wolfenstein parameterization, the mixing is written in the form of the matrix:

$$\begin{pmatrix} 1 - \lambda^2 / 2 & \lambda & \lambda^3 A(\rho - i\eta) \\ -\lambda & 1 - \lambda^2 / 2 & \lambda^2 A \\ \lambda^3 A(1 - \rho - i\eta) & -\lambda^2 A & 1 \end{pmatrix}$$

$\lambda$, $A$, $\rho$ and $\eta$ are defined, principally, as experimental parameters, but $\lambda$ is the Cabibbo parameter for the first and second generation mixing, and $\eta$ defines the *CP* violating phase.

The matrix in this formulation is largely empirical, but presumably has some basis in the electroweak splitting, which, according to our previous arguments, has an idealised 1 in 4 ratio. We may imagine as a working hypothesis that, ideally, the Cabibbo mixing is 1/4 for the first and second generations ($\lambda$) and 1/16 for the second and third ($\lambda^2$), and use this to devise an 'idealised' CKM matrix approximately of the form:

$$\begin{pmatrix} 1 & \lambda & 0 \\ -\lambda & 1 & \lambda^2 \\ 0 & -\lambda^2 & 1 \end{pmatrix} = \begin{pmatrix} 1 & 0.25 & 0 \\ -0.25 & 1 & 0.0625 \\ 0 & -0.0625 & 1 \end{pmatrix}$$

The CKM matrix was originally produced to derive the weak eigenstates of quarks from the mass eigenstates, but, in a fully unified theory, with parity between quarks and leptons, it is difficult to believe that it does not apply equally to leptons, especially as the electroweak mixing, the mechanism actually producing the mass, is blind to the presence or absence of the strong charge. Of course, as Halzen and Martin write of the quark matrix, 'a more involved mixing in both the $u$, $c$ and $d$, $s$ sectors can be used but it can always be simplified (by appropriately choosing the phases of the quark states) to the one parameter form'.[21] They also ask: 'Why is there no Cabbibo-like angle in the leptonic sector?' And answer: 'The reason is that if $v_e$ and $v_\mu$ are massless, then lepton mixing is unobservable. Any Cabbibo-like rotation still leaves us with neutrino mass eigenstates.'

Lepton masses, of course, unlike quark masses, are fixed, with no 'running' aspect, and so, if the CKM matrix applies to leptons, we might expect to find it in a particularly pure form, its values approaching more closely to the idealised ones. Let us suppose, therefore, that our hypothetical 'pure' matrix acts upon a set of lepton mass eigenstates $e$, $\mu$, $\tau$ to produce a mixed set of weak eigenstates $e'$, $\mu'$, $\tau'$. That is, we assume that, though there is no compulsion or mechanism for leptons to be mixed in the same way as quarks, the symmetry determining the masses of $e$, $\mu$, $\tau$ requires a set of mixed states $e'$, $\mu'$, $\tau'$, such that

$$\begin{pmatrix} 1 & 0.25 & 0 \\ -0.25 & 1 & 0.0625 \\ 0 & -0.0625 & 1 \end{pmatrix} \begin{pmatrix} e \\ \mu \\ \tau \end{pmatrix} = \begin{pmatrix} e' \\ \mu' \\ \tau' \end{pmatrix}$$

Applying the principle that the fermion masses are generated through the perfectly random rotation of $\hat{\mathbf{p}}_a$, $\hat{\mathbf{p}}_b$, and $\hat{\mathbf{p}}_c$, we might expect that the intrinsic masses of the fermions are related in some way to the constant $\alpha_3$, which provides the 'unit' mass under ideal conditions. Using the accepted values for the respective masses of $e$, $\mu$ and $\tau$ at $0.511 \times 10^{-3}$, $0.10566$ and $1.770$ GeV, we obtain the respective mass values of $e'$, $\mu'$ and $\tau'$ as $0.0269$, $0.216$ and $1.76$ GeV, with the corresponding mass ratios of $\tau' / \mu' \approx 8.1$ and $\mu' / e' \approx 8.0$. These values are essentially equal to $1 / \alpha_3$ at the energy of the electroweak splitting represented in the CKM matrix (with $\alpha_3$ possibly decreasing slightly at the higher energy required in the third generation).

So, continuing the parallel between the lepton and quark sets, we imagine that the separation between the mass values for $e'$ and $\mu'$ is determined by the 'strong' factor $\alpha_3$ (at the energy of $M_W$ - $M_Z$), with the first generation mass being $\alpha_3 \approx 1 / 8$ times that of the second, and that the same applies to the separation between the mass values for $\mu'$ and $\tau'$. Again, the connection with $\alpha_3$ occurs through the connection between the strong interaction potential and the perfectly random rotation of the angular momentum operators, rather than due to the necessary presence of any strong charge; so, perfect randomness applied to lepton angular momentum operators has the same structure as that applied to those defined for the quarks in baryons and mesons. In principle, it is the perfectly random rotation of the angular momentum states, $\hat{\mathbf{p}}_a$, $\hat{\mathbf{p}}_b$, and $\hat{\mathbf{p}}_c$, which *determines the behaviour of the strong interaction*, with its linear potential and asymptotic freedom, and the value of its fine structure constant, $\alpha_3$, and associated unit mass; and not the strong interaction which determines the rotation of the angular momentum states.

The result of the CKM calculations seems to suggest that the masses of $e'$, $\mu'$, $\tau'$ might be determined as though in a quark mixing, though there is no actual mixing between $e$, $\mu$ and $\tau$. If, then, as is highly probable, the mass of $e$ is determined uniquely in the form of $m_e / \alpha$ for 'unit charge', then the masses of $e$, $\mu$ and $\tau$ would be, in principle, determined absolutely.

Identical considerations should apply to the quarks $d$, $s$, $b$ and their CKM-rotated equivalents $d'$, $s'$, $b'$, as to the leptons $e$, $\mu$ and $\tau$. At grand unification their masses would be the same as those of the free fermions or leptons. (For example, $d$, $b$ and $s$ quark masses of something like $6 \times 10^{-9}$, 0.25 and 4 GeV would fit the same pattern.) However, at other energies, the mass values associated with $d$, $s$, $b$ and $d'$, $s'$, $b'$ would become variable, along with the fine structure constants, and, presumably, the mixing angles. The exact CKM parameters would be similar to the idealized ones but would diverge from them according to the necessity of fulfilling such conditions, from the renormalization of $\alpha_3$, as the quark masses at measurable energies being approximately 3 times the lepton masses; and of fixing the sum of fermion masses at 182 GeV.

Neutrino mixing is currently a major topic in particle physics, and has a distinct bearing upon the concept of neutrino mass, and the status of the neutrino as a 'Dirac' or 'Majorana' particle. Ideally, of course, we might expect weak mixing or oscillation between neutrino states whose charge structures are either $+w$, $-z_P k w$, or $-z_T k w$. The last two structures, in particular, look virtually identical if we realize that there is no observable difference between parity- and time reversal-violation, if no other type of measurement can be made. The structures of both neutrinos and antineutrinos as composed purely of $+w$ and $-w$ values which are indistinguishable via the weak interaction might point to Majorana-type behaviour for massive neutrinos, with left-handed neutrinos mixing with right-handed antineutrinos (and possibly an embedding of the $SU(5)$ grand unified gauge group into something like $SO(10)$). In addition, with only weak charges present, the parity and time-reversal violations required to distinguish between $-z_P k w$ and $-z_T k w$ are, in themselves, indistinguishable and suggest maximal mixing of the muon and tau neutrino states. However, neutrino observations cannot, at present, be made outside of their interactions with the other leptons; and the issue of their mixings and oscillations cannot be considered separately from the possibility of mixings between these other lepton states, and the parallels they suggest with the already-observed mixings between the quarks.

Though the observation of neutrino mixing might suggest, at first sight, the existence of 'physics beyond the Standard Model', it needs to be looked at in connection with the parallel mixing in the quark sector, where the $u$, $c$, $t$ mixing is effectively 'gauged' away. The mixing of $d$, $s$, $b$, rather than $u$, $c$, $t$, is represented as a convention in standard theory, as of course it is, but there may be a reason for the convention if we attribute the introduction of mass to the presence of $e$ alongside $w$. The fact that we need only one isospin state to be mixed must reflect the observation that only one such state has a nonzero value of *e* for any lepton or colour of quark. If one argues that the neutrinos are mixed, then one should also argue that, by comparison with the quark sector, the other leptons should also be mixed, and that, by symmetry with the quarks, we may transform away any mixing in one of the isospin states for the leptons, and, again by symmetry, this should be that of the neutrino states, which parallel *u*, *c*, *t*. Presumably, we could not tell physically whether it is neutrinos that are 'really' mixed or the other leptons. In this sense, there seems to be no reason why we could not *choose* an appropriate lepton mixing matrix to restore the mathematical basis of the Standard Model.

## 33 A summary of the mass calculations

Though the suggestions for mass calculations given in these final sections cannot be claimed to have equal status with most of the qualitative results that precede them, or even to have equal status with each other, taken together, they do provide a strategy for calculating particle masses which can be put forward as a working hypothesis, and conveniently presented as a unified approach, though the different arguments used largely stand or fall by themselves. Some of the arguments, I believe, are relatively certain – definitely the grand unification calculation, probably the masses of the composite baryons and mesons; others are strongly suggestive – the ones relating to the Higgs and weak gauge bosons and the total mass of the twelve known fermions. Some other arguments are necessarily more tentative, particularly those involving in setting up an idealised CKM matrix for predicting the weak eigenstates of leptons, but most are available to testing by experiment. So it will be useful to collect the arguments in a brief summary.

The electric, weak and strong couplings ($\alpha$, $\alpha_2$, $\alpha_3$) unify at the Planck mass, a quantity which is in effect a pure number, formed from the purely dimensional constants, $\hbar$, *c* and *G*. The electric and weak couplings are related by a factor, $\sin^2\theta_W$ = 0.25, calculable from first principles, at the energy of the intermediate bosons, *Z* and *W*, though the actual production of such massive bosons may be thought to slightly reduce the value as measured. The running of the couplings is well known from equations derived from their respective $U(1)$, $SU(2)$ and $SU(3)$ symmetries. Together with the value for $\sin^2\theta_W$, they produce absolute values for each of the couplings at any given mass-energy, with no other empirical input. The strong coupling represents a perfect gauge-invariant rotation between its three phases, and its value may be thought to become unit ($\hbar c = 1$) at some fundamental value ($m_f c^2$) of mass-energy appropriate to a 'pure' unit charge. Given such a value, the mass of the electron ($m_e$), which is assumed to be determined solely from the electric charge, may be supposed to represent a reduction to that determined by the ratio of electrical to unit coupling ($m_f \alpha$). That is, $m_f = m_e / \alpha$. The perturbation calculation from $\alpha_3$ produces a value of the right order but is not yet exact.

The masses of particles connected ultimately with vacuum states may be assumed to come from the removal of units of charge (the reverse process to creation). Imagining a bosonic state created from the zeroing of all possible charge structures (as specified in the tables A, B, C, L) suggests a possible Higgs mass of 2592 $m_e$ / $\alpha$ = 182 GeV, exactly as would be necessary if supersymmetry were implicit, as in this theory, rather than explicit. Eliminating the strong states (i.e. making A = B = C) would also indicate a purely electroweak interacting boson with maximal coupling ($Z$) with a mass (from A / L) of half this value, i.e. 91 GeV. Electroweak theory then equates $M_W$ with $M_Z \cos\theta_W$, which, using the effective value of $\sin^2\theta_W$, makes $M_W \sim 80.4$ GeV. From the standard theory, the expectation value of the vacuum Higgs field ($f$) is determined by $M_W = gf$ / 2, with $g$ the weak coupling constant expressed in charge units. Taking the charge unit as the one expected for a quark-type system with the charge divided between three phases, we obtain $f \sim 3\,M_W$ or 241 GeV, which compares well with the 246 GeV obtained empirically from the Fermi constant. In generating fermion mass by coupling the Higgs boson to the twelve known fermion states, we assume that the Higgs mass is partitioned proportionately via weak coupling so that its value is also the total value of the masses of the fermion states.

The electroweak splitting with $\sin^2\theta_W = 0.25$ is also assumed to operate, ideally, in the weak mixing between quark and lepton generations which converts mass eigenstates to weak eigenstates. So the same ratio determines the idealised Cabibbo angle in a postulated idealised CKM matrix applied to the leptons, $e$, $\mu$, $\tau$, with neutrino mixing transformed away mathematically in the same way as $u$, $c$, $t$ mixing for quarks. Starting with the electron mass, and the assumption that the weak eigenstates for the three generations are separated from each other by the factor which indicates pure coupling ($\alpha_3$), and mass generation through completely random rotation of the angular momentum operators (as assumed in making $m_f$ the fundamental mass unit), we can, in principle, derive mass eigenstates for $\mu$ and $\tau$.

The $SU(2)$ symmetry creates paired states of weak isospin. The general formula derived for fermion charge states, however, suggests that the masses relating to the splitting of the isospin states do not come from the perfectly random rotation which determines the masses of all other states, but from the creation of a full or empty 'electromagnetic vacuum'. We therefore separate the generation masses determined this way by the electromagnetic factor $\alpha$, making the quark, or, more probably the quark-lepton, generations, partition the total Higgs boson mass in this way. Taking into account values found for the lepton states $e$, $\mu$, $\tau$, we obtain 179, 1.3, and 9.5 × $10^{-3}$ GeV for the masses of the three quark generations.

The $t$ quark, at least is taken to be that required for maximal coupling to the Higgs field, $f$ / $\sqrt{2} \sim 174$ GeV, which fixes $m_b$ in the region of 4 GeV. We may also apply $SU(n)_f$ symmetries to the zero charge values produced by the less massive composite baryon and meson states, assuming that the masses of the heavier quarks are of sufficient size to create mass degeneracies within the three-quark and quark-antiquark sets of states. If $SU(4)_f$ is valid in this context, as well as $SU(3)_f$, then the second generation quark masses may be partitioned so that $m_c \sim 1.2$ GeV and $m_s \sim 0.1$ GeV. Model-dependent theories also give us information about the 'running' values of the masses of $m_b$ and $m_s$ which relate them to the values of $m_\tau$ and $m_\mu$ at certain energies. After putting all this information together, we are left with the small problem of $\sim 9.5 \times 10^{-3}$ GeV for the total mass of $u$ and $d$, with no immediate method of partitioning, except to note that, empirically, $m_d \sim 2m_u$, in the reverse proportion to their electric charges. The neutrinos are here assumed here to have relatively negligible masses, and no direct suggestions can be offered as yet on finding more exact values beyond the proposals already made by theorists.

## 34 Conclusion

The algebraic form of the Dirac equation leads to a more fundamental understanding of the symmetry breaking between the three nongravitational interactions which incorporates an explanation of the Higgs mechanism for generating the masses of fermions and bosons. The explanation is quantitative, as well as qualitative, leading to a considerable reduction in the number of arbitrary parameters in the Standard Model. In particular, exact grand unification is shown to occur at the Planck mass, by a process which can be tested by experiment, and some derivations of particle masses are suggested.

## References

1   P. Rowlands, An algebra combining vectors and quaternions, *Speculat. Sci. Tech.,* 17, 279-282, 1994.

2   P. Rowlands, Some interpretations of the Dirac algebra, *Speculat. Sci. Tech.*, 19, 243-51, 1996.

3   P. Rowlands, The physical consequences of a new version of the Dirac equation, in G. Hunter, S. Jeffers, and J-P. Vigier (eds.), *Causality and Locality in Modern Physics and Astronomy: Open Questions and Possible Solutions. Fundamental Theories of Physics*, vol. 97, Kluwer Academic Publishers, Dordrecht, 1998, 397-402.

4   P. Rowlands and J. P. Cullerne, The connection between the Han-Nambu quark theory, the Dirac equation and fundamental symmetries. *Nuclear Physics* A 684, 713-5, 2001.

5   P. Rowlands and J. P. Cullerne, The Dirac algebra and its physical interpretation, arXiv:quant-ph/00010094.

6   P. Rowlands and J. P. Cullerne, Applications of the nilpotent Dirac state vector, arXiv:quant-ph/0103036.

7   P. Rowlands and J. P. Cullerne, The Dirac algebra and grand unification, arXiv:quant-ph/0106111.

8   P. Rowlands and J. P. Cullerne, QED using the nilpotent formalism, arXiv:quant-ph/0109069.

9   P. Rowlands, The factor 2 in fundamental physics, arXiv:physics/0110069.

10  P. Rowlands, Why gravity acts instantaneously at a distance, in A. E. Chubykalo, N. V. Pope, and R. Smirnov-Rueda (eds.), *Instantaneous Action at a Distance in Modern Physics: Pro and Contra* (*A Volume in the Contemporary Fundamental Physics Series*), Nova Science Publishers, Inc., New York, 1999, 157-166.

11  P. Rowlands and B. Diaz, A universal alphabet and rewrite system, arXiv:cs.OH/0209026.

12  P. Rowlands, The fundamental parameters of physics. *Speculat. Sci. Tech.*, 6, 69-80, 1983.

13  P. Rowlands, *The Fundamental Parameters of Physics*, PD Publications, Liverpool, 1991.

14  P. Rowlands, Quantum uncertainty, wave-particle duality and fundamental symmetries, in S. Jeffers, S. Roy, J-P. Vigier and G. Hunter (eds.), *The Present Status of the Quantum Theory of Light: A Symposium in Honour of Jean-Pierre Vigier. Fundamental Theories of Physics*, vol. 80, Kluwer Academic Publishers, Dordrecht, 1997, 361-372.

15  P. Rowlands, A foundational approach to physics, arXiv:physics/0106054.

16  P. Rowlands, J. P. Cullerne and B. D. Koberlein, The group structure bases of a foundational approach to physics, arXiv:physics/0110092.

17  T. T. Takahashi, H. Matsufuru, Y. Nemoto, and H. Sugunama, H., *Phys. Rev. Lett.*, **86**, 18, 2001.

18  S. J. Brodsky, J. Ellis, and M. Karliner, Chiral symmetry and the spin of the proton, *Physics Letters*, **B 206**, 309-15, 1988.

19  H. Georgi and S. L. Glashow, Unity of all elementary-particle forces, *Phys. Rev. Lett.*, **32**, 438-41, 1974.

20  I. J. R. Aitchison and A. J. G. Hey, *Gauge Theories in Particle Physics*, second edition, Adam Hilger, 1989.

21. F. Halzen and A. D. Martin, *Quarks and Leptons*, John Wiley, 1984.

22  M. Kaku, *Quantum Field Theory*, Oxford University Press, 1993.

23  M. Y. Han and Y. Nambu, Three-Triplet Model with Double SU(3) Symmetry, *Phys. Rev*, **B 139**, 1006, 1965.

24  F. E. Close. *An Introduction to Quarks and Partons*, Academic Press, 1979.

25  R. B. Laughlin, *Phys. Rev. Lett.*, **50**, 1395, 1983.

26  P. W. Anderson, *Science*, **177**, 393, 1972.

27  R. B. Laughlin, *Rev. Mod. Phys.*, **71**, 1395, 1999.

28  M. E. Peskin and D. V. Schroeder, *Introduction to Quantum Field Theory*, Addison-Wesley, 1995.

29  P. Rowlands and J. P. Cullerne, Can gravity be included in grand unification?, in R. L. Amoroso, G. Hunter, M. Kafatos and J-P. Vigier (eds.), *Gravitation and Cosmology: From the Hubble Radius to the Planck Scale*, Kluwer, 2002.

30  S. Weinberg, *The Quantum Theory of Fields*, 2 vols., Cambridge University Press, Vol. II, 327-32, 1996.

31  U. Amaldi, W. de Boer, and H. Fürstenau, *Phys. Lett. B*, **260**, 447, 1991.

32. C. Kounnas. Calculational schemes in GUTs, in C. Kounnas, A. Masiero, D. V. Nanopoulos, and K. A. Olive. *Grand Unification with and without Supersymmetry and Cosmological Implication*, World Scientific, 145-281, 1984.

33 D. T. R. Jones, personal communication, 2000.

34 J. Ellis and D. Ross, A light Higgs Boson would invite Supersymmetry, arXiv:physics-p/0012067.

35 C. Ford, D. T. R. Jones, P. W. Stephenson, and M. B. Einhorn, The effective potential and the renormalisation group, *Nuclear Physics*, **B 395**, 17-34, 1993.